\documentclass{article}

\usepackage{arxiv}

\usepackage[utf8]{inputenc} 
\usepackage[T1]{fontenc}    
\usepackage{hyperref}       
\usepackage{url}            
\usepackage{booktabs}       
\usepackage{amsfonts}       
\usepackage{nicefrac}       
\usepackage{microtype}      
\usepackage{lipsum}
\usepackage{tikz}
\usepackage{multicol}
\usepackage{ulem}

\usepackage{cancel}
\newtheorem{Def}{Definition}
\newtheorem{pro}{Proposition}
\newtheorem{cor}{Corollary}
\newtheorem{lem}{Lemma}
\newtheorem{thm}{Theorem}
\newtheorem{ex}{Example}
\usepackage{xcolor}
\usepackage[font={itshape,small}]{quoting}
\usepackage{lipsum}
\usepackage{babel}
\newtheorem{proof}{Proof}
\newtheorem{rem}{Remark}

\usepackage{color}   
\usepackage{hyperref}
\hypersetup{
    colorlinks=true, 
    linktoc=all,     
    linkcolor=blue,  
}

\usetikzlibrary{arrows}
\tikzset{
  treenode/.style = {align=center, inner sep=0pt, text centered,
    font=\sffamily},
  arn_n/.style = {treenode, circle, white, font=\sffamily\bfseries, draw=black,
    fill=black, text width=1.5em},
  arn_r/.style = {treenode, circle, red, draw=red, 
    text width=1.5em, very thick},
  arn_x/.style = {treenode, rectangle, draw=black,
    minimum width=0.5em, minimum height=0.5em}
}

\title{Point-dimension theory (part I):\\ The point-extended box dimension}

\author{Nadir Maaroufi\\ 
  TICLab, College of Engineering and\\ Architecture,
  Université Internationale de Rabat\\
  Sala Al Jadida, Maroc\\
  \texttt{nadir.maaroufi@uir.ac.ma} \\
   \And
 El Hassan Zerouali \\
  Faculté des sciences de Rabat\\
  Université Mohammed V\\
  Rabat, Maroc \\
  \texttt{zeroualifsr@gmail.com} \\
}

\begin{document}

\maketitle

\begin{abstract}\footnote[1]{We would like to extend our thanks to the College of Engineering and Architecture of the International University of Rabat for supporting this project through one-year teaching discharge granted to professor Maaroufi Nadir to finalize the drafting of this article.}
This article is an introductory work to a larger research project devoted to pure, applied and philosophical aspects of dimension theory. It concerns a novel approach toward an alternate dimension theory foundation: the point-dimension theory. For this purpose, historical research on this notion and related concepts, combined with critical analysis and philosophical development proved necessary. Hence, our main objective is to challenge the conventional zero dimension assigned to the point. This reconsideration allows us to propose two new ways of conceiving the notion of dimension, which are the two sides of the same coin. First as an organization; accordingly, we suggest the existence of the \textit{Dimensionad}, an elementary particle conferring dimension to objects and space-time. The idea of the existence of this particle could possibly adopted as a projection to create an alternative way to unify quantum mechanics and Einstein's general relativity. Secondly, in connection with Boltzmann and Shannon entropies, dimension appears essentially as a comparison between entropies of sets. Thus, we started from the point and succeeded in constructing a point-dimension notion allowing us to extend the principle of box dimension in many directions. More precisely, we introduce the notion of point-extended box dimension in the large framework of topological vector spaces, freeing it from the notion of metric. This general setting permits us to treat the case of finite, infinite and invisible dimensions. This first part of our research project focuses essentially on general properties and is particularly oriented towards establishing a well founded framework for infinite dimension. Among others, one prospect is to test the possibility of using other types of spaces as a setting for quantum mechanics, instead of limiting it to the exclusive Hilbertian framework. \end{abstract}


\newpage
\tableofcontents

\section{Introduction}

It is indisputable that Henri Poincaré was one of the greatest mathematicians of history. It could hardly be put better than Emile Picard when he wrote in \cite{picard}:
\begin{quoting}[font+=bf,begintext= ,endtext=]
\textit{Henri Poincaré qu'il ne fut pas seulement un grand mathématicien, mais la Mathématique elle-même.}
\end{quoting}
Many Poincaré's articles were particularly concerned with geometry foundation and a profound conception of space. Before coming, in one of his latest articles \cite{poincare1} entitled "Space and its three dimensions", to the conclusion that the common point between all geometries is the more difficult notion of the dimension: 
\begin{quoting}[font+=bf,begintext= ,endtext=]
\textit{Dans les articles que j'ai précédemment consacrés à l'espace, j'ai surtout insisté sur les problèmes soulevés par la géométrie non-euclidienne, en laissant presque complètement de côté d'autres questions plus difficiles à aborder, telles que celles qui se rapportent au nombre des dimensions. Toutes les géométries que j'envisageais avaient ainsi un fond commun, ce continuum à trois dimensions qui était le même pour toutes...}
\end{quoting}
Furthermore, in his last article \cite{poincare2} entitled "Why does space have three dimensions?", he seems to suggest that a good definition of the notion of dimension should be considered adequate for both the mathematician and the philosopher: 
\begin{quoting}[font+=bf,begintext= ,endtext=]
\textit{Ces définitions, irréprochables, nous l'avons dit, au point de vue mathématique, ne sauraient satisfaire le philosophe.}
\end{quoting}
Pavel Urysohn, a great mathematician who died prematurely at the age of 26, goes further than Poincaré in asserting in \cite{ury1} that the problems related to the definition of dimension have higher philosophical interest:
\begin{quoting}[font+=bf,begintext= ,endtext=]
\textit{Il est d'ailleurs à remarquer que l'intérêt purement mathématique de ces problèmes est bien moindre que leur intérêt philosophique (qui est très considérable): tant qu'on a pas trouvé de bonne définition, on emploie, et avec succès, la mauvaise qui est logiquement irréprochable.}
\end{quoting}
Hence, the main objective of this ambitious article is to combine philosophy and mathematics to contribute, as much as possible, to the enlightenment of the dimension notion. We are well aware that this approach may not be to the taste of some readers. Accordingly, we wanted to formulate a mathematical part (see section \ref{5}) that can be read independently from the rest of the article. We completely agree with Claude Tricot when he denounces the dictatorship of Mathematical Logic in \cite{tri} by talking about the denigration of Georges Bouligand's works because of his style:
\begin{quoting}[font+=bf,begintext= ,endtext=]
\textit{Avec son style littéraire, il a été généralement accusé de manque de rigueur. On peut se demander maintenant si c'est un péché si grave, après un demi-siècle de dictature de la Logique mathématique, au détriment de la créativité.}
\end{quoting}

The other ambition of the present article is to include physics in the reflection and the possible prospects because physics is also concerned in the first place with the notion of dimension. In particular, this work can be considered in the opposite direction of the quote of the great physicist Richard Phillips Feynman pronounced during his lecture at the \textit{Messenger Lectures} in 1965:
\begin{quoting}[font+=bf,begintext= ,endtext=]
\textit{Mathematicians are only dealing with the structure of the reasoning and they do not really care about what they're talking about. They don't even need to know what they're talking about, as they themselves say, or whether what they say is true...The physicist has meaning to all the phrases...}
\end{quoting}

Nevertheless, we must point out that this statement is not valid for the pre-twentieth century mathematicians and even for some mathematicians after the 20th century like, John Von Neuman. In order to avoid any confusion of the scope of our thought, we have to specify our purpose more succinctly. We are not saying that mathematics must necessarily have as an objective and finality the servitude of other sciences like physics. It is, in our opinion, indispensable that mathematicians continue to explore the mathematical field guided only by the abstraction of the mathematical reasoning and logic. Nevertheless, they must keep in mind what Kurt Godel taught us through one of the consequences of his first incompleteness theorem of 1931. Roughly speaking, mathematical axiomatic theories can not be perfect since any axiomatic theory (capable of formalizing arithmetic) cannot be both complete and consistent. We are simply arguing that the one, does not exclude the other; with respect to the question: is it still possible to use something other than logic and mathematical reasoning to think in a mathematical article even if it is pure? Our answer to this question is, without any hesitation, Yes. We have thought for a long time before deciding on the necessity to present this work in this form. One of the reasons that made us take this step is the risk analysis. Indeed, what do we risk to present this work in this way? The greatest risk would be the \textit{indifference}, which we do not hope for, but at least we are determined to go all the way. While the least would be the \textit{criticism}, which would be beneficial for the debate and possible improvements.

In the same speech, Feynman said about physics and physicists:
\begin{quoting}[font+=bf,begintext= ,endtext=]
\textit{But the physicist has meaning to all his phrases... In physics, you have to have an understanding of the connection of words with the real world.}
\end{quoting}

We can oppose the current popular formula adopted in quantum mechanics: \textit{"Shut up and calculate"} due to David Mermin, to the above quote. An equivalent motto in fundamental mathematics should be: \textit{"Shut up and demonstrate"}. In this work we want to free ourselves from this obsession of silencing, we request the right to the use of language as an indispensable vector of human thought. We do not say that we should not calculate or demonstrate, or that discussion can in any case replace mathematical demonstration or calculation. Nevertheless, the necessity of demonstrating and calculating does not exclude in any way the need and the interest of discussing and philosophizing,  as we have already started to undertake in \cite{maar1}. Of course, more intention must be taken to avoid thoughtfully any metaphysical reference that has nothing to do with science in general and with mathematics in particular. In summary, we take a stand against the famous sentence of the great physicist Stephen Hawking: \textit{"Philosophy is dead"}. In our opinion, philosophy cannot be dispensed with, for by conveying thought in language, creativity is encouraged and boosted. Indeed, the very structure of the language and its handling favors inventiveness through analogies, associations of ideas and metaphors, for example. 

Admittedly, ever since Poincaré there have been many developments concerning the dimension theory; nevertheless, we strongly believe that it remains much more to do regarding dimension. In reality, even if this notion seems at first sight intuitive and therefore accessible, in looking a bit closer it becomes very quickly complex to formalize, quantify and interpret. Moreover, dimension plays a central role not only in mathematics, but also in many other disciplinary fields. There is no doubt that the need of a better understanding and an adequate use of this notion is particularly felt today.  For example, string theories assume extra-dimensions of space-time in order to be mathematically consistent and tractable, 26 dimensions in bosonic string theory, 11 dimensions in $M$-theory and 10 dimensions in superstring theory. These extra-dimensions are still not experimentally observed, which simply means that all existing theories trying to explain these dimensions remain speculative. Another basic example concerns non-integer dimensions, which completely escapes our spatial intuition. There are many other examples such that \textit{Curse of dimensionality} and also its opposite \textit{Blessing of dimensionality} in data science or even some arising dimension problems in biology.

We do not presume through this labour to completely elucidate these issues; however, we aspire to put our shoulder to the wheel and contribute by investigating other ways of looking at things. Historically speaking, one can be certain; however, that every time researches into dimension notion not only initiated deep reflections but also allowed creating new scientific fields. For example, it can reasonably be argued that there would have been no Einstein's \textit{general relativity} theory without the mathematical framework and the philosophical debates concerning the counter-intuitive $n$-dimensional ($n>3$) space, see \cite{fra}, \cite{sw}, \cite{dur} and \cite{pat}. In particular, at the time, the four-dimensional geometry was widely investigated by mathematicians. Another example is the emergence and development of the \textit{point-set topology} (\cite{john}, \cite{john1}, \cite{john2}, \cite{tay}, \cite{tay1}, \cite{arbo}, \cite{arbo1} and \cite{dur}). Actually, great mathematicians renowned for having significantly participated to topology growth such as Bernard Bolzano, Bernhard Riemann, George Cantor, Henri Poincaré, Giuseppe Peano, Henri-Léon Lebesgue, Frigyes Riesz, Hermann Minkowski, Felix Hausdorff, Luitzen Egbertus Jan Brouwer, Waclaw Sierpinski, René Maurice Fréchet, Kazimierz Kuratowski, Georges Bouligand, Heinrich Franz Friedrich Tietze, Lev Pontryagin, Abram Besicovitch, Pavel Aleksandrov, Pavel Urysohn, Karl Menger, Eduard \v{C}ech, Stefan Banach, Andrey Kolmogorov, Ryszard Engelking, Claude Tricot, Kenneth Falconer and many others, were widely involved with dimension notion works. We will analyze in the sequel several important and key contributions from some of them.

The main contributions of this article concern several aspects. First, a deep and critical analysis of the history of the dimension notion. Hence, we have managed to investigate part of both the philosophical and mathematical history of this notion. Secondly, with this analysis in mind, we have developed our own approach to the notion of dimension which we have called point-dimension theory. This allowed us in particular to suggest the existence of an elementary particle conferring dimension to objects and space-time that we named Dimensionad. Also, to propose a new conception of the mathematical space. As well as to establish a clear link with the notion of entropy. Thirdly, to introduce a new mathematical definition of the notion of dimension starting from the point. This notion allows to find the classical dimensions of some mathematical sets like vector spaces. It also permits to conceive more twisted sets, where the dimension depends on the position. Moreover, it offers an adequate framework to develop the theory of dimension in infinite dimension. This article is organized as follows: section \ref{glance} will deal with the history of the notion of dimension until Cantor. Section \ref{over} will present an overview of the different notions of dimension introduced in the twentieth century. In section \ref{pointdim} we will present our approach and the point-dimension theory. Our definition and the mathematical developments will be detailed in section five \ref{5}. While in the last section \ref{persp} we will have a discussion about the mathematical, philosophical and physical prospects of this first work. 

\section{A glance at the history of dimension}\label{glance}
Dimension is a fundamental concept involved in many theories and scientific fields. The \textit{nature} of such notion must be continuously questioned taken into consideration scientific progress and philosophical thought evolution. For this purpose a holistic view is required. Therefore, a meticulous and critical scrutiny of its development throughout history is needed beforehand. Obviously, there exist several detailed research articles and books on dimension history (see \cite{john}, \cite{john1}, \cite{john2}, \cite{whi} and \cite{jam} for example). We nonetheless wish to express our own perspective, interpretation and reflections regarding its progress. In this section we present our reading of dimension history without pretending to be exhaustive. 

\subsection{Geometrically intuitive but hardly discernible}\label{2.1}
\subsubsection{Euclide ($\approx$300 BC)}\label{2.1.1}
 Monuments from ancient civilizations, like Egyptian pyramids, reflect a real empirical and technical know-how in geometry. It is not surprising that geometry should surely be one of the earliest knowledges in mathematics, since we can reasonably argue that this last relies in part on our intuition and sense experience. One of the oldest fairly formalised, rigorously built and sufficiently complete mathematics which come down to us is \textit{classical Greek geometry}, is based on the study of plane and solid figures. It is well known that \textit{Euclid's Elements} are the major and most influential textbooks gathering Greek geometry knowledge. In view of Euclid's definitions in the Elements we can draw interesting conclusions in link with dimension notion. First of all, let us recall Euclid's definitions in connection with dimension.
\begin{multicols}{2}
\centering{Book I (see \cite{he1})} 
\begin{enumerate} 
\item A point is that which has no part.
\item A line is breadthless length. 
\item The extremities of a line are points.
\setcounter{enumi}{4}
\item A surface is that which has length and breadth only. 
\item The extremities of a surface are lines. 
\end{enumerate}  

\centering{Book XI (see \cite{he2})} 
\begin{enumerate}
    \item A solid is that which has length, breadth, and depth.
 \item An extremity of a solid is a surface.
\end{enumerate}
\setcounter{unbalance}{2}
\end{multicols}
These definitions introduce the basic Euclidean geometric entities such as \textit{point}, \textit{line}, \textit{surface} and \textit{solid}. It turns out that these same definitions are also considered the oldest definitions of dimensions 0, 1, 2 and 3. In other words, there is a kind of closed loop between the definitions of these fundamental geometric objects and those of dimensions. The line define the 1-dimensional and conversely, the surface define the 2-dimensional and vice versa... That is, the distinction between a line and a surface is done through their dimensions, and inversely. Philosophically speaking, we can claim that dimensions 0, 1, 2 and 3 could be considered as the \textit{essence} of the point, line, surface and solid respectively. Most intriguing is that, as mentioned in \cite{john}, \textit{Euclid's Elements} does not include any dimension notion explanation or even a general term that designates the concept of dimension. At this stage, there are two main findings. On the one hand, we can conclude that any attempts of a rigorous geometry formalisation or conceptualisation requires at least, even implicitly, an intuitive dimension notion. One can even advance with saying there can not be mathematics development without the dimension notion. On the other hand, despite its intuitive character, we can stress the difficulty to discern or extract the notion of dimension, which can be regarded as the main obstacle to formalize and capture such a concept.   

First of all, we want to mention that several great names in modern dimension theory like Poincaré and Menger (see \cite{poincare1} and \cite{john}), there may be many others, have surely reviewed the different Greeks definitions of dimension and based their own conceptions on these last. Indeed, looking these definitions more closely, we can stress that they already carry within themselves the seeds of two very modern ways of conceptualising dimension notion, which will be presented in the sequel: \textit{measured} and \textit{topological} dimensions. Let's start by examining the first type which we shall call the measured dimensions definitions. More precisely, we can see definitions I.1, I.2, I.5 and XI.1 as a straightforward link between measure and dimension notions since, according to these definitions, a geometrical object's dimension is completely indexed on the fact of possessing a length (L) or a length and width (area, L$\times$W) or a length, width and height (volume, L$\times$W$\times$H). In modern mathematical language, "A line is breadthless length" it can be seen as: the 1-dimensional measure of a line is different from zero while its 2-dimensional measure is equal to zero. In some way, this is the same principle to evaluate the Hausdorff dimension $s$ of a set $X$ since it will be totally based on the $s$-dimensional Hausdorff measure of this last \cite{hau}.           

At this stage, we want to shed light on a shortcut which confers some persisting confusions concerning dimension notion. More precisely, from definition I.1: "A point is that which has no part", we traditionally deduce that a point is zero-dimensional. This must surely provide the following argumentation: having no part basically implies possessing neither length, width and height, nor length and width, nor only length and then its dimension shall be zero. But, this does not logically mean having zero dimensional, according to the previous definitions of dimensions. It only implies not possessing neither three dimensional, nor two, nor one. In other words, there is a confusion between having no dimension and possessing zero dimension. Saying that the point is zero dimensional is attributing a dimension to it, which is contradictory with not having dimension. In any case, this reasoning shows the existence of some ambiguity of attributing a dimension to a point from the beginning. In our conception, the dimension of the point depends on the geometric object to which it belongs; its dimension is therefore variable. More generally, having a null measure certainly doesn't imply possessing zero dimension. For example, the Cantor set possess neither length nor width nor height since its Lebesgue measure is equal to zero, while its Box or Hausdorff dimension are different from zero. We can naturally argue that it is a matter of adequate measure since the Hausdorff measure of the Cantor set is not zero. More deeply, this raises the fundamental following question: should an adequate dimension notion whether or not be indexed on a measure notion?        

\subsubsection{Aristotle (384-322 BC)}\label{2.1.2}
Now let's turn to the second type which we shall call the topological dimension definitions. To begin with, there are definitions I.3, I.6 and XI.2 mentioned above which can be seen as a kind of "recursive" definition of dimension, it surely occurred involuntary since the Greeks had never defined anything recursively (see \cite{john}). These definitions are very interesting for the reason that they define the basic Euclidean geometric entities by linking them together via their extremities (or boundaries in reference to topology). Aristotle completed these definitions by adding to extremities, sections, divisions and limits, see \cite{hea}:   
\begin{quoting}[font+=bf,begintext= ,endtext=]
\textit{If we suppose lines or what immediately follows them (I mean the primary surfaces) to be principles, these are at all events not separable substances but are sections and divisions, the one of surfaces, the other of bodies (as points are of lines); they are also extremities or limits of the same things; but all of them subsist in other things, and no one of them is separable.}
\end{quoting}
This is relevant since these definitions no longer make sense when we consider geometrical objects without extremities such closed lines (as a circle) or closed surfaces (as a sphere). Of course, in these cases, it is more suitable to consider sections or divisions because their extremities are empty sets.
\begin{rem}
We would point out that, in this passage Aristotle stresses that all basic euclidean geometric objects (points, lines and surfaces) are not separable substances. But, they can be considered as sections, divisions, extremities and limits. We can interpret this distinction as follows: Aristotle accepts these entities in \textbf{potency} but not in \textbf{act}. In other words, he consents the formal existence of these entities abstractly, but he intuitively refuses their existence as real substances dislocated from a body. We will develop this idea more deeply in the subsection \ref{4.2.1}.             
\end{rem}
This second type of Greeks definitions can naturally bring about an inductive definition of dimension: starting with attributing dimension three to solids, a surface should have two dimensions (3-1) as a solid's extremity, a line is one-dimensional (2-1) as a surface's extremity and a point is zero-dimensional (1-1) as a line's extremity. This approach sounds like many modern definitions (see subsection \ref{3.1}). Indeed, modern topological definitions of dimension start from Poincaré's idea in his article \cite{poincare1}, which was taken up by Luitzen Egbertus Jan Brouwer to develop in \cite{brou} his definition named "Dimensiongrad" becoming in its final version the "Large inductive dimension" \cite{eng} also called Brouwer-\v{C}ech dimension \cite{cech}. In the same spirit, the "Small inductive dimension" \cite{eng} was formulated independently by Pavel Urysohn \cite{ury} and Karl Menger \cite{men}.            

Nevertheless, Euclid's definitions were criticised by both Aristotle \cite{hea} and Bernard Bolzano \cite{john} for the same reason. Their disagreement with concerns the ordering of the concepts, since they are top down definitions from solid to point. More precisely, a point is designed as the extremity of a line, which itself is the extremity of a surface, which itself is the extremity of a body. As it was mentioned in \cite{hea}: "All these definitions explain the prior by means of the posterior". On another side, modern topological definitions of dimension like the Large and the Small inductive dimensions avoided this issue by well ordering things. But, this seems to require attributing a dimension to the empty set ($\emptyset$) in order to well initialise the inductive definition. For example, regarding Euclid's definitions, to ensure that one must start from a point with zero-dimensional one have to begin by assigning minus one as dimension to the empty set ($dim(\emptyset)=-1$); this provides dimension zero (-1+1) to the point since its extremity is an empty set. And so on, a line is one-dimensional since its extremity is a point (0+1=1)... There is no reasonable argument in favor of assigning dimension minus one to the empty set. Of course, we should accept that as a convention, only once again, in our opinion, this conceals some ambiguity of attributing a dimension to a point. Especially since initializing the definition by the point does not really pose a problem, as we have done for the modified definition \ref{mbdim} of Bolzano. In any case, a second time we can formulate the fundamental following question: should an adequate dimension notion whether or not be indexed on topology?  

In addition, we want to highlight some other aspects in relationship with Greek conceptions of the dimension notion. First of all, although Aristotle (384-322 BC) criticized Euclid's dimension definitions, he lived long before the appearance of Euclid's Elements ($\approx$300 BC). In fact, the mathematics contained in Euclid's elements were developed well before Euclid itself \cite{bernard}. In reality, the Elements is the most complete and well organised textbook bringing together Greek's geometry and arithmetics. It therefore makes sense that Aristotle knew dimensions definitions. Furthermore, Aristotle provides his own definition of dimension in "On the Heavens" Book I, Part 1, see \cite{ari}:     
\begin{quoting}[font+=bf,begintext= ,endtext=]
    \textit{Now a continuum is that which is divisible into parts always capable of subdivision, and a body is that which is every way divisible. A magnitude if divisible one way is a line, if two ways a surface, and if three a body. Beyond these there is no other magnitude, because the three dimensions are all that there are, and that which is divisible in three directions is divisible in all....But if it is divisible in three dimensions it is every way divisible, while the other magnitudes are divisible in one dimension or in two alone: for the divisibility and continuity of magnitudes depend upon the number of the dimensions, one sort being continuous in one direction, another in two, another in all.}
\end{quoting}
In this passage Aristotle uses clearly the term "dimension", in Greek he wrote "$\Delta\iota\acute{\alpha}\sigma\tau\alpha\sigma\eta$" which is generally called the distance between two boundary points. It is derived from the verb "$\delta\iota\acute{\iota}\sigma\tau\eta\mu\iota$", which in ancient Greek means: place separately or to separate. To the best of our knowledge, it is the first apparition of a term designating the concept of dimension. It may be observed that the confusion between measure and dimension already exists at the origin of its naming. Otherwise, it is almost certain through the Elements that Euclid was widely influenced by Aristotle's thought (see \cite{sw}). The question becomes, why doesn't Euclid designate the concept of dimension by its name? Assuming that Euclid was surely aware of the parts where Aristotle speaks about dimension, only two options remain. Either Euclid was not capable of discerning or capturing this notion and thus he does not name it, or he had not devoted more attention to this concept because he probably thought that it is unnecessary to distinguish it in mathematics since he likely judged, at the time, that such concept was more philosophical than mathematical. We must remember that Aristotle was and still is considered more as a philosopher than a mathematician. Of course, all these hypotheses remain purely speculative. 

We want to add three important remarks. Firstly, we must insist that for Aristotle the notion of dimension concerns exclusively bodies ("On the Heavens" Book I, Part 1 \cite{ari}) and not the space or the spatial extent since these last made no sense for ancient Greeks. Secondly, he clearly stressed that there may not be more than three dimensions.Thirdly, according to Aristotle's passage given above (and "On the Heavens" Book I, Part 1 \cite{ari}), one can observe that its dimension definition was not accurate and not exploitable, in particular what he intend by 'directions'. But, we can remark that trying to provide a definition of dimension he, naturally in our view, attempted to define the continuum. The same process, in some sense, can be clearly identified in the mathematician and philosopher Bolzano \cite{john} and many others. We will return to this point later on.      
\subsubsection{Nicomachus of Gerasa ($\approx$100 AD)}\label{2.1.3}
To finish this subsection we move to approximately four hundred years after Euclid's Elements. One could reasonably say that the mathematician and philosopher Nicomachus of Gerasa have surely drawn conclusions from both Aristotle's work and Euclid's Elements in his book \textit{Introduction to Arithmetic} ($\approx$100 AD) (see \cite{gu}, \cite{john1}):
\begin{quoting}[font+=bf,begintext= ,endtext=]
    \textit{
...Indeed, when a point is added to a point, it makes no increase, for when a non-dimensional thing is added to another non-dimensional thing, it will not thereby have dimension. ... Unity, therefore, is non-dimensional and elementary, and dimension first is found and seen in 2, then in 3, then in 4, and in succession in the following numbers; for 'dimension' is that which is conceived of as between two limits. The first dimension is called 'line', for a line is that which is extended in one direction. Two dimensions are called 'surface', for a surface is that which is extended in two directions. Three dimensions are called 'solid', for a solid is that which is extended in three directions... The point, then, is the beginning of dimension, but not itself a dimension, and likewise the beginning of a line, but not itself a line; the line is the beginning of surface, but not surface, and the beginning of the two-dimensional, but not itself extended in two directions. Naturally, too, surface is the beginning of body, but not itself body, and likewise the beginning of the three-dimensional but not itself extended in three directions. Exactly the same in numbers, unit is the beginning of all number that advances unit by unit in one direction; linear number is the beginning of plane number, which spreads out like a plane in more than one dimension; and plane number is the beginning of solid number, which possesses a depth in the third dimension besides the original ones.}
\end{quoting}
For Nicomachus of Gerasa the number of dimensions can be deduced according to the minimum points required to obtain the basic geometrical figures. Hence, one dimension can be associated with the line which we can draw between two independent points, two dimensions is associated to the surface delimited by a triangle drawn from three independent points and finally three dimensions is associated to the volume delimited by the tetrahedron drawn from four independent points. Like Aristotle, he designates the concept of dimension by its name and he used 'directions' without more explanation. As Euclid's Elements, he consider a sort of inductive definition using the extremities of different geometrical objects and looking more closely we can capture a notion of measure behind its definition of dimensions. He clearly argues that a point does not have dimension. Nevertheless, we cannot decide if that is essentially due to his argument presented below or if that is more generally due to the ancient Greeks doubt concerning the status of zero as a number. In reality, attributing zero dimension to a point appeared long after ancient Greek mathematicians. Finally, trying to establish a link between the arithmetic progression and the different dimensions, the reasoning of Nicomachus of Gerasa seems to be the first in history to make us think about dimensions greater than three.  
\begin{rem} The point's definition in Euclid's Elements, "that which has no part" can also work for the empty set. The definition should be "A point is a non-empty set that has no parts". But, the notion of empty set also didn't have any meaning at the time.      

\end{rem}

\subsection{A counter-intuitive extension forced by the arising algebraic formalism}\label{2.2}

To the best of our knowledge, since Greek's dimension definitions, we could not notice any significant progress of this notion until the advent of Cartesian coordinate system. We strongly believe that the dimension notion has always suffered from its intuitive character. The latter gives the impression of naturally capturing this notion of, which pushes one to be satisfied with existent definitions and any new investigations remain superfluous. In reality, all developments of the dimension notion until the twentieth century, apart from Bolzano's contribution, were simple consequences of other arising formalisms and theories and never been done for its own interest. As we will discuss in the forthcoming subsection, it was only after Cantor's discovery that the need arose to establish a real theory of dimension, since former conceptualisations seem to have been passed by the new arising questions. Now, let us turn to the analysis of Descartes' contribution to the progress of dimension notion and the scrutiny of some consequences of his new Cartesian coordinate system.    

\subsubsection{René Descartes (1596-1650)}\label{2.2.1}
Nowadays, Cartesian coordinate system seems to be very natural, but it should be kept in mind that its introduction by Descartes comes subsequent to one of the most famous and influential philosophical text in the history: "Discours de la méthode". Furthermore, it is indisputable that Cartesian coordinate system has brought mathematics in particular and science in general to a new era, since it has permitted developing \textit{Analytic Geometry} used in many scientific fields. In fact, by his system, Descartes succeeded in creating a fruitful communion between geometry and algebra, which in particular gave rise to modern mathematical analysis. For example, it is well known that Leibniz and Newton were largely influenced by "La Géométrie" published as an appendix to "Discours de la méthode" where Descartes introduced Cartesian coordinate system as one of a practical examples of the successful achievement of his philosophical methodology. Thus, infinitesimal calculus developed after and was assigned to both Leibniz and Newton (independently) who were widely impacted by this system. Moreover, Descartes pointed the way forward by combining different mathematical fields to reach an unexpectable wealth such that we can observe in Poincare's Algebraic Topology for instance.

In fact, looking more closely we can claim that Cartesian coordinate system is the fruit of combining two ingredients. On the one hand,  the use of the philosophical methodology and rules described in "Discours de la méthode". On the other hand, Descartes' conception of \textit{space} arose from his philosophical principle \textit{Res Extensa}. In fact, Descartes’s contribution represents a profound paradigm shift from Aristotle’s conception, since he distinguishes matter (substance) forming the body from its spatial extent. This way of thinking allowed him to retain from a body only its geometrical figure \cite{bar}, \cite{mor} and \cite{pat}. For Descartes, the \textit{extent} is an intuitive notion emerging from our sense experience, it is the main attribute of bodies \cite{gab} and even the matter's essence. Descartes takes his reasoning a step further reaching an even more abstract conceptualization by conceiving the no limited three dimensional continuum: the notion of \textit{mathematical space} is born! By his coordinate system, Descartes succeeded in creating an abstract but a quantitative geometrical space where all bodies are reduced exclusively to their geometrical extent which will be represented by pure quantities or numbers (coordinates), permitting in particular the mathematization of physics.              

In Descartes' construction, dimensions are of particular importance because they are related to the body's geometrical extent, i.e. its length, width and height. More generally, dimension is for Descartes essentially associated with the way in which a subject can be measured \cite{pat}. In this sense, for a geometrical object, Descartes' position closely resembles the Greek's measured dimensions definitions discussed in the previous subsection. Furthermore, the axes of Cartesian coordinate system permit the creation a framework for what Aristotle and Nicomachus of Gerasa perhaps would have meant by 'directions' (see the previous subsection). Descartes' divergences with Greek's conception of dimension concern essentially two aspects. First, dimensions are not exclusively confined to bodies since they can be associated with the whole ambient space in particular and an abstract mathematical space in general. The dimension of a mathematical space is an idea which will be expanded in many algebraic formalisms much later, as vector spaces for instance. Secondly, reducing dimensions to the number of parameters or coordinates allowed Descartes to think about dimension outside of the sole spatial framework. Dimensions can be associated with all that is measurable, such as velocity,  weight, times... This permits the extraction of the dimension notion from the exclusive geometrical field bringing it also into algebra.              

This abstraction forces to begin thinking about the counter-intuitive possibility of dimensions higher than three. Indeed, in this setting, dimension is indexed to the number of coordinates which may be arbitrarily large. But, taking this step turned out to be more difficult, perhaps at the psychological level because mathematicians must now do without geometrical representation due to these higher dimensions. We must remember that in mathematics, Descartes intends to apply his Method to geometry with the support of the algebraic language. As it was exposed in \cite{char} with reference to \cite{shea}, in Descartes' mind, it certainly does not mean the supremacy of algebra over geometry since Descartes was above all a geometer. More precisely, algebra is essentially a tool permitting to clarify geometrical reasoning. In fact, Descartes began with transforming his geometrical problem into algebraic language to get back after to a geometrical resolution. Therefore, geometrical representation is crucial in this process. This means that the lack of geometrical representation in higher dimensions surely braked rigorous mathematical development of these dimensions for more than two centuries. This view is supported by the beginning of Camille Jordan's article entitled "Essai sur la géométrie à $n$ dimensions" published in 1875 \cite{jor}:   

\begin{quoting}[font+=bf,begintext= ,endtext=]
\textit{On sait que la fusion opérée par Descartes entre l'algèbre et la géométrie ne s'est pas montrée moins féconde pour l'une de ces sciences que pour l'autre. Car, si d'une part les géomètres ont appris, au contact de l'analyse, a donner à leurs recherches une généralité jusque-là inconnue, les analystes, de leur côté, ont trouvé un puissant secours dans les images de la géométrie, tant pour découvrir leurs théorèmes que pour les énoncer sous une forme simple et frappante. Ce secours cesse lorsqu'on passe à la considération des fonctions de plus de trois variables; aussi la théorie de ces fonctions est-elle relativement fort en retard. Le moment semble venu de combler cette lacune en généralisant les résultats déjà obtenus pour ce cas de trois variables. Un grand nombre de géomètres s'en sont déjà occupés d'une manière plus ou moins immédiate. Nous ne connaissons cependant aucun travail d'ensemble sur ce sujet.}
\end{quoting}
Before proceeding with drawing certain conclusions from Descartes' contribution we formulate our third remark. 
\begin{rem}\label{rem3}
It is important to insist that when undertaking a historical review of a notion, we can not be holistic. On the one hand, we strongly believe that the individual genius of certain historical thinkers like Descartes is an important ingredient, but, there is a temporality and maturity of notions which we must take into account. For example, Descartes' contributions are based on crucial developments initiated by El Khawarizmi (783-850), Ibn Qurra (826-901), Ibn El Heythem (965-1039) and El Kheyyam (1048-1131) who used geometrical tools to solve some algebraic problems thereby demonstrating the interest of mixing different mathematical fields. Or even the use of letters to designate numerical quantities initiated by François Viete (1540-1603). By the way, the paternity of Cartesian coordinate system is also attributed to Pierre de Fermat (1601-1665) since he developed the same idea independently from Descartes, this proves that this idea was mature at the time. There are many examples proving that such situation of finding a new idea or concept occurs at the same time independently by different researchers. On the other hand, they often introduced ideas and concepts which are not exactly as they come down to us, there are additional developments operated by successors to reach the version known today. For example, Descartes always managed to use non-negative coordinates since they still had not the status of number at the time. Thus, to designate negative solutions of equations Descartes talks about: "racines fausses ou moindres que rien". 
\end{rem}

Let us turn to take some deductions from Descartes contribution. First, according to Descartes, the point possesses no dimension since it has no extent. We can even go further by claiming that the point itself does not exist as a body if we want to respect his philosophical principle \textit{Res Extensa}. And still further, the point clearly does not respect Cartesian dualism. Indeed, let us recall that Descartes defended a strict dualism claiming that substances are split into two separate types, the mind \textit{Res Cogitans} and the body \textit{Res Extensa}. Thus, according to Cartesian dualism, what substance to attribute to the point? A mind or a body? 
We may now say that the point has won an algebraic identity with the Cartesian coordinate system. Traditionally and even nowadays, points forming a line, a plane or a body are the same since they possess a similar nature. We want to point out that the number of coordinates provides a proper identity to points. More precisely, as we will develop through this article, all points possess in absolute terms the same \textit{individual identity} when considered individually without precising their belonging. However, they have a different \textit{contextual identity} according to where they are. 

Indeed, a point with two coordinates automatically lives into a surface while a point with one coordinate belongs to a line, but this algebraic identity will turn out to be insufficient with respect to Cantor's result, which we will discuss in subsection \ref{2.3.2}. The point can also possess a topological identify, according to the concept of basis of neighbourhoods, they are intervals for a point into a line while they are surfaces for a point in a plane, this identity will appear to be more appropriate. We strongly believe that this aspect is fundamental to take into account in mathematical reasoning. The very natural question is then: what is the status of a point into a line which itself is an element of the plane? In this case the point must be regarded \textit{relatively} to the desired context. As we shall explain in details in the sequel, our goal is not to modify the individual identity of the point or its nature as 'which has no part' but we want to take into account its contextual identity as much as possible, dimension is a notion permitting this approach.               

It follows then that the Cartesian coordinate system allows us to view dimension also as a \textit{quantity of information}. Indeed, in a conventional way until the 20th century and even today, the dimension of an object is regarded as the least number of real parameters needed to describe or identify all its points. In other words, the minimal quantity of information is required for tracing all points of an object. For example, points of a line can be completely described by one parameter; while those of a plane by two parameters and three parameters for a body. This stresses an important propriety of the notion of dimension, its \textit{relativity}. Indeed, looking more closely, we need one parameter to completely control the positions of all line's points, but this tacitly presupposes that we are inside this line, the same goes with plane. Thereby, which quantity of information is needed to describe these geometrical object's points once we are outside of them? Assume that we possess a line contained in a plane, therefore, we have to find the equation of this line before claiming that one parameter is enough to identify all its points. In reality, relatively to that plan we do need at least two information to describe all this line's points, either two coordinates if we do not know this line's equation or the Cartesian equation of the latter, plus one parameter. If we consider now this line into the three dimensional space, we will need at least three information, either three coordinates if we do have no equations of this line or the two Cartesian equations of planes whose intersection produces this line plus one parameter.

These aspects of contextual identity of points, quantity of information and relativity must be taken into account rigorously in dimension's definition we should introduce.

We shall now succinctly proceed to highlight more explicitly the link between philosophy and mathematical development of higher dimensions. For more details on mathematical history of these dimensions' introduction, readers are referred to \cite{john1}, \cite{whi} or \cite{jam}. We have already mentioned above that mathematicians' lack of geometrical representation has hampered the acceptance of higher dimensions since they are disconnected from the three dimensional sensible space. Thus, they are naturally cautious about the risk of speaking nonsense. In addition, the fact that the algebraic framework of Cartesian coordinate system seems to be able to carry within it dimensions higher than three, expresses an indisputable advantage of algebra over geometry. Thereby, accepting this situation would implicitly mean that algebra is superior to geometry which was not necessarily to mathematicians' liking at the time. At this stage, to move forward and start building an $n$-dimensional hyperspace geometry, philosophers and mathematicians clearly identified a brake at the philosophical level. Indeed, the philosophical questions arising from these considerations become increasingly obvious but hard to tackle: what is the status or the nature of our sensible space? can we conceive an abstract geometry disconnected from our intuition and sense experience of the ambient physical space? Should we revise our conception of geometry and its link with sensible space? And if we can not be based either on our intuition or the sensible space, what and how can we know anything about geometry? 

\subsubsection{Immanuel Kant (1724-1804)}\label{2.2.2}
To begin with, let us briefly recall two opposite concepts concerning space defended by two great mathematicians/philosophers successors of Descartes; namely, Isaac Newton and Gottfried Wilhelm Leibniz. For Newton (\cite{whi}, \cite{jam}, \cite{pat}, \cite{pit} or \cite{den}) space is an \textit{absolute} real homogeneous tridimensional infinite physico-mathematical continuum. Space is a reality in itself, it is a sort of permanent substance, which is independent from matter within it. For Newton, according to his famous \textit{bucket argument}, space is indubitably absolute. On the other side, Leibniz (\cite{whi}, \cite{jam}, \cite{boi} or \cite{bru}) had a very modern conceptualisation of space: the \textit{Leibnizian relationism}. For him space is purely \textit{relative} exactly like time. More precisely, he drew an analogy between the relation "time-motion" and "space-matter". Therefore, space does not exist as an absolute reality in itself independently from matter. This does not mean that space and matter are the same things, but, there is no space without matter. In fact, Leibniz conceives of space as the collection of a spatial relation between objects. An interesting metaphor as given in \cite{den}, is to see space as the existing relations between family members. the latter are interconnected but their relationships do not exist if they do not exist themselves. Furthermore, according to his philosophy of \textit{monadology}, Leibniz conceives of space as discrete and not continuous. We will develop further this idea in subsection \ref{4.2.1}.                   

The next pathfinder for some of the questions formulated above was a great successor of Leibniz and Newton, philosopher but not mathematician: Kant. We think that Kant's thoughts certainly contributed to the debate concerning higher dimensions. And this, even if his thought and his philosophical positions varied during his life, as it was mentioned in \cite{cout}, \cite{de} or \cite{the}. Or that there are some of his thoughts which are somewhat ambiguous as the philosopher/mathematician Louis Couturat confessed in a letter addressed to Russel in 1904 \cite{cout}:
\begin{quoting}[font+=bf,begintext= ,endtext=]
    \textit{Mon article sur Kant a été très diversement jugé : M. Boutroux, M. Hannequin disent que je n’ai pas compris Kant; à quoi je réponds en demandant si Kant s’est bien compris lui-même.}
\end{quoting}

There are two aspects in Kant's conception of space which we want to focus on in this passage. First of all, imbued by Leibniz' (\cite{cout}) relationism, at a very early stage (twenty three years old) Kant was interested in the possibility of higher dimensions existence. In fact, the questioning about space had a central place in Kant (see \cite{bes} for example). His hope, expressed in his first publication 1749 (\cite{kant1}), of a science capable of conceiving a hyperspace theory turned out to be premonitory since it will be realised a century later: 
\begin{quoting}[font+=bf,begintext= ,endtext=]
    \textit{A science of all these possible kinds of space would undoubtedly be the highest enterprise which a finite understanding could undertake in the field of geometry. The impossibility, which we observe in ourselves, of representing a space of more than three dimensions seems to me to be due to the fact that our soul receives impressions from without according to the law of the inverse square of the distances, and because its nature is so constituted that not only is it thus affected but that in this same manner it likewise acts outside itself.}
\end{quoting}
In this passage, Kant does not only defend his conviction of the relevance of creating a geometry of hyperspaces, but also makes an attempt to give an explanation to the question: why do our senses receive an impression of three dimensional space? Kant believed that the sensation of three dimension is due to Newton's law stating that the gravitational attraction force between two masses is proportional to the inverse-square of their separation distance. This clearly suggests that in other configurations, one should think about other dimensions. To the best of our knowledge, this is the first apparition of a thought linking gravitation force to the three dimensional space. This idea can be seen as the infancy of Einstein's General relativity. Later, Kant will adopt Newton's conception of an absolute space \cite{bes}. He intended to defend that position by considering the famous paradox, also analyzed by Leibniz and Newton, of \textit{symmetric figures}, which are congruent but not superposable. Kant goes back to that paradox in many writings before arriving at his own thesis concerning the purely intuitive character \cite{bes} of space in \cite{kant} in 1783:

\begin{quoting}[font+=bf,begintext= ,endtext=]
 \textit{Those who cannot yet rid themselves of the notion that space and time are actual qualities inherent in things in themselves may exercise their acumen on the following paradox. When they have in vain attempted its solution and are free from prejudices at least for a few moments, they will suspect that the reduction of space and time to mere forms of our sensuous intuition may perhaps be well founded.}
 \end{quoting}

This paradox, also named left and right (\cite{johnd}, \cite{miz}), concerns the impossibility of superposing the left and right hands (or ears...), even if they are completely similar and equal. For Kant, this geometrical aspect cannot be explained by any \textit{concept}, but only by \textit{intuition}. For him, it proves the intuitive nature of geometrical figures and that of space itself. To be more precise, Kant thinks that the intuitively accessible propriety of being left or right depends on the observer. Thereby, it is not an intrinsic property related to the object itself but an interaction between the latter and the observer. This can be seen as the foundation of his doctrine of transcendental idealism formulated in his famous \textit{Critique of Pure Reason} in 1787 book II, Ch 1 \cite{kant0}:    
\begin{quoting}[font+=bf,begintext= ,endtext=]
\textit{I understand by the transcendental idealism of all appearances, the doctrine that they are all together to be regarded as mere representations and not as things in themselves, and accordingly that space and time are only sensible forms of our intuition, but not determinations given for themselves or conditions of objects as things in themselves.}
\end{quoting}
 More precisely, Kant made a clear distinction between reality \textit{per se} and experienced reality. For him the objective reality, if it exists, is completely inaccessible to us. We can only access a subjective reality limited by our perceptions. Furthermore, space and time are \textit{a priori} forms of perception existing in our spirit (conscious or unconscious). They are pure intuitions independent from the experience. Yet, their are necessary requirements for the acquisition of any empirical knowledge. Thus, for Kant, what preconditions our geometrical knowledge is our pure intuition of space and not our experience of the sensible world, because the idea of space is already in us. Just as our \textit{a priori} intuition of time constitutes the basis of the arithmetic. In sum, Kant's transcendental aesthetics provides a framework for the possibility of pure mathematics, which are independent of all experiences, nevertheless they describe relations involved in all experiences. Kant goes further by introducing the doctrine of \textit{transcendental logic}, which should be the quest for certain elements of this \textit{a priori} structure of our mind. He was very critical of logic in general which he thought had reached saturation since Aristotle. According to \cite{cout}, it is a pity that he did not believe in the attempts of his predecessors like Leibniz to go beyond Aristotelian logic.       

Let us go back to the left and right paradox in order to highlight its link with the higher dimensions. Briefly, imagine two flat left and right hands i.e. without either palm or reverse side. Putting them in a two dimensional plan they are distinguishable as left or right hand since there exist no transformation in the plane able to superpose them. Now, consider these same hands in a three dimensional space: they are no longer distinguishable because with an adequate rotation we can superpose them. The same goes with a no flat hands possessing palm and reverse side; in our three dimensional space, they are either left or right hand. But, if we assume that these hands live in a four dimensional space, this should no longer be the case, there should exist an adequate transformation able to make them superposed. At this moment, they will be no more discernible.  

We strongly believe that this type of reflection helped unlock mathematicians' imagination concerning higher dimensions. For example, we can easily lead the following reflection. Let us start by granting Kant the fact that our geometrical representations are bounded by our pure intuitions. Indeed, we can represent a ball or a cube in dimension two or three quite easily, even in our mind. Nevertheless, nobody is able to represent a ball or cube in four dimensions. Our representations are completely corrupted by our pure intuitions about a three dimensional ambient space which stops our geometrical representation in higher dimension. Nevertheless, our imagination and reasoning combined with more abstraction and the use of adequate analogies can allow us to conceptualize and develop these hyperspaces' properties. Indeed, the above example shows us that we can ask ourselves geometrical questions even if we are unable to have a representation of them. The question whether there is a left-right direction for three-dimensional objects immersed in a space of dimension four is completely legitimate.  

\subsubsection{Bernhard Riemann (1826-1866) and Hermann G$\ddot{\bf u}$nther Grassmann (1809-1877)}\label{2.2.3}

Kant's thought about space \textit{per se} as outside from any experience in our sensible space was widely criticised by some of his contemporary mathematicians such as 
Carl Friedrich Gauss, to name but one. Kant himself wrote about the opposition of Gauss \cite{john1} concerning this view. Initially for Gauss, space is palpable making it an important empirical basis for constructing our knowledge concerning geometry. Nevertheless, one of the earliest mathematicians introducing a rigorous framework of higher dimensions was his favorite student Riemann in Habilitationsvortrag in 1854 \cite{mcc} and \cite{pont}, influenced by his mentor Gauss. Riemann extend the works of his doctoral advisor Gauss to n-dimensions. It should be noted that Gauss had moved from his initial position \cite{john1} by defending the possible complete abstraction for geometrical theories building. In general, mathematicians investigating higher dimensions has changed their views concerning geometry at this time. 

Beside Riemann, Grassmann was one of the pioneering mathematicians introducing rigorously hyperspaces in his monograph " Die lineale Ausdehnungslehre"  1844, and both of them proceeded in their articles with the support of philosophy. The influence of Leibniz's and Kant's thoughts \cite{ott} become indisputable through Grassmann's philosophy. The latter thinks mathematics as a science of pure abstraction which should be exclusively guided by logic \cite{john1}, that he continued to designate by philosophy instead of logic in his manuscript \cite{ott}. In this sense, geometry can be divorced from our spatial intuition since it has to be a pure form of thought governed by logic. It will be the gateway toward an axiomatization of geometry. This can be seen as a fundamental step towards developing the schools of thought named \textit{formalism} or \textit{logicism}, which will be widely defended respectively by Hilbert and Frege. Towards the second half of the nineteenth century, the subject of higher dimension became a very fashionable trending subject for mathematicians.

It is worth noting that all attempts of introducing higher dimensions were based on the Descartes' idea of Cartesian coordinate system. We want to conclude this subsection by focusing in particular on vector spaces' dimensions, initially introduced by Grassmann and developed later by Giuseppe Peano. The same principles govern the degree or dimension of algebraic extensions arising from Galois theory. In this framework, in the beginning, dimension concerns exclusively unlimited spaces and not limited objects except the singleton $\{0\}$. Thus, in this setting, the dimension of a vector space $E$ is equal to the cardinality of its basis. First, vector spaces' dimensions are \textit{relative}. For example, the field $\mathbb{C}$ of complex numbers has dimension one if it is considered as a $\mathbb{C}$-vector space, while its dimension is equal two when it is regarded as a $\mathbb{R}$-vector space. Now that the notion of hyperspaces is completely accepted, the next step is obvious but hardly acceptable: infinite-dimensional spaces. Indeed, according to this relativity, we can consider $\mathbb{R}$ as a $\mathbb{Q}$-vector space, its dimension is clearly infinite. 

Secondly, in this framework, the dimension of a singleton is still ambiguous. Indeed, it is admitted that the smallest vector space $\{0\}$ has zero dimension. Hence, its basis' cardinal should be zero and then its basis is an empty family. This automatically means that the empty family is both spanning and linearly independent. Even if it is conceptually not clear how the empty set can generate zero, we can roughly convince ourselves, either with the convention that the summation over an empty set is equal to zero, or one can justify that $span(\emptyset)=\{0\}$ using the definition of the linear span as the smallest linear subspace that contains $\emptyset$, or even that $span(\emptyset)$ is contained in the intersection of all subspaces that contain $\emptyset$. In any case, the linear independence of $\{\emptyset\}$ must be admitted as a convention. In reality, we do not wish to call into question that $\{0\}$ is zero-dimensional. Indeed, this is adequate because it generates no contradiction in linear algebra. But, we think that we have to take into account a subtlety here. More precisely, as we will explain more in detail in the sequel, we have to distinguish if we have to considered dimension of $\{0\}$ relatively to it self in which case it is zero, or relatively to another vector space in which case it is not zero.

\subsection{Dimension's crisis as catalyst for point-set topology development}\label{2.3}

As aforementioned, looking for an adequate definition of dimension, Aristotle naturally attempted to begin by defining what continuum is. Bolzano will go the opposite way, i. e. trying to define adequately the continuum, he will focus on the notion of dimension. In fact, we could easily believe that there would be a kind of closed loop between wondering about dimension notion and the notion of continuum. Indeed, as well as was the case to Aristotle at the time, trying to understand the notion of dimension becomes essentially a matter of finding a relevant differentiation between three continuums; namely, line, surface and solid to which we intuitively assigned respectively dimensions one, two and three. In this study it is obvious that a better understanding of continuum is highly desirable. Conversely, any attempt to clarify the notion of continuum leads to reflect on the distinction between known continuum like line, surface and solid. In other words, finding a good conceptualisation of the continuums presupposes understanding its essence which should be present in all types of continuum even with different dimensions. In retrospect, this is not true since one can define the continuum without any reference to dimension as Cantor's conception (see discussion in subsection {2.3.2}). Conversely, one can compute the dimensions of sets which are not continuous via the notions of measured dimension (Box or Hausdorff dimensions). In any case, capturing the notion of dimension had become central for the precursor Bolzano. 
\subsubsection{Bernard Bolzano (1781-1848)}\label{2.3.1}
Nowadays, it is well known that Bolzano was a brilliant and \textit{avant-garde} philosopher, mathematician and logician \cite{russ}, \cite{lap}, \cite{sin}, \cite{sin1} and \cite{sin2}. We would also like to emphasize this fact; that is why we strongly believe that it is quite appropriate to mention Bolzano's contribution to the notion of dimension in this section dedicated to the modern history of dimension. In fact, when the mathematicians of his time were content with their intuitions concerning the notion of dimension, or at best with the Greek definitions or those based on the Cartesian coordinate system, Bolzano conceived a very modern definition of it. A definition which is really close to that of the small inductive dimension introduced independently by Urysohn \cite{ury} and Menger \cite{men} in 1922 and 1923 respectively (see section \ref{3.1}). In other words, he did not need the problem raised by Cantor in 1877 to embark on the conceptualization of his notion of dimension, as was the case for other mathematicians who tried to find an adequate definition after Cantor's counter intuitive result (see next subsection). In fact, Bolzano's work on the notion of dimension, like all his work in mathematics, logic and philosophy, received little interest from his contemporaries and even from the next generation of mathematicians as it was noted in \cite{john}. His first attempt to conceptualize the notion of dimension was made in his manuscript \cite{bol1} 1817, while his final version of 1843-44 was not published until a century later in 1948 \cite{bol}.  

Like Descartes, Bolzano undertakes mathematics as a philosopher, he mentioned this fact in his autobiography of 1836 \cite{john}:
\begin{quoting}[font+=bf,begintext= ,endtext=]
    \textit{My special pleasure in mathematics therefore rests only in its purely speculative part, or, in other words, I value in it only that which is at the same time philosophical.}
\end{quoting}

By and large, he thought that the philosophy of science had an important role to play in recognizing and dealing with philosophical issues within a science, and even contributing to scientific discoveries. In our experience, we have also been able to develop an aspect of this fruitful communion between science and philosophy of science through what we have named \textit{Experimental Philosophy of Science} introduced in \cite{maar1}. Additionally, Bolzano had developed a doctrine of \textit{philosophical realism}, the doctrine of objective sense as opposed to Kantian subjectivity. He considered himself to be an "anti-Kant" because his philosophy was strictly opposed to Kantian idealism \cite{ben}. Briefly, Bolzano agrees with Kant on the fact that all knowledge is made either by concept or by intuition. Also, on the fact that our knowledge does not concern directly the things but the representations that we have of them. However, their divergence lies in the distinction between concept and intuition. Unlike Kant, Bolzano defends the doctrine that concepts are \textit{a priori} and intuitions are a posteriori. According to his arguments, space and time are concepts and not intuitions (for more details see \cite{sin1} and \cite{ben}). As long as Bolzano is concerned, geometry is not the science of space but rather the science of the concept of space. He was the only mathematician of his time to dissociate geometry from intuition and experience. For him, geometric space is clearly a mathematical concept and not a representation of sensible space \cite{sin1}. Nevertheless, to the best of our knowledge we do not know of any reference where Bolzano deals with the question of hyperspace.

As a matter fact, Bolzano had deep reflections on geometry and its definitions which surely contributed to build his conceptions. Indeed, as mentioned in \cite{john}, the earliest Bolzano’s aim since he has 16 years old and throughout his entire life, is finding adequate definitions of continuum in general and basic geometrical objects like, line, surface and solid in particular. This will lead him to take a closer look at the notion of dimension. Initially geometrical, this problem will lead him to invent a still non-existent discipline, the topology or to be more precise the \textit{point set topology}. Bolzano's erudition combined with his philosophy probably helped him to better pose the problems. For example, at the beginning he was very critical about considering a geometrical figure as a set of points. He reconsidered this position by assuming in his work \cite{bol1} that every geometric figure is a set of points \cite{john}. We can see here the link with his philosophy since he had in the meantime conceived geometry as a science of the concept of space and not of space itself. Thus, the distinction becomes very clear since an object of sensible space is a real physical object whereas a geometrical object is a system of points in the space of concepts. Therefore, geometric theorems are not about real objects of sensible space but about systems of points. In other words, the truths expressed by a geometric theorem concern the links or relations between concepts and not the experiences or the intuitions.  

With this conception, he will succeed in having a set-theoretical approach to geometry, which will lead to the point set topology. He was surely aware of the problems and paradoxes underlying the assumption that the continuum of a line is a set of points, since he had begun by criticizing this conception. Nevertheless, he circumvented this difficulty by focusing on the rigorous formalization of the most general properties of the points constituting a given set through their organisations and the existing links or relations between them. We will come back to this aspect in the next subsection. It took him several attempts until he managed to: first formulate a primitive notion of metric space through his concepts of direction and distance in 1804. Then, he conceived a notion of neighbours of a point based on his concept of distance in 1817. Then, he finally defined clearly his concept of isolated point in 1843-44. 

Thus, Bolzano defines the continuous or the \textit{extension} as a set that has no isolated point. His definition of the continuum is very accurate but not correct. Moreover, Cantor will criticize the latter but partly for the wrong reason since he had confused Bolzano's definition of isolated points with his own definition, which is the one used today. Nevertheless his definitions are very profound and unquestionably far ahead of his time. It is important to highlight that his guideline was the notion of dimension, since his goal was to characterize typical geometrical objects of dimensions one, two and three. For more details on the history of Bolzano's mathematical and philosophical investigations on dimension, continuum and topology, readers are referred to reference article \cite{john}, see also the book \cite{seb}.

Moving on to more details, we will present Bolzano's ideas in a modern way. We are inspired by the presentations given in books \cite{kat}, \cite{aul} and \cite{seb}. First, by \textit{neighbours} (B-neighbours) of a point $x$ in a spatial object $X$ with a distance $\delta$ he means the set of points belonging to the intersection of $X$ with the 2-sphere $\mathbb{S}^2(x,\delta)$ of $\mathbb{R}^3$ with center $x$ and radius $\delta$: $neighbours(x,\delta)=X\cap \mathbb{S}^2(x,\delta)$. In fact, Bolzano restrict his definition to spherical neighbourhoods. Now according to Bolzano, we will say that $x$ is a Bolzano's isolated (B-isolated) point when $$\forall \epsilon >0, \exists \delta\in ]0,\epsilon[\ such\ that; neighbours(x,\delta)=\emptyset.$$   

In the modern sense, the point $x$ in a metric space $(X,d)$ is said to be isolated when $\exists \delta>0$ such that $B(x,\delta)\cap X=\{x\}$, while $B(x,\delta)$ denote the ball centred in $x$ with radius $\delta$. It is obvious that an isolated point in the modern sense is also an isolated point in Bolzano's conception, but the converse is false since Bolzano's neighbour of his isolated point may contain points when in the modern sense it is not the case. Let us focus on Bolzano's concept of continuum or extension. First of all the negation of being a B-isolated point can be written as $$\exists \epsilon >0, \forall \delta\in ]0,\epsilon[; neighbours(x,\delta)\ne \emptyset.$$  

In fact as pointed by Cantor, Bolzano's continuum is not necessary connected. But, he was already aware of this fact and accepted it since he raised this objection himself through the following example \cite{seb}:

\begin{ex}\label{ex1}
Consider an interval $X=[a,b]$ and construct a sequence of midpoints starting by the midpoint between $a$ and $b$ denoted $c_0$, $c_1$ the midpoint between $c_0$ and $b$, $c_2$ the midpoint between $c_1$ and $b$, and so forth. Now, if we consider the set $X'$ constituted by the set $X$ without the constructed sequence of midpoints $c_0, c_1, c_2,...$ and also without the point $b$, then we can easily check that $X'$ still a Bolzano continuum. However, if we decide to consider the set $X''$ where the sequence of midpoints $c_0, c_1, c_2,...$ is removed  but not the point $b$ then it will be no longer a Bolzano's continuum since the point $b$ is a B-isolated point.   
\end{ex}

This example clearly demonstrates that Bolzano was aware of the limitations of his definition of the continuum. While this construction does not demonstrate how Bolzano recognized the imperfections of his definition of the continuum, it implicitly shows that he had to accepted them, for lack of no other alternative. In particular, this example highlights the subtleties that can be revealed by a pointillistic approach to sets. Now, Bolzano's dimension, or should we say Bolzano's classification of geometrical objects, of a nonempty subset $X$ of $\mathbb{R}^3$ is defined inductively as follows: 

\begin{Def}[Bolzano's dimension 1843-44]
\begin{enumerate}
\item If for any $x\in X$ and any $\epsilon>0$ there exists a positive $\delta <\epsilon$ such that $neighbours(x,\delta)=\emptyset$ we will say that $X$ is discontinuous or not an extension (dimension 0).
\item If for any $x\in X$ there exists $\epsilon>0$ such that for any positive $\delta <\epsilon$ the set $neighbours(x,\delta)$ is discontinuous we will say that $X$ is a line or a simple extension (dimension 1).
\item If for any $x\in X$ there exists $\epsilon>0$ such that for any positive $\delta <\epsilon$ such that the set $neighbours(x,\delta)$ is a line we will say that $X$ is a surface or a twofold extension (dimension 2).
\item If for any $x\in X$ there exists $\epsilon>0$ such that for any positive $\delta <\epsilon$ such that the set $neighbours(x,\delta)$ is a surface we will say that $X$ is a solid or a threefold extension (dimension 3).
\end{enumerate}
\end{Def}

The principle of this definition is simple, roughly speaking we define the dimension of a non-empty set $X$, dimensionally homogeneous, inductively as the dimension of the set defined by the section of $X$ with a 2-sphere (B-neighbours) plus one. We would like to point out a subtlety contained in these definitions. Indeed, the first definition is different from the three other ones because it states that for any point there exists an empty B-neighbours as close as we want to the point in question. Whereas the other definitions, which concern the continuums of one, two and three dimensions, stipulate that from a certain rank all the B-neighbours are of the same type: discontinuous for the dimension one, line for the dimension two and surface for the dimension three. All in all, we must emphasize here that to the best of our knowledge, this was the first formalization that rigorously made the set of rationals numbers $\mathbb{Q}$ a discontinuous set. 

This way of defining the dimensions reminds us of the Euclidean definitions, which we called topological dimension definitions. Bolzano had not only read them, but had studied and analyzed them in depth, since he had formulated the same criticism of them as Aristotle (see subsection \ref{2.1.2}). The correspondence lies both in the inductive way of stating them and in the fact that his definitions do not contain the term dimension. In fact, with these definitions, Bolzano rather defines the discontinuous, the line, the surface and the solid. In this sense, he proceeds like the Greeks in defining the different basic objects of geometry through their dimensions. However, it is undeniable that his definitions are deeper and that he chooses this way of linking the geometrical objects with their dimensions deliberately. 

More precisely, as noted in \cite{seb}, Bolzano was aware of the revolutionary scope of conceiving the different geometric objects as a set of points, which can be considered in itself as the founding act of point-set topology as a fundamental geometric discipline. This constituted a break with the 17th and 18th century geometers' conception of a line as an assembly of infinitesimal elements ($dx, dy...$) having locally the specific shape of the latter as a signature (depending on whether it is a straight line, circle, parabola... for example). Similarly, a surface is the assembly of infinitesimal surfaces ($dS$), also, a volume is made up of infinitesimal volumes ($dV$). Thus, reducing all geometric objects to a set of points makes them all similar, which poses the following problem: how can we distinguish between the different geometrical objects since they are all considered equally as sets of points? Bolzano brings the answer to this question through the characterization of their dimensions.

Let's now move on to analyze these definitions a little further. First, these definitions are well initialized since when $X=\{x\}$ then it is an isolated point. Secondly, these definitions are easily extensible to higher dimensions. More deeply, even if this way of defining the dimension seems to be local since it concerns each point constituting the object $X$, it is not exactly the case. Indeed, this is due to the fact that these definitions indicate that the dimension of $X$ is indexed on the fact that all its points possess the same type of neighbours' sets. This implies in particular that the definition does not allow to evaluate the dimension of a geometrical object which is not dimensionally homogeneous like the previous example \ref{ex1} concerning the set $X''$. Indeed, all its points are 1-dimensional except the point $b$ which is a B-isolated point and then is zero-dimensional. In this case we must rely on Bolzano's conception of the continuum as a set without B-isolated points, and not on these definitions. Thus, since $X''$ admits one point which is B-isolated then it is not a continuum, which is correct. 

Here we come a cross the limit of introducing a notion of dimension with an exclusive purpose and a main function which are: the concordance with the basic geometrical objects (one-dimension $\iff$ continuum of the line). It is necessary at this point to accept that the notion of dimension should be certainly not adequate in other situations. For example, in this case of the set $X''$, it would be better if its dimension was one like the set $X$, even if it is not a line. In reality, with his definition, Bolzano wanted above all to characterize the discontinuous and the different dimensions 1, 2 and 3 of the continuum. Any mixture of these dimensions cannot be taken into consideration. Another basic example that will not be taken into account by his definitions is: a disc with a segment that stands out from it.


In other words, though the core of Bolzano's definitions is local, he gives them an exclusively global scope. We can say that Bolzano had the \textit{avant-gardism} to consider geometric objects as a set of points, which is clearly reflected in his definitions. Nevertheless, he either did not know or did not want to push this way of seeing further by clearly and directly attributing dimensions to the points. And this, even if its definitions are capable, as we will explain, of taking this step. One could possibly see a psychological or philosophical brake due to its own conception of the point. This is understandable in the sense that, as explained above, Bolzano will use the notion of dimension to distinguish between different geometric objects. At this stage, to say that a point could have the dimensions one, two or three would go against what he wanted to do. Indeed, this may seem to create a lot of confusion because for example: what dimension would a line formed by points of dimension three have at that moment? These types of questions are the main objective of this article. In fact, for Bolzano, all the points have a single nature and that is the same for all. However, Bolzano's work is consistent with the distinction we made in subsection \ref{2.2.1} between the individual and the contextual identities of the point since, according to his definitions, we can easily deduce that points can have different dimensions according to which set they form. 

Here we want to go further by proposing a modified Bolzano definition focusing on the dimension of the point, but that both respect the spirit of his definitions and do not use mathematical tools beyond those of the time. More precisely, Bolzano's definitions are able to be oriented towards definitions of the dimensions of the points constituting $X$. Only after evaluating all the dimensions of all the points of $X$, not necessarily similar, we assign a dimension to $X$. We formulate a definition that is not restricted to the dimensions zero, one, two and three. Here the 2-spheres used to define Bolzano's neighbours will be the boundaries of the balls of a metric space $(X,d)$.   

\begin{Def}[Maaroufi-Zerouali modified dimension of Bolzano 2022]\label{mbdim}
Let $(X,d)$ be a nonempty metric space and $x\in X$.
\begin{enumerate}
\item $MBdim(x)=0$ if for all $\epsilon>0$, there exists a positive $\delta <\epsilon$ such that $neighbours(x,\delta)=\emptyset$. $MBdim(X)=0$ if for all $x\in X$; $MBdim(x)=0$.
\item For $n\in \mathbb{N}^{\ast}$, $MBdim(x)=n$ if there exists $\epsilon>0$ such that for all positive $\delta <\epsilon$; $MBdim(neighbours(x,\delta))= n-1$. $MBdim(X)=n$ if for all $x\in X$; $MBdim(x)\leq n$ and there exists $x_0$ such that $MBdim(x_0)=n$.
\item If there is no integer $n$ such that $MBdim(x)=n$, we posit that $MBdim(x)=\infty$. If for all non-negative $n$, there exists $x\in X$ such that $MBdim(x)\geq n$ then $X$ is infinitely dimensional.
\end{enumerate}
\end{Def}

Obviously, unlike Bolzano, here we define the dimensions of higher order and even infinite. We want to point out that we can have $MBdim(X)=\infty$ without having any infinite dimensional point in $X$, like for example the space of sequences that vanish from a certain rank. It can be said that this way of defining dimension respects the spirit of the Bolzanian conception. We strongly believe that if Bolzano had taken the conceptual step of directly attributing a dimension to the points, he could have formalized a definition close to this one. This way of defining the dimensions allows not only to obtain the dimension of each point of a set, but also to compute the dimension of non dimensionally homogeneous sets. For sets that are not dimensionally homogeneous, there is obviously a problem of choosing the dimension assigned to the whole set between the minimum and maximum dimensions of its points. According to this definition, we chose to assign to a non-dimensionally homogeneous set $X$ the largest of the dimensions of its points. This is consistent because, on the one hand, it is justified by the problem of formalizing an inductive definition choosing to take the minimum of the dimensions of the points. On the other hand, an adequate notion of dimension must at least respect the \textit{monotonicity} property stating that: $A \subset B \Rightarrow dim(A)\leq dim(B)$, which will be violated by choosing the minimum of the dimensions of the points. Of course, the definition we introduce here also allows, if we wish, to compute the minimal dimension of a set since it evaluates the dimension of each point of the set. 

For example, the sets $X$, $X'$ and $X''$ of the previous example \ref{ex1}, have the same dimension; namely, one. Thus, if like Bolzano's goal, we wish to define the continuum through dimension, the continuum would no longer be defined as a set that does not have an isolated point, but rather a set that contains a subset (or a point) of dimension greater than or equal to one. A more twisted example is the set $Y=([-1,0]\cap\mathbb{Q})\cup ([0,1]\textbackslash \mathbb{Q})$. This set possess dimension one according to the given modified definition of Bolzano's dimension even if it is not covered by Bolzano's definition. This set $Y$ is also not dimensionally homogeneous; it has one point of dimension one (the point 0) and all its other points have dimension zero. This proves that one point of dimension one is enough for the whole set to be of dimension one. We will come back, in the following, to the notion of Bolzano's dimension and more particularly to a modified definition introduced by Miroslav Kat\u{e}tov in 1983 when we will treat the case of the small inductive dimension. We have also managed to formalize another definition based on this idea of Bolzano, even more elaborate and relevant than the one we have introduced here. We will present this definition in the next article.    

\subsubsection{Georg Cantor (1845-1918)}\label{2.3.2}

Cantor is recognized as the founding father of the set theory in general and point set topology in particular \cite{john1}. The decisive contribution of Bolzano to both aspects is undeniable today. Indeed, well aware of Bolzano's work, Cantor was able to rely on the latter's conceptual advances. On the one hand \cite{sin1}, while the mathematicians and philosophers of the time were still following the trend initiated by Aristotle of accepting potential infinity but not actual infinity, in his famous book \textit{Paradoxes of the Infinite} published after his death, Bolzano had defended the acceptance of actual infinity as a mathematical concept. Of course, he could not create a coherent arithmetization of it like Cantor's; nevertheless, he accepted the criterion of the existence of a bijection between the infinite sets for their equipotence. He went further to argue that properties of infinity that seem paradoxical in a finite framework should be adopted as a definition of infinity. Hence the definition of an infinite set as being a set equipotent to one of its proper parts.  More than that, he managed to affirm the impossibility of the existence of a single infinite but of several infinities. On the other hand, as we explained above, Bolzano was the first to take the step of having an infinite set-theoretical approach to geometry, thus founding the point set topology. In this sense, and according to the previous subsection, we can reasonably argue that the point set approach and the search for an adequate quantification of the notion of dimension through a definition are the essential ingredients for the creation of point set topology. 
\paragraph{Cantor and continuum.}
Bolzano's influences do not, in any way, call into question the genius or the immeasurable contribution of Cantor. His breakthrough in the field of mathematics is undeniable since, among other things, infinity clearly becomes an object of rigorous mathematical study with Cantor. Even if, as we have explained above, research on dimension contributed to the creation of point set topology, Cantor's formalization of the latter, to his own surprise, was independent of the notion of dimension. More precisely, Cantor began by constructing the continuum of the number line $\mathbb{R}$ as the completion of $\mathbb{Q}$ via Cauchy's sequences in 1872 \cite{cant3}, which surely helped him to develop his own conception of continuum. This procedure is standard since it can be used for any metric space $E$ to obtain a complete metric space $E'$ regardless of its dimension. Of course, completeness is not sufficient to characterize the continuum because it lacks the connectedness. In fact, the construction of the number line $\mathbb{R}$ was an essential step for a better understanding of what the continuum can be since this last had become, in the minds of mathematicians, the ideal representative of the continuum. In particular, understanding the role of the density property in forming a continuum had long been an obstacle to a better formalization of the latter.

As mentioned in \cite{sha} and \cite{john1}, the series of six articles Cantor published between 1879 and 1884 can be considered the pinnacle of his life work, the "quintessence of Cantor's lifework" as it was written by Ernst Zermelo \cite{john1}. In them he develops in a clear and mature way his conception of point set topology theory combining his finding on continuum and set theory. The 1883 article, described as a masterpiece in \cite{sha}, was divided into two parts: mathematical \cite{cant2} and philosophical \cite{cant1}. In the philosophical essay, Cantor not only specifies his conception of the continuum and its place in the history of philosophical thought, but also others of his philosophical positions. It appears that the thorny question of the continuum was central for Cantor. It is, for him, of major importance for science in general, as he wrote in \cite{cant1} (see the translation in \cite{brag}). 

\begin{quoting}[font+=bf,begintext= ,endtext=]
    \textit{The concept of the 'continuum' has not only played an important role everywhere in the development of the sciences but has also always evoked the greatest differences of opinion and even vehement quarrels.}
\end{quoting}

In the following passage, Cantor underlines the admission of impotence of his predecessors to rigorously analyze the continuum. He explicitly criticizes the Kantian position of pure intuition by pointing out its failure to produce conclusive results. He also implicitly criticizes Kronecker's constructivist and finitist position, relegating it to religious dogma.   

\begin{quoting}[font+=bf,begintext= ,endtext=]
    \textit{Here we see the medieval-scholastic origin of a point of view which we still find represented today, in which the continuum is thought to be an unanalysable concept or, as others express themselves, a pure a priori intuition which is scarcely susceptible to a determination through concepts. Every arithmetical attempt at determination of this mysterium is looked on as a forbidden encroachment and repulsed with due vigour. Timid natures thereby get the impression that with the 'continuum' it is not a matter of a mathematically logical concept but rather of religious dogma.... It is far from my intention to conjure up these controversial questions again; and besides, I lack the space for a detailed discussion of them. I feel obliged only to develop the concept of the continuum as soberly and logically and briefly as possible, and only with regard to the mathematical theory of sets. This treatment has not been easy for me because, among those mathematicians whose authority I should gladly invoke, not a single one has occupied himself with the continuum in the exact sense which I find necessary here... has not been able to produce any fruitful, incontestable success, although since Kant the time for this has not been lacking.}
\end{quoting}

Afterwards, Cantor explains that mathematicians have focused on the concept of a continuous function while completely ignoring the continuum itself. As usual Cantor questions here concepts that seem to need no further mathematical investigation.
\begin{quoting}[font+=bf,begintext= ,endtext=]
    \textit{By taking as a basis one or several real or complex continuous quantities (or, more precisely, sets of continuous quantities) one indeed developed the concept
of a single-valued or many-valued continuum dependent upon them, i.e. the concept of a continuous function; and in this way the theory of the so-called analytic functions arose, as well as of more general functions with remarkable properties (such as non-differentiability and the like); but the independent continuum itself has been assumed by mathematical writers only in its simplest guise, and has not been subjected to any more thorough inspection.}
\end{quoting}
By rejecting the Kantian idea of pure intuition concerning the continuum and by affirming that the latter must be analyzed from a mathematical point of view in a logical, exact and sober way as a concept. Hence, Cantor joins the philosophical position of Bolzano without mentioning it. Nevertheless, he will criticize, as we pointed out in the previous subsection, the flaws of the Bolzanian conception of the continuum in the continuation of the same article. Moreover, he clearly displays his desire to define the continuum in the most general way possible by relying on the concept of real numbers. This is the coronation of the idea of the continuum as a set of points, as stated in the following passage: 

\begin{quoting}[font+=bf,begintext= ,endtext=]
    \textit{It is likewise my conviction that with the so-called form of intuition of space one cannot even begin to acquire knowledge about the continuum. For only with the help of a conceptually already completed continuum do space and the structure thought into it receive that content with which they can become the object, not merely of aesthetic contemplation or philosophical cleverness or imprecise comparisons, but of sober and exact mathematical investigations. Consequently, there is nothing else remaining for me than, with the help of the concept of real number as defined in §9, to look for a pure arithmetical concept of point-continuum which will be as general as possible.}
\end{quoting}

Let us turn to Cantor's mathematical conception of the continuum in his work \cite{cant2}. According to him, to say that a metric point set $X$ is a continuum it must have the two properties of being \textit{perfect} and \textit{connected}. The set $X$ is perfect if it is closed and dense in-it-self when the latter means that each of its points is a limit point of it. Here we see Cantor's inspiration due to his construction of the real line $\mathbb{R}$. An equivalent definition of a perfect set is to be closed and without any isolated points. Through this equivalent definition, the influence of Bolzano concerning the continuum is palpable, even if for Cantor the concept of isolated point is defined in the modern sense and not in the Bolzanian sense (see subsection \ref{2.3.1}). In fact, Cantor's first definition of a perfect set is based on the notion of \textit{derived set} that he introduced in 1872 in the same article where he constructed $\mathbb{R}$ \cite{cant3}. In particular, he understood that the derived set of rational numbers $\mathbb{Q}$ is $\mathbb{R}$. Thus, he posit that a set $X$ is perfect if $X=X'$ when $X'$ denote the derived set of $X$. The set $X'$ consists of all the limit points of $X$ when his definition of the limit point is (see \cite{sha}):        

\begin{quoting}[font+=bf,begintext= ,endtext=]
    \textit{By a limit point of a point-set P I understand a point of the line whose position is such that in any neighbourhood [Umgebung], infinitely many points of P are
found, whereby it can happen that the same point itself also belongs to the set. By a neighbourhood of a point one should understand here any interval which contains the point in its interior. Accordingly, it is easy to prove that a point-set consisting of an infinite number of points always has at least one limit point.}
\end{quoting}

We want to stress here a strong link between this conception and Bolzano's ideas. Indeed, the notion of the limit point for Cantor involves an infinity of points, which means that between two limit points as close to each other there exists an infinity of points. Indeed, Bolzano had managed to capture a notion close to this one via his B-neighbours and the non B-isolated points without succeeding in explicitly formalizing what a limit point is, let alone a derived set. Nevertheless, it is reasonable to think that all these notions derive from the justification or the contribution he had given to an old paradox. 

Indeed, in order to defend his conception of the continuum as a set of points, Bolzano had to overcome the following old objection: if we accept that the point is that which has no part, then how can a set of points form the continuum? More precisely, if the continuum is an extension without holes, then the successive points should touch each other, but then how would they touch each other since they have no part? If they would touch, they would necessarily be confused. This implies that there are only two configurations for points forming a continuum, either they are merged or they are separated by a non-zero distance. Thus, the distance between any two non-identical points forming the continuum is non-zero which implies that there are holes in the continuum, which is contradictory to our conception of the continuum. This was one of the major obstacles to thinking of the continuum as a set of points across the times. Of course, this difficulty is completely circumvented in the infinitesimal calculus by introducing the undefined fiction $dx$ so criticized in particular by the philosopher George Berkeley in his famous 1734 pamphlet "The Analyst".  

The solution that Bolzano found is (see \cite{gran}): if we want to conceive the continuum as a set of points, we must suppose that between any two points there are an infinity of intermediate points. This apparently simple answer is very profound and will have immeasurable consequences. Indeed, this constitutes, for us, the starting point or the spark that will allow the development of both set theory and point-set topology theory. This proves how the philosophical and mathematical questions concerning infinity have a central role for the development of mathematics. Hermann Weyl goes further in making the notion of infinite as the fundamental object of mathematical science through the first sentence of his philosophical essay "Levels of Infinity", 1930 \cite{hw}: "Mathematics is the science of the infinite". We strongly believe that Bolzano's assumption concerning the infinity intermediate points existing between any two points, allowed him to develop the innovative concepts that we have exposed in the previous sub-section. In particular, it forces to completely discard the need to apprehend the direct neighbor of a point forming the continuum. More precisely, this hypothesis highlights the relevance of considering the key notion of a point's \textit{neighborhood} instead of unnecessarily trying to understand how a point's direct neighbor should be arranged so that they can form the continuum. Moreover, this notion of neighborhood of a point is easily quantifiable via the set of points defined by their distances to the considered point.   

To be more precise, it is reasonable to think that the innovation of the Bolzanian conception of the continuum begins by considering the latter as a set of points having the property that between any two points there exist an infinity of other points. This can be considered as the genesis of the Cantorian notion of the limit point and the derived set. In other words, Bolzano probably thought that this property was the essence of the continuum before he realized by himself that it lacks connectedness, which he nevertheless accepted, even though he had succeeded in his time in formalizing a notion of connectedness \cite{seb}. In any case, Bolzano's approach in introducing the B-neighbours was a consequence of this conception. As we mentioned above, Bolzano wanted to distinguish the different geometric objects according to the type of neighborhood of their constituent points, which will give rise to his notion of dimension. Thus, he starts with the B-isolated points whose B-neighborhood is empty. We want to emphasize here that Bolzano's work implicitly contains Cantor's idea of a limit point through the notion of non-B-isolated point. However, in Cantor's work this takes on a greater significance in the sense that he directly defines the notion of a limit point positively and not negatively by the implicit deduction of the opposite of what an isolated point is. In fact, the opposite of being an isolated point is completely equivalent to the other definition of limit point (also called accumulation point) if Bolzano had used a good notion of neighborhood i.e., via balls instead of spheres. 

Indeed, let $(X,d)$ be a metric space, if we denote by $B(x,r)=\{y\in X/ d(x,y)<r \}$. Let $P$ be a subset of $X$, the point $x$ is said to be a limit or accumulation point of $P$ if for all $r>0$ the intersection of $B(x,r)$ with $P$ contains at least one point of $P$ different from $x$ itself. Then in the case, of the metric spaces, the idea of a limit point is clearly contained in Bolzano's work. We should also point out that it is very easy to show that this definition is equivalent to Cantor's even under weaker conditions; namely, when the space $X$ in question is just $T_1$ i.e., when every finite subset of $X$ is closed (see \cite{mun} for example). Basically, if any neighbourhood of a point $x$ possess an infinite points of $P$, then in particular the intersection of every neighbourhood of $x$ and $P$ contains an other point than $x$. Conversely, if we suppose that there exists a neighbourhood $U$ of $x$ containing only a finite number of points then we can easily construct another neighbourhood $U'$ of $x$ by excluding these points, in which case the intersection of $U'$ and $P$ contains only $x$ which would be contradictory.     

In fact, we should mention that Cantor's definition of the limit point was the first definition to be published. But, it is well known that the notion of accumulation point is also attributed to Karl Weierstrass through his famous lectures of the 1860s. Lectures renowned for their rigor and innovative aspect, where he built the foundation of modern analysis (see \cite{dug} and \cite{dug1}). On the one hand, it is recognized that Weierstrass was widely influenced by Bolzano's work (see \cite{dug} and \cite{dug2}). On the other hand, as mentioned in \cite{dug}, Cantor not only attended some of Weierstrass' lectures, but Weierstrass was also the second exterminator of his thesis. In fact, Cantor's sentence: "Accordingly, it is easy to prove that a point-set consisting of an infinite number of points always has at least one limit point.", clearly refers to the Bolzano–Weierstrass theorem, if we add the condition that the point set is bounded. This fundamental theorem shows the great influence of Bolzano on both Weierstrass and Cantor. This theorem was stated and demonstrated by Bolzano in 1817, re-demonstrated 50 years later by Weierstrass for whom Bolzano's geometrical proof cannot be considered rigorous \cite{dug}, and then stated by Cantor through this sentence as an obvious result. We must also point out that Bolzano was the pioneer in identifying the concept of interior and boundary point as well as closed and bounded set by relying on his notion of B-neighbours (see \cite{seb}). 

This notion of limit or accumulation point plays a central role in the genesis and the development of the point-set topology. Indeed, it generalizes the notion of a limit, in particular, if $x$ is a limit point of a set $X$ there exists a subsequence of $X$ which converges toward $x$. It permits to highlight the weaker notion of adherent point. It allows Cantor to introduce the notion of a closed set as a set $X$ which contains its limit points, i.e its derived set $X'$ is included in itself. This paves the way to define the notion of an open set as the complementary of a closed set and also helps to define the topological closure of a set. The highlighting of the notion of compactness is also essentially due to the notion of limit point. Indeed, it arises from the generalisation of Bolzano-Weierstrass theorem which basically expresses that if an infinity of points is confined in a bounded region then they admit at least one accumulation point. This notion of limit point is also behind the birth and development of Cantor's set theory. At the time, defining the derived set $X'$ as the set of limit points of a set $X$ was an important conceptual leap that Cantor was the only one to make. As was also pointed out in \cite{sha}, this is the first formalization of operations on infinite sets. It can be considered as the founding act of the concrete use of \textit{actual} infinity. We can say that this is the starting point for Cantor to understand the arithmetics of infinity. More precisely, considering the successive derived sets of a set allowed Cantor to conceive the notions of ordinal numbers and cardinal numbers.

Returning to Cantor's concept of continuum which is a perfect and connected point set, as mentioned above, the perfect character of a set which means being without isolated points and containing its limit points (closed) is in the spirit of the Bolzanian definition. Where Bolzano is completely outdated, is in the fact that Cantor managed to capture through the property of a set to be perfect a characteristic and not the least: \textit{the power of the continuum}, a notion introduced by himself. In other words, Cantor had succeeded in showing that any perfect subset of $\mathbb{R}$ has the power of the continuum $[0,1]$ i.e., it is uncountable. The question that remained for him was the problem of connectedness that Bolzano had also encountered. More precisely, is a perfect set necessarily connected? The answer is no. This is the reason why Cantor will introduce for the first time his famous \textit{Cantor ternary set} which is perfect but not connected. Cantor's definition of connectedness is: a point set $X$ is connected if for every two of its points $x$ and $x'$, and an arbitrary given positive number $\epsilon$, there exists a finite system of points $x=x_1, x_2,..., x_n=x'$ of $X$ such that the distances $x_1x_2, x_2x_3,..., x_{n-1}x_n$ are
all smaller than $\epsilon$. This was the first very accurate mathematical formalization of the notion of connectedness. 

Cantor's notion of connectedness is completely equivalent to the modern notion in the compact metric space case as noticed in \cite{kur}, where Kuratowski established the modern definition of the continuum as a compact, connected Hausdorff space (see also \cite{sha}). The compactness is equivalent in finite dimension to a closed and bounded subset. Through Cantor's sentence above "...always has at least one limit point", it is reasonable to think that Cantor believed that the compactness property, not yet evidenced  at that time, was automatic since he considered only finite dimensional bounded subsets of $\mathbb{R}^n$. In \cite{mich}, the explanation of the substitution of the perfect set by the compact set is justified by the intrinsic character of the compactness contrary to the closed set contained in the definition of a the perfect set. In any case, the notion of continuum has always been a source of questioning and a challenge to rigorous concept formalization. Cantor was the first to distinguish and dissociate the two properties that can be considered as the essence of the continuum: the density of the points forming the continuum and the connectedness formed from their arrangement. Not only did he clearly distinguish these two properties, but he also succeeded in formalizing them rigorously by introducing the two new mathematical concepts for the time of perfect set and connected set.

\paragraph{Cantor and dimension.}
In the same 1883 article \cite{cant2}, Cantor emphasizes the fact that his definition of the continuum is completely independent of any notion of dimension:
 
\begin{quoting}[font+=bf,begintext= ,endtext=]
    \textit{Observe that this definition of a continuum is free from every reference to that which is called the dimension of a continuous structure; the definition includes also continua that are composed of connected pieces of different dimensions, such as lines, surfaces, solids, etc. On a later occasion I shall show how it is possible to proceed in an orderly fashion from this general continuum to the special continua with definite dimension. I know very well that the word 'continuum' has previously not had a precise meaning in mathematics; so my definition will be judged by some as too narrow, by others as too broad. I trust that I have succeeded in finding a proper mean between the two.}
\end{quoting}

Through this passage, Cantor seems somewhat upset that in his definition of the continuum there is no reference to the notion of dimension. He promises to address the notion of dimension in future works, which he will never do. In fact, according to Cantor's conception, it is understandable that the dimension notion is needed for at least two reasons. On the one hand, as for Bolzano (see subsection \ref{2.3.1}), the fact of having a pointillist conception of the different geometric objects raises the problem of the criterion to be adopted in order to differentiate between them, since they are all similarly conceived as sets of identical points. The number of dimensions seems to be the most appropriate characterization to make this distinction. On the other hand, as mentioned above, it is easy to believe that there would be a closed loop between the notion of dimension and continuum. All those who were interested in the notion of the continuum were automatically interested in the notion of dimension and vice versa. Cantor is one of them, as we will explain in the following subsection. This is why obtaining a notion of the continuum completely independent of the notion of dimension makes him somewhat uncomfortable, as he expresses it in this passage.  


In fact, a better understanding of the notion of dimension was always a challenge for Cantor, as evidenced by his correspondence with Dedekind (see \cite{brag} or \cite{sha}). More precisely, quite early in his career, he quickly came to question the notion of dimension. This is probably due to at least two reasons. The first one is his pointillistic conception of sets, which we  have already discussed above. The second one, which we will develop now, is his research which aims to show different cardinalities of infinite sets. Cantor had already given birth to his theory of transfinite sets in his famous article \cite{cant4} of 1874. Indeed, Cantor's proof concerning both the countability of the algebraic real numbers and thus in particular of the rational numbers $\mathbb{Q}$, and the uncountability of the real numbers $\mathbb{R}$, can be considered as the founding act of the theory of transfinite sets. In his continuous quest to find different types of infinite cardinalities, the natural question that arose for him was: do sets of different dimensions have different cardinalities? In his letter to Dedekind of 1874, Cantor formulates the question precisely in the following terms (see \cite{brag} or \cite{sha}): 
\begin{quoting}[font+=bf,begintext= ,endtext=]
    \textit{Can a surface (say a square including its boundary) be one-to-one correlated to a line (say a straight line including its endpoints) so that to every point of the surface there corresponds a point of the line, and conversely to every point of the line there corresponds a point of the surface?}
\end{quoting}
In reality, for Cantor, the answer is \textit{a priori} strongly no if we refer to what he writes in the next paragraph:       
\begin{quoting}[font+=bf,begintext= ,endtext=]
    \textit{It still seems to me at the moment that the answer to this question is very difficult—although here too one is so impelled to say no that one would like to hold the proof to be almost superfluous.}
\end{quoting}

For more details on the correspondence between Cantor and Dedekind concerning this subject, we refer the reader to \cite{brag}, \cite{john1} and \cite{sha}. We will nevertheless briefly underline the aspects which seem to us the most important for the goal we have set ourselves; namely, a better understanding of the notion of dimension. As usual, Cantor is led by his research to formulate questions which, as he says himself, seem at first sight not to need further mathematical investigation. In this regard, he follows the Bolzanian philosophical tradition forbidding to base mathematics on intuition that only provides certainty but rather on demonstration to reach the truth \cite{seb}.  For example, Bolzano was the first to underline the necessity of a demonstration concerning the intuitive statement known today as the Jordan curve theorem. Not only does Cantor do this questioning, but one can say that Cantor is the archetype of the scientist who goes where his research leads him, going beyond his own preconceptions. The question of the existence of a bijection between an interval and a surface is a concrete example of this, since he intuitively began by believing that there is no such correspondence before demonstrating the opposite. Moreover, he was the first one to be surprised as shown by his famous sentence ""I see it, but I don't believe it!" concerning this discovery that can be found in his correspondence with Dedekind in 1877 \cite{brag}:

\begin{quoting}[font+=bf,begintext= ,endtext=]
    \textit{Please excuse my zeal for the subject if I make so many demands upon your kindness and patience; the communications which I lately sent you are even for me so unexpected, so new, that I can have no peace of mind until I obtain from you, honoured friend, a decision about their correctness. So long as you have not agreed with me, I can only say: je le vois, mais je ne le crois pas. And so I ask you to send me a postcard and let me know when you expect to have examined the matter, and whether I can count on an answer to my quite demanding request.}
\end{quoting}

Let us briefly recall how Cantor managed to prove this unexpected and counter-intuitive result. In order to exhibit a bijection between the interval $[0,1]$ and the square $[0,1]\times [0,1]$, Cantor will try to find a way to rearrange all points of $[0,1]$ into points of $[0,1]\times [0,1]$. More precisely, in the first proof sent to Dedekind he considered the infinite decimal expansions of all elements of $[0,1]$ to avoid the non uniqueness of the writing in finite numbers of decimal; for example, 0,3 which can be written also 0.29999.... Then the basic idea consists of interlacing the decimals in the sense that if $x=0,\alpha_1 \beta_1 \alpha_2 \beta_2\alpha_3\beta_3...\in [0,1]$ we can set $(y,z)=(0,\alpha_1\alpha_2\alpha_3...;0,\beta_1\beta_2\beta_3...)\in [0,1]\times [0,1]$. Dedekind's objection was the cases like $x=0.642120809010\beta_7 0\beta_8 0\beta_90...$ which provide a $z$ of the form $z=0,41$ and replace it by $z=0,409999...$ will no longer correspond to the initial $x$. Cantor's final version was to use the well known, at the time, uniqueness of the irrational numbers' representation as a continued fraction. Hence, by interlacing the continued fraction representation, the question for Cantor becomes: can we put the interval $[0,1]$ into one-one correspondence with the irrational numbers $[0,1]\setminus{\mathbb{Q}}$? The answer is of course yes since removing a countable set from an uncountable one does not affect its cardinality, Cantor will demonstrate this with an easy elegant proof. This result was so surprising and unexpected that the acceptance of the article in question took longer than usual \cite{john1}. This article also contains the first appearance of the famous continuum hypothesis that Cantor will conjecture in view of his result.

Let us first focus on the conclusions made by Cantor and Dedekind regarding this result through their correspondence. In an admirable long letter to Dedekind in June 1877 \cite{brag}, Cantor explains, among other things, the supposed link between the number of coordinates and the dimension:      
\begin{quoting}[font+=bf,begintext= ,endtext=]
    \textit{For several years I have followed with interest the efforts that have been made, building on Gauss, Riemann, Helmholtz, and others, towards the clarification of all questions concerning the ultimate foundations of geometry. It struck me that all the important investigations in this field proceed from an unproven presupposition which does not appear to me self-evident, but rather to need a justification. I mean the presupposition that a $\rho$-fold extended continuous manifold needs $\rho$ independent real coordinates for the determination of its elements, and that for a given manifold this number of coordinates can neither be increased nor diminished.}
\end{quoting}
In the following passage, Cantor draws two conclusions from his discovery. First, he deduces that his result invalidates the fact that dimension can be identified with the number of independent coordinates, which implies that there is a need to look for other more adequate ways to capture this notion. Second, for him this finding automatically discredits all philosophical and mathematical conclusions that appeal to this error. From the preceding passage, it is clear that this is particularly true of the works of Gauss, Riemann and Helmholtz. 

\begin{quoting}[font+=bf,begintext= ,endtext=]
    \textit{Now it seems to me that all philosophical or mathematical deductions that use that erroneous presupposition are inadmissible. Rather the difference that obtains between structures of different dimension-number must be sought in quite other terms than in the number of independent coordinates—the number that was hitherto held to be characteristic.}
\end{quoting}

To our knowledge, this is the first appearance in history of a mathematical problem arising from the absence of a good understanding and adequate formalization of the concept of dimension. Thus, to go further in the development of mathematics, there is a real need to clarify this notion. The debate around the concept of dimension was born. Dedekind replied to Cantor on July 2, 1877 \cite{brag}, with a letter that reads:

\begin{quoting}[font+=bf,begintext= ,endtext=]
    \textit{Your words make it appear—my interpretation may be incorrect—as though on the basis of your theorem you wish to cast doubt on the meaning [BedeutungJ or the importance of this concept...Against this, I declare...my conviction or my faith...that the dimension-number of a continuous manifold remains its first and most important invariant, and I must defend all previous writers on this subject. To be sure, I gladly concede that the constancy of the dimension-number is thoroughly in need of proof, and so long as this proof has not been furnished one may doubt. But I do not doubt this constancy, although it appears to have been annihilated by your theorem. For all authors have clearly made the tacit, completely natural presupposition that in a new determination of the points of a continuous manifold by new coordinates, these coordinates should also (in general) be continuous functions of the old coordinates, so that whatever appears as continuously connected under the first set of coordinates remains continuously connected under the second. Now, for the time being I believe the following theorem: 'If it is possible to establish a reciprocal, one-to-one, and complete correspondence between the points of a continuous manifold A of a dimensions and the points of a continuous manifold B of b dimensions, then this correspondence itself, if a and b are unequal, is necessarily utterly discontinuous.'...similarly it seems to me that in your present proof...you are compelled to admit a frightful, dizzying discontinuity in the correspondence, which dissolves everything to atoms, so that every continuously connected part of the one domain appears in its image as thoroughly decomposed and discontinuous...I hope I have expressed myself with sufficient clarity; the intent of my letter is only to ask you not to engage in public polemics against the article of faith that has hitherto been regarded as a fundamental truth of the theory of manifolds until you have thoroughly examined my objection.}
\end{quoting}

We must emphasize Dedekind's relevance, foresight and wisdom in his conclusions with respect to Cantor's result. More precisely, in spite of Cantor's theorem, Dedekind continues to consider the dimension as the most important invariant of a continuous manifold. For him, Cantor's finding is indeed remarkable and thought-provoking. Nevertheless, it does not call into question the concept of dimension used by Gauss, Riemann and Helmholtz and the results they obtained in connection with this quantity. For his part, there is a tacit and completely natural assumption made by these authors in the transformations of coordinate changes. Thus, he points out that the transformation used by Cantor has a 'dizzying discontinuity' which completely scatters the starting set making it lose its original property of being composed of connected parts. This analysis will allow him to formulate the question that will keep mathematicians in check for 24 years, which can be formulated as: is there a homeomorphism between $\mathbb{R}^n$ and $\mathbb{R}^m$ if $n=m$?  Known today as \textit{Invariance of domain}, it was proved by Brouwer in 1911. Dedekind concludes his letter by asking Cantor to be more measured in the conclusions he will draw from his discovery in view of the light he has brought to him.   

In his reply of July 4, 1877 \cite{brag}, Cantor clarified his statements and recognized the relevance of Dedekind's objection:

\begin{quoting}[font+=bf,begintext= ,endtext=]
\textit{In the conclusion of my letter of 25 June I unintentionally gave the appearance of wishing by my proof to oppose altogether the concept of a $\rho$-fold extended continuous manifold, whereas all my efforts have rather been intended to clarify it and to put it on the correct footing. When I said: 'Now it seems to me that all philosophical and mathem. deductions which use that erroneous presupposition—' I meant by this presupposition not 'the determinateness of the dimension-number' but rather the determinateness of the independent coordinates, whose number is assumed by certain authors to be in all circumstances equal to the number of dimensions. But if one takes the concept of coordinate generally, with no presuppositions about the nature of the intermediate functions, then the number of independent, one-to-one, complete coordinates, as I showed, can be set to any given number. I am also of your opinion that if we require that the correspondence be continuous, then only structures with the same number of dimensions can be related to each other one-to-one; and in this way we can find an invariant in the number of independent coordinates, which ought to lead to a definition of the dimension-number of a continuous structure.\\
    However, I do not yet know how difficult this path (to the concept of dimension-number) will prove, because I do not know whether one is able to limit the concept of continuous correspondence in general. But everything in this direction seems to me to depend on the possibility of such a limiting [Begrenzung].\\
    I believe I see a further difficulty in the fact that this path will probably fail if the structure ceases to be thoroughly continuous; but even in this case one wants to have something corresponding to the dimension-number all the more so, given how difficult it appears to be to prove that the manifolds that occur in nature are thoroughly continuous.}
\end{quoting}

In summary, Dedekind maintains his support for the mathematical edifice based on the notion of dimension indexed on Cartesian coordinates. After all, at the time, mathematicians only had the latter at their disposal because the question of obtaining an accurate definition of the notion of dimension never really arose.  His defense of this notion is legitimate in view of his pertinent remark on the tacitly assumed continuity concerning Cartesian coordinates. Cantor recognizes that continuity plays an important role for the notion of dimension based on Cartesian coordinates. Nevertheless, he seems to question such a notion in view of the difficulties it creates. In particular, in the last sentence of the above passage, he seems to take the notion of dimension explicitly out of the exclusive framework of continuity. This could imply the existence of non-integer dimensions. To the best of our knowledge, this is the first time that a mathematician hypothesizes the possibility of a quantity representing the notion of dimension outside the setting of continuity. He goes even further by questioning the existence of continuous manifolds in nature. It is curious that the one who has best succeeded in giving a mathematical definition of a continuous set, doubts the existence of a continuous structure in the sensible world. This shows his philosophical position concerning the notion of space. Indeed, this could indicate that for Cantor, space is above all a concept that can be completely divorced from the sensible world. In this sense, he joins the Bolzanian position.

At this stage, we want to emphasize once again the influence of the 'Master' Bolzano on the contributions of Weierstrass, Dedekind and Cantor. Indeed, in this work we have tried as much as possible to point out precisely some of Bolzano's key contributions, which have allowed decisive steps to be taken in the development of analysis, point-set topology and set theory. For example, as pointed out in \cite{dug1}, his book "Paradoxes of the Infinite" was so important to Cantor that he sent it to Dedekind who was also inspired by it. Nevertheless, on the one hand, it is a pity that his relevance and his vanguardism are only recognized and rediscovered long after his death. For us, this affirms continuity with remark \ref{rem3}, that there must exist a temporality and maturity of the notions without which it is difficult that the community realizes the conceptual brake realized by a thinker. It was the case for the Bolzano brake. On the other hand, in view of the uneasiness of mathematicians with the notion of dimension and the need they had for an adequate formalization of the latter at the time. It is reasonable to think that if mathematicians, in particular Cantor and Dedekind, had been aware of Bolzano's work on the notion of dimension \cite{bol} published a century after (in 1948), the progression of mathematics would have been significantly richer.

\paragraph{Analysis of Cantor's result.} Let us draw some conclusions according to Cantor's apparently paradoxical result. It is worth asking at this stage: what is shocking at first sight in Cantor's theorem? First of all, this theorem invalidates the fact that the dimension is, by convention, the smallest number of real parameters necessary to describe or identify the points of an object. More precisely, the bijection between the interval and the square allows to parameterize all the points of the square by a single parameter going against the minimal character of the initial parametrization of the square. Then, is the square not of dimension two but rather one, or is this way of indexing the dimension on Cartesian coordinates wrong? One cannot imagine that for any $n\in\mathbb{N}^{\ast}$, the dimension of $\mathbb{R}^n$ is 1. We think that, for mathematicians, this has never questioned the dimension as an intuitive concept, the square is of course of dimension two, but it is rather the way used to quantify the dimension which is not adequate. 

Let's move on to Cantor's reasons for questioning this concept in order to obtain more information. It seems that initially Cantor thought that there was a link between cardinality and dimension. This in itself is not contradictory for finite sets as we will develop in the rest of this article (see seubsection \ref{4.3.2}). Nevertheless, for sets of infinite cardinals this is not the case. We need another way to quantify the dimension. Cantor surely believed that the difference in dimension between an interval and a square is directly related to the difference in their cardinality as a set of points, but it turns out that this is not the case. The square has the same cardinality of the continuum as the interval. Concerning the fact that Cantor's result indicates that an infinite set can have the same cardinality as an infinite number of copies of itself, this is not so shocking, it is of the same order as Hilbert's Grand Hotel paradox.

Knowing that Cantor probably thought at the beginning that the dimension should be indexed on the cardinality, the question is therefore what dimension he would assign to $\mathbb{Q}$ or more generally to countable sets at the beginning of his investigations? Indeed, he had already proved at that time that $|\mathbb{R}|=2^{|\mathbb{Q}|}$ which underlines the gap of cardinality between the countable and the uncountable. Thus, on the one hand, since Cantor did not question the fact that the dimension of $\mathbb{R}$ is equal to one, then the dimension of $\mathbb{Q}$ should necessarily be less than one. On the other hand, it is reasonable to think that finite sets are of dimension zero, so $\mathbb{Q}$ should then have, for Cantor, a dimension between zero and one. He surely changed his mind on the subject in view of his discovery which implies in particular that the dimension of continuous objects has no connection with their cardinality. Indeed, the square and the interval have the same cardinality but different dimensions. Nevertheless, it appears from the above passage that he must have wondered about the possibility of the notion of dimension for the non-continuous and traced in favor of this idea. Of course this development is purely speculative, we have not found any writing in this sense.  

Forget cardinality and think geometrically. What is shocking about Cantor's discovery at this level? We have explained the obstacles that Bolzano encountered in taking the step of considering the different geometric objects as sets of points. The most important one was that all points are identical, which raises the problem of distinguishing between different geometric objects. Cantor's result can be seen as an elaborate mathematical formulation of the initial Bolzanian reluctance. Indeed, Cantor's theorem proves that if we could take the points of the interval one by one and arrange them in a certain way, we could construct a square and vice versa. More generally, we can use the points obtained from a curve to fill a surface and conversely. The problem here is: if we accept the pointillist representation of these different geometric objects, then this seems to make us lose their intrinsic characteristic of being a line, a surface or a body. Knowing that we cannot accept the confusion between the different basic geometrical objects, the set of possibilities is reduced to either we need to take a step back from considering the different geometrical objects as sets of points or something is missing in this new way of looking at geometric objects, which leads to this kind of result. Of course, given the advances allowed by point-set topology, it is natural to choose the second option. The question is: what is wrong in this  process introduced by Cantor?

Continuity is what is missing, as Dedekind underlined. It is indeed an important element, but it is not the only one and not always. We think, and this is the basis of our approach, that putting the notion of point back into question, in particular through its dimension according to its contextual identity, is central to this problem. If we take Cantor's demonstration again, the points look all similar and bear in themselves no mark allowing to distinguish them. Cantor's idea consists precisely of renaming or relabeling the points of the interval $[0,1]$ by the set of all labels of the square $[0,1]\times [0,1]$. In this sense, the only ingredient that matters is the cardinality of the two sets, no information on their geometry is needed in this process. One can very easily construct, by a process similar to that of Cantor, even more shocking examples like a bijection between the Cantor ternary set and the square. At this stage, we will not only have lost the dimension through this correspondence, but also even the continuity and more precisely the connectedness. Even more twisted, we can create a bijection between the square $[0,1]\times [0,1]$ in itself via the Cantor transformation which will make us lose the information about its dimension while we are still on the same two-dimensional geometrical object. 

This indicates that it is not valid to continue to see dimensions roughly as degrees of freedom. Cantor's discovery explains that one can have this conception but not in any way. Concretely, the nuance that Cantor brings warns against the risk that this conceptualization of the dimension entails. One can for example use Cantor's correspondence and generate completely false results in physics or other sciences. In summary, if we want to capture the geometrical properties of the starting object we are obliged to perform an adequate labeling with a certain coherence between the starting geometrical object and the labels assigned to its own points.           

This highlights the basic fact that points are not coordinates and conversely. This difference between the object 'square' and the labeling attributed to its points 'coordinate' imposes more caution concerning the correspondence to choose in order to link them. Once considering the square as a set of points, it is not enough to be able to assign unique coordinates to each point to obtain a good characterization of the latter. Dedekind's pertinent remark, and later demonstrated by Brouwer, requires that for a given continuum, the coordinates must be defined in a continuous manner in terms of the points on that continuum. Now, if the starting object is not continuous, is it still the continuity of the correspondence that comes into play? The answer is no. Indeed, all compact metric spaces which are totally disconnected and without isolated points are homeomorphic to each other. This implies that they can all be uniquely labeled using only one of them, such as the Cantor set for example. However, the way these homeomorphic sets fill the space is not the same. If we consider the Smith-Volterra-Cantor set, it is of non-zero Lebesgue measure and of dimension Hausdorff and box one. Thus, if we compute their box or Hausdorff dimensions they are not necessarily equal. In this case, we need a correspondence of another type (bilipschitz will be enough).  In other words, if the labeling is required to respect the dimension of the set under consideration, in the case of non-continuous sets there may be conditions in play other than the continuity of the labeling.           

Cantor had accepted Dedekind's objection concerning continuity; nevertheless, he seems to argue for the necessity of another notion of dimension. It is reasonable to see this as Cantor's will to free the notion of dimension from its exclusive link to Cartesian coordinates. It is totally justified, since the labeling that one gives to an interval is artificial in the sense that it is subjective. We need a notion of dimension which transcends the framework of coordinates and which is linked to the intrinsic nature of the objects in question. If we look at things independently of the coordinates assigned to the points, the contextual identity of the points comes into consideration. Indeed, intrinsically a point inside a line does not play the same role as a point inside a square in terms of connectedness. More precisely, removing a point from the interior of a line forms two disjoint parts, which is not the case for the square. This is a simple proof of Dedekind's conjecture in the case interval/square since a continuous transformation must respect the connectedness. This argument allows us to highlight the contextual identity that a point can acquire according to the links it forms in its neighborhood. In this sense, we strongly believe that the basis of neighbourhoods of the points give them an identity that can be captured through a notion of dimension assigned to the point. 

In the same line of argument, the \textit{invariance of domain theorem} proved by Brouwer, one of whose consequences is the non-existence of a homeomorphism between $\mathbb{R}^n$ and $\mathbb{R}^p$ if $n\neq p$, has in particular a local scope. The latter precisely states that any continuous and injective application $f$ from an open neighborhood $U$ of a point $x\in\mathbb{R}^n$ onto $\mathbb{R}^n$, is an open map in the sense that it maps the open neighborhood of $x$ into an open neighborhood of $f(x)$.  
\begin{thm}[Brouwer 1911]
If $\mathbb{U}$ is an open subset of $\mathbb{R}^{n}$ and $f:U\rightarrow \mathbb{R}^{n}$ is an injective continuous map, then $V=f(U)$ is open in $\mathbb {R}^{n}$ and $f$ is a homeomorphism between $U$ and $V$.
\end{thm}
We will focus more deeply on several aspects of this theorem in a forthcoming article. The central idea of this article is to try, as much as possible, to bring this localization to the point by assigning a dimension to it. This allows to add to the point a signature or an information of its environment and therefore an identity. At this moment, it will be necessary to find the adequate conditions that a transformation should have to make two points correspond between them by respecting their dimensions. We have already, through the modified Bolzano's definition \ref{mbdim} of dimension presented above, emphasized that one can have a single point in a set with a different dimension from the others ones (see the end of subsection \ref{2.3}). In what follows, we will present another interesting example and also the case of a set consisting only of points of different dimensions. 

In a forthcoming article, we will analyze in depth the cat's throw among the pigeons performed by Peano through his space-filling curve. Motivated by Cantor's result, Peano will construct a continuous surjective curve of the interval $[0,1]$ in the square $[0,1]\times [0,1]$. This result, which challenges the notion of a curve by shaking the conviction that it can fill a surface because of the dimensional problem, has long been considered a pathology. It is nevertheless an example with a central role in any attempt to better understand the notion of dimension.

\section{An overview of the different dimensional concepts in the 20th century}\label{over}

 After the problem posed by Cantor through his identification of the dimension problem, mathematicians focused both on finding a proof of Dedekind's conjecture and on searching for new ways to define the notion of dimension. For the sake of brevity, we will develop in more detail some fundamental philosophical and mathematical aspects of the post-Cantor period in a coming article. Nevertheless, we will present in this section a brief account of our reading of the evolution of the notion of dimension through the most important and significant stages. This is a first overview of some different notions, we will come back to more details in future works. Thanks to the previous section, we have acquired an overview allowing us both to identify the reasons behind the particular subsequent evolution of this notion and to establish a clear framework for our contribution to this subject.

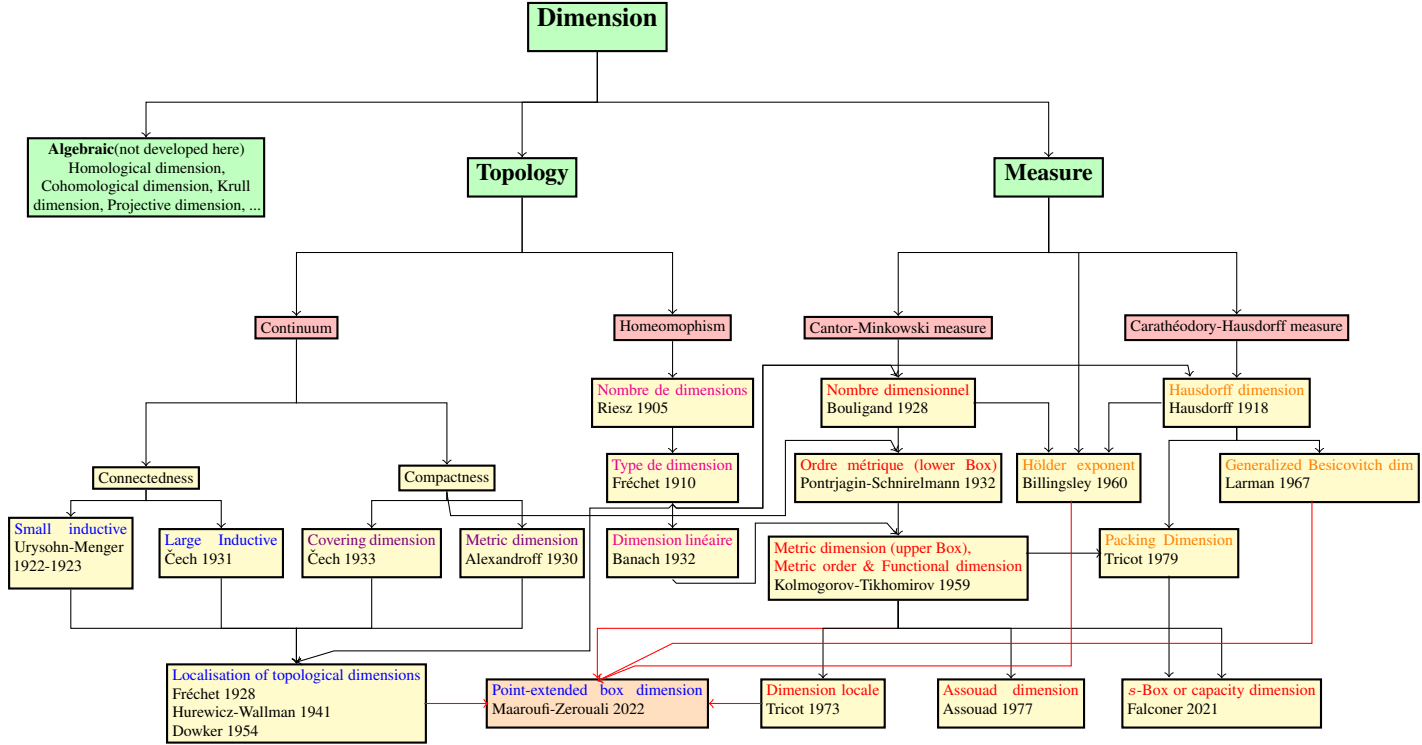
\begin{figure}[h!]
    \centering
    
\begin{tikzpicture}\tiny{

\tikzset{individu/.style={draw,thick,fill=#1!25},
individu/.default={green}}
\node[individu] (dim) at (-4,2)
{\begin{minipage}[t][0.5cm][t]{1.7cm} \begin{center}
     \normalsize{\textbf{Dimension}}\end{center}
                 \end{minipage}};
\node[individu] (top) at (-5,0) {\begin{minipage}[t][0.37cm][t]{1.3cm} \begin{center}\small {\textbf{Topology}}\end{center}
                 \end{minipage}};
\node[individu] (alg) at (-10,0) {\begin{minipage}[t][0.9cm][t]{3cm} \begin{center}{\textbf{Algebraic}(not developed here)\\ Homological dimension, Cohomological dimension, Krull dimension, Projective dimension, ...}\end{center}
                 \end{minipage}};
\node[individu] (mes) at (2,0) {\begin{minipage}[t][0.37cm][t]{1.3cm} \begin{center}\small{\textbf{Measure}}\end{center}
                 \end{minipage}};
\node[individu=red] (mescant) at (0,-2) {Cantor-Minkowski measure};
\node[individu=red] (meshaus) at (4.5,-2) {Carathéodory-Hausdorff measure};
\node[individu=yellow] (ord) at (0,-3) {\begin{minipage}[t][0.5cm][t]{1.9cm} \color{red}{Nombre dimensionnel} \color{black}{Bouligand 1928}
                 \end{minipage}};
\node[individu=yellow] (haus) at (4.5,-3) {\begin{minipage}[t][0.5cm][t]{1.8cm} \color{orange}{Hausdorff dimension} \color{black}{Hausdorff 1918}
                 \end{minipage}};
\node[individu=yellow] (hold) at (2.4,-4) {\begin{minipage}[t][0.5cm][t]{1.5cm} \color{orange}{H\"{o}lder exponent} \color{black}{Billingsley 1960}
                 \end{minipage}};  
\node[individu=yellow] (besi) at (5.6,-4) {\begin{minipage}[t][0.5cm][t]{2.5cm} \color{orange}{Generalized Besicovitch dim} \color{black}{Larman 1967}
                 \end{minipage}};       
\node[individu=yellow] (pack) at (3.6,-5) {\begin{minipage}[t][0.5cm][t]{1.7cm} \color{orange}{Packing Dimension} \color{black}{Tricot 1979}
                 \end{minipage}};                 
\node[individu=red] (conti) at (-8,-2) {Continuum};
\node[individu=red] (hom) at (-3,-2) {Homeomophism};
\node[individu=yellow] (nbr) at (-3,-3) {\begin{minipage}[t][0.5cm][t]{2cm} \color{magenta}{Nombre de dimensions} \color{black}{Riesz 1905}
                 \end{minipage}};
\node[individu=yellow] (typ) at (-3,-4) {\begin{minipage}[t][0.5cm][t]{1.6cm} \color{magenta}{Type de dimension} \color{black}{Fréchet 1910}
                 \end{minipage}};
\node[individu=yellow] (lin) at (-3,-5)  {\begin{minipage}[t][0.5cm][t]{1.6cm} \color{magenta}{Dimension linéaire} \color{black}{Banach 1932}
                 \end{minipage}};
\node[individu=yellow] (connex) at (-10,-4) {Connectedness};
\node[individu=yellow] (compac) at (-6,-4) {Compactness};
\node[individu=yellow] (ind) at (-11,-5) {\begin{minipage}[t][0.8cm][t]{1.5cm} \color{blue}{Small inductive} \color{black}{Urysohn-Menger 1922-1923}
                 \end{minipage}};
\node[individu=yellow] (Ind) at (-9,-5) {\begin{minipage}[t][0.5cm][t]{1.5cm} \color{blue}{Large Inductive} \color{black}{\v{C}ech 1931}
                 \end{minipage}};
\node[individu=yellow] (localind) at (-8,-7) {\begin{minipage}[t][0.9cm][t]{3.3cm} \color{blue}{{Localisation of topological dimensions}} \color{black}{Fréchet 1928\\Hurewicz-Wallman 1941 \\ Dowker 1954}
                 \end{minipage}};;                 
\node[individu=yellow] (cov) at (-7,-5) {\begin{minipage}[t][0.5cm][t]{1.7cm} \color{violet}{Covering dimension} \color{black}{\v{C}ech 1933}
                 \end{minipage}};
                
\node[individu=yellow] (met) at (-5,-5) {\begin{minipage}[t][0.5cm][t]{1.5cm}  \color{violet}{Metric dimension} \color{black}{Alexandroff 1930}
                 \end{minipage}};
\node[individu=yellow] (ordmet) at (0,-4) {\begin{minipage}[t][0.5cm][t]{2.6cm} \color{red}{Ordre métrique (lower Box)} \color{black}{Pontrjagin-Schnirelmann 1932}
                 \end{minipage}};
\node[individu=yellow] (metdim) at (0,-5.2)  {\begin{minipage}[t][0.7cm][t]{3.3cm} \color{red}{Metric dimension (upper Box),\\ Metric order \& Functional dimension} \color{black}{Kolmogorov-Tikhomirov 1959}
    \end{minipage}};
\node[individu=orange] (notre) at (-4,-7)  {\begin{minipage}[t][0.5cm][t]{2.8cm} \color{blue}{Point-extended box dimension} \color{black}{Maaroufi-Zerouali 2022}
    \end{minipage}};
    \node[individu=yellow] (loctri) at (-1,-7)  {\begin{minipage}[t][0.5cm][t]{1.5cm} \color{red}{Dimension locale} \color{black}{Tricot 1973}
    \end{minipage}};
     \node[individu=yellow] (assouad) at (1.5,-7)  {\begin{minipage}[t][0.5cm][t]{1.8cm} \color{red}{Assouad dimension} \color{black}{Assouad 1977}
         \end{minipage}};
         \node[individu=yellow] (falco) at (4.3,-7)  {\begin{minipage}[t][0.5cm][t]{2.5cm} \color{red}{$s$-Box or capacity dimension} \color{black}{Falconer 2021}
         \end{minipage}};
\draw[->=latex] (dim) |- (-5, 1) -| (top);
\draw[->=latex] (dim) |- (-10, 1) -| (alg);
\draw[->=latex] (dim) |- (2, 1) -| (mes);
\draw[->=latex] (top) |- (-8, -1) -| (conti);
\draw[->=latex] (mes) |- (0, -1) -| (mescant);
\draw[->=latex] (mes) |- (4.5, -1) -| (meshaus);
\draw[->=latex] (mescant) |- (0, -2.5) -| (ord);
\draw[->=latex] (meshaus) |- (4.5, -2.5) -| (haus);
\draw[->=latex] (mes) |- (2.4, -1) -| (hold);
\draw[->=latex] (haus) |- (5.6, -3.5) -| (besi);
\draw[->=latex] (haus) |- (3.6, -3.5) -| (pack);
\draw[->=latex] (ord) |- (0,-3.5) -| (ordmet);
\draw[->=latex] (ordmet) |- (0,-4.7) -| (metdim);
\draw[->=latex] (top) |- (-3, -1) -| (hom);
\draw[->=latex] (hom) |- (-3, -2.5) -| (nbr);
\draw[->=latex] (nbr) |- (-3, -3.5) -| (typ);
\draw[->=latex] (typ) |- (-3, -4.5) -| (lin);
\draw[->=latex] (conti) |- (-10, -3) -| (connex);
\draw[->=latex] (conti) |- (-6, -3) -| (compac);
\draw[->=latex] (connex) |- (-11, -4.3) -| (ind);
\draw[->=latex] (Ind) |- (-8, -6) -| (localind);
\draw[->=latex] (ind) |- (-8, -6) -| (localind);
\draw[->=latex] (cov) |- (-8, -6) -| (localind);
\draw[->=latex] (met) |- (-8, -6) -| (localind);
\draw[->=latex] (connex) |- (-9, -4.3) -| (Ind);
\draw[->=latex] (compac) |- (-7, -4.3) -| (cov);
\draw[->=latex] (compac) |- (-5, -4.3) -| (met);
\draw[red][->=latex] (metdim) |- (-4, -6) -| (notre);
\draw[->=latex] (metdim) |- (-1, -6) -| (loctri);
\draw[->=latex] (metdim) |- (1.5, -6) -| (assouad);
\draw[->=latex] (metdim) |- (4.3, -6) -| (falco);
\draw[black] [->=latex] (1,-3)--(2,-3) -- (2,-3.66);
\draw[black] [->=latex] (3.5,-3)--(2.8,-3) -- (2.8,-3.66);
\draw[black] [->=latex] (3.6,-5.3)--(3.6,-6.65);
\draw[black] [->=latex] (-3,-4.35) -- (-2.95,-4.4)-- (-1.8,-4.4) -- (-1.8,-2.5) -- (-0.1,-2.5) -- (-0.02,-2.65);
\draw[black] [->=latex] (-3,-4.35) -- (-3.05,-4.4) -- (-4.1,-4.4)-- (-4.1,-6.3)-- (-7.5,-6.3)-- (-8,-6.45);
\draw[black] [->=latex] (-3,-4.35) -- (-2.95,-4.4)-- (-1.8,-4.4) -- (-1.8,-2.5) -- (3.85,-2.5) -- (3.88,-2.65);
\draw[black] [->=latex] (-6,-4.2) -- (-5.95,-4.5) -- (-1.5,-4.5)-- (-1.5,-3.5)-- (-0.1,-3.5)-- (-0.02,-3.65);
\draw[black] [->=latex] (-3,-5.35) -- (-2.95,-5.4)-- (-1.9,-5.4) -- (-1.9,-4.6) -- (-0.15,-4.6) -- (-0.02,-4.75);
\draw[black] [->=latex] (1.7,-5) -- (2.7,-5);
\draw[red] [->=latex] (-6.3,-7) -- (-5.45,-7);
\draw[red] [->=latex] (-1.8,-7) -- (-2.52,-7);
\draw[red] [->=latex] (2.3,-4.3) -- (2.3,-6.5)--(-3.5,-6.5)--(-4,-6.7);
\draw[red] [->=latex] (5.5,-4.3) -- (5.5,-6.2)-- (2.3,-6.2)--(-3,-6.2)--(-4,-6.7);
}
\end{tikzpicture}

    \caption{Diagram of the notions of topological and measured dimension}
    \label{dia1}
\end{figure}

The main objective of this section is to comment on the diagram (figure \ref{dia1}), except the localized notions which will be discussed in the next section. At the outset, it should be noted that the arrows in this diagram can have a conceptual, technical, inspirational or mind-expanding meaning. We do not pretend to present a holistic diagram, there must be many dimensional notions left, that are not included in it. Moreover, apart from the explicitly formalized notions of dimensions, there are certainly notions of dimension implicitly involved in several theories in view of the central role of this notion. As announced in the introduction of this article, we strongly believe that a deep state of the art on the notion of dimension should be conducted. We present through this scheme a starting point expressing our way of seeing things, which can be reviewed, completed and improved in the future. To be brief, we will not develop all the definitions of dimensions' notions involved in the diagram. Instead, we will focus only on some of them and we will indicate the references for the others. We will surely analyze these notions in more depth in forthcoming works.

Let us take up the historical thread that we have developed in the previous section. After the deep questioning initiated by Cantor's result, followed by Peano's curve and Dedekind's conjecture, the mathematical world was in turmoil. It was necessary to prove Dedekind's conjecture and at the same time to try to introduce an adequate notion of dimension, to succeed in doing both at the same time would be any mathematician's grail. Thus, the greatest mathematicians of the time had taken up the challenge. The new notions of dimensions turned out to be important tools to both analyze finely the topological and measured properties of sets and at the same time generate deep, difficult and crucial questions for the development of these theories. We want to focus on the notions of dimension related to topological and measured aspects, hence the two main branches of the scheme. This follows the distinction we have identified since Euclid's elements (see sub-section \ref{2.1.1}). Moreover, this is justified since the families of definitions of these branches are based either on topological notions or on measured notions. Concerning the branch related to algebra, it will not be developed in this article.   

\subsection{Topological dimensions related to connectedness: \small{Poincaré (1854–1912), Brouwer (1881-1966), \v{C}ech (1893-1960), Urysohn (1898-1924) and Menger (1840-1921)}}\label{3.1}
We have explained why we think there is an intimate connection between research on the notion of dimension and the notion of continuum. In the diagram (figure \ref{dia1}), it seems natural to place the continuum in the branch of topology, the latter being characterized by the fact that the notions of dimensions found therein take only integer values. The essence of the modern notion of the continuum, as we explained above, being connectedness and compactness, we have chosen to classify the notions of dimensions by their link with one of these two characteristics of the continuum.

Poincaré was one of the first to tackle the problem of dimension. In two philosophico-mathematical articles at the end of his career in \cite{poincare1} in 1903 and \cite{poincare2} in 1912, he dealt with the notion of continuum, dimension and the question of the link between human conceptualization of geometry and the sensible world. We must point out that Poincaré had already started this questioning in two previous articles, \cite{poincare3} in 1895 and \cite{poincare4} in 1898 where he had tried a first way to introduce the notion of dimension based on Lie groups. We will not deal with this notion in this article. Briefly, we can convene that for Poincaré the notion of continuum is prior to the conceptualization of any kind of geometry (\cite{poincare1}, 1903) :   

\begin{quoting}[font+=bf,begintext= ,endtext=]
\textit{L'espace euclidien n'est pas une forme imposée à notre sensibilité, puisque nous pouvons imaginer l'espace non-euclidien; mais les deux espaces euclidien et non-euclidien ont un fond commun, c'est ce continuum amorphe dont je parlais au début; de ce continuum nous pouvons tirer soit l'espace euclidien, soit l'espace Lobatchewski...}
\end{quoting}
In particular, he implicitly criticizes in this passage the Kantian conception of space as a pure intuition. As pointed out by the quote of Poincaré in the introduction of this article, for him, the common point between all geometries is the continuum and its dimensions. He certainly did not conceive the notion of dimension outside the continuum as he wrote in \cite{poincare2} in 1912:    
\begin{quoting}[font+=bf,begintext= ,endtext=]
\textit{...la question du nombre des dimensions est intimement liée à la notion de continuité et elle n'aurait aucun sens pour celui qui voudrait faire abstraction de cette notion.}
\end{quoting}
Poincaré initiated the reflection on the notion of dimension by introducing the notion of 'cuts' in relation to connectedness. He had succeeded in making the link with the definition of the Greeks that we have presented above (subsection \ref{2.1.2}). Basically, he points out that to disconnect a line into two disjoint parts, one point is enough. To disconnect a surface into two disjoint parts, all that is needed is a line. And finally, for the body it is enough to make a cut with a surface. We can see Poincaré's definition as an attempt to quantify the 'degree of connectedness' of a set. Even if this way of introducing dimension was criticized by both Riesz in 1909 and Brouwer in 1913 independently (see \cite{john1}) and via the same counter example of the double inverted cone which would be of dimension one with respect to Poincaré's definition, this way of introducing dimension in an inductive way allowed the development of the so-called 'inductive' definitions. The first one, inspired by Poincaré, was the Dimensionsgrad introduced by Brouwer in 1913 (in \cite{brou}) based on connectedness and his own research on the invariance of domain problem. Followed independently on the same idea by Urysohn (\cite{ury}) and Menger (\cite{men}) respectively in 1922 and 1923 by the introduction of the small inductive dimension (ind). Later, in 1931, \v{C}ech (\cite{cech}) introduced the large inductive dimension (Ind). 

\begin{Def}[Urysohn-Menger dimension or small inductive dimension 1922-1923]\label{ury-men}
let $X$ be a regular space, the small inductive dimension of $X$, denoted by $ind(X)$, can be defined as follows:  
\begin{enumerate}
\item $ind(X) = -1 \iff X = \emptyset$;
\item $ind(X) \le  n\in\mathbb{N}$, if for every point $x \in  X$ and every neighborhood $V$ of the point $x$
there is an open set $U\subset X$ such that $x\in U \subset \overline{U}\subset V$ and $ind( \partial U) \leq
n -1$;
\item $ind(X)=n$ if $ind(X)\leq n$ and the inequality $ind(X)\leq n - 1$ does not hold;
\item $ind X =\infty $ if the assumption $ind(X) \le n$  does not hold for each $n=-1,0,1,2...$.
\end{enumerate} 
\end{Def} 

\begin{Def}[Brouwer-\v{C}ech dimension or large inductive dimension 1931]\label{brou-cech}
let $X$ be a normal space, the large inductive dimension of $X$, denoted by $Ind(X)$, can be defined as follows:
\begin{enumerate}
\item $Ind(X) = -1 \iff X = \emptyset$;
\item $Ind(X) \le  n\in\mathbb{N}$, if for every closed set $V \subset  X$ and every open set $U\subset X$ which contain $V$ there is an open set $W\subset X$ such that $V\subset W\subset \overline{W}\subset U$ and $Ind(\partial W)\leq n - 1$;
\item $Ind(X)=n$ if $Ind(X)\leq n$ and the inequality $Ind(X)\leq n - 1$ does not hold;
\item $Ind X =\infty $ if the assumption $Ind(X) \le n$  does not hold for each $n=-1,0,1,2...$.
\end{enumerate} 
\end{Def}
We can notice that the notion $Ind$ is the same as $ind$ where we change the point in its definition by a closed set. Of course, these notions of dimension transcend the sole framework of connectedness, but it is really this property of the continuum that inspired both their introduction by Poincaré and their development by his successors as clearly expressed by Urysohn in \cite{ury1} in 1925: 
\begin{quoting}[font+=bf,begintext= ,endtext=]
\textit{Il est à remarquer que la notion de connexite qui fut mon point de départ, n'intervient pas (ou, du moins peu s'en faut) dans l'exposé actuel.}
\end{quoting}
Moreover, these definitions allow a fine analysis of the property of connectedness. For example, we have the following implication $ind(X)=0\Rightarrow X\ is\ totally\ disconnected$ with equivalence if $X$ is a compact metric space. We should also mention that these notions are invariant by homeomorphism and that in general $ind(X)\leq Ind(X)$ with equality in the case of separable metric spaces. For more development in connection with these concepts, we refer the reader to reference book on the subject \cite{eng}. Let us now present the definition of the Bolzano's dimension modified by $\check{K}$atetov in 1983 in \cite{kat1} to give it a modern form. We use the notions of the subsection \ref{2.2.3}.  
\begin{Def}[Bolzano inductive dimension 1983]\label{bol-ind}
let $X$ be metric space, the Bolzano inductive dimension of $X$, denoted by $Bind(X)$, can be defined as follows:
\begin{enumerate}
\item $Bind(X) = -1 \iff X = \emptyset$;
\item $Bind(X) = 0$ iff $X\neq \emptyset$ and, for any $x\in X$ and any $\epsilon> 0$, there exists a positive $\delta < \epsilon$ such that $neighbours(x,\delta)=\emptyset$.
\item $Bind(X)=n\in\mathbb{N}^{\ast}$, iff 
\begin{itemize}
    \item $Bind(X)\leq n - 1$ does not hold,
    \item for every $x \in X$ there is an $\epsilon > 0$ such that $Bind(neighbours(x,\delta))\leq n-l$ whenever $0< \delta < \epsilon$.
\end{itemize}
\item If $Bind(X) = n$ for no $n = -1,0, 1,2 ...$, we put $Bind(X)=\infty$. 
\end{enumerate} 
\end{Def}
We note that the difference between the definition \ref{mbdim} we introduced and this one is the explicit localization at the point, see subsection \ref{2.2.3} for the development concerning the consequences of the localization. We have already mentioned that Bolzano's notion of dimension was not published until 1948 posthumously, well after the independent introduction of the notion of small inductive dimension. Here is the comment of $\check{K}$atetov in \cite{aul} who has examined in depth its properties and the link it has with the notion of $ind$: 
\begin{quoting}[font+=bf,begintext= ,endtext=]
\textit{Thus, essentially, this Theorem says that in the case of metric spaces Bolzano's definition in a sense gives the same as the small inductive dimension.\\
Bolzano's approach is remarkable in many aspects. The intuitive idea, close to the approach of P. Urysohn and K. Menger, is immediately converted to a definition, precise enough even by contemporary standards. He selects a definition, which is the most efficient one in the situation when the notions of topological space and homeomorphism are still unknown. He presents examples, which show that he works - roughly speaking - with arbitrary subsets of a threedimensional space, and not only with subsets defined by means of sufficiently "nice" functions.\\
It may be said that - with several reservations - B. Bolzano is the discoverer of the dimension $ind$. He was definitely the first one who discovered the basic idea of that dimension and formulated it the best way possible in his times}
\end{quoting}
In this passage we can see that $\check{K}$atetov recognizes the genius of Bolzano, and the fact that he was well ahead of his time. Nevertheless, we do not agree with him on the intuitive aspect of the discovery of Bolzano. Knowing the steps Bolzano went through before formalizing his definition of dimension and also his position concerning the subject of intuition, we think that even Bolzano would not have agreed with him. We have already underlined Bolzano's genius, but this situation raises the unavoidable and difficult question of the determinism of the discoveries of mathematical structures. More generally, this question concerns the mathematical objects' objectivity; namely, the question of \textit{mathematical realism}. To be more precise: how is it that the definition of dimension introduced independently by Uryshon and Menger, and independently of Bolzano's definition, about a century later, with all the mathematical progress made in the meantime, is exactly in the same spirit, not to say the same, as Bolzano's? 

\subsection{Topological dimensions related to compactness: \small{Lebesgue(1875-1941), Mazurkiewicz(1888-1945), Alexandroff(1896-1982) and \v{C}ech(1893-1960)}}\label{3.2}
To begin with, Henri Lebesgue was the first mathematician to see a link between the dimension of a set and its covering, which is far from being obvious. First, it is important to know that Lebesgue was one of the great specialists in the properties of covering. For example, there is the Borel-Lebesgue covering theorem which states that from any open covering of a bounded closed set of $\mathbb{R}^n$ one can extract a finite subcovering. This theorem was first proved by Borel in his 1894 Ph.D. thesis \cite{bor} for the countable case, and extended to the uncountable case by Lebesgue in 1904 \cite{leb}. One can also mention the Lebesgue measure theory or even the famous Lebesgue's number lemma. Secondly, as soon as René Maurice Fréchet formalized the notion of compactness in the general framework of abstract sets in his 1906 Ph.D. thesis \cite{fre1}, this notion raised a lot of interest from mathematicians. Indeed, we have explained above the link between the notion of compactness and the Bolzano-Weierstrass theorem. Fréchet was the first mathematician to make the link between the Bolzano-Weierstrass theorem and the Borel-Lebesgue covering theorem (see \cite{pier} and \cite{ram}). It is therefore natural that the fact that there is a link between the extraction of convergent sub-sequences and the finite covering of sets has automatically interested mathematicians. 

In his attempt to prove the invariance of domain theorem in 1911, Lebesgue will highlight a property of the covering directly related to the notion of dimension: all $n$-dimensional domain having a finite sufficiently small closed cover $A_1, A_2 ... ,A_p$ has at least one point contained in at least $n+1$ of these sets. Lebesgue's proof \cite{leb1} of this property contained an error that Brouwer will point out, which contributed to the famous quarrel between the two mathematicians. It is only in 1921 \cite{leb2} that Lebesgue will give a correct demonstration. In the case of compact metric spaces (see \cite{eng}), Stefan Mazurkiewicz was the first mathematician (in 1915 in \cite{maz}) to introduce a notion of dimension on the basis of the ideas developed by Lebesgue in connection with this property of covering. The latter has not had much influence. It is only in 1930 in \cite{alex} that Pavel Allexandroff introduced the notion of metric dimension in the framework of metric spaces, transcending the sole setting of compactness. Three years later, in \cite{cech1}, Eduard \v{C}ech introduced the notion of covering dimension in the framework of normal spaces. Nevertheless, these two notions are based on the property of covering put forward by Lebesgue and they are equivalent in the case of compact metric spaces. In order to present \v{C}ech's covering dimension, we have to begin first by the notion of the order of a family of sets.
\begin{Def} Let $X$ be a set and $\mathcal{A}$ a family of subsets of $X$. The order of the family $\mathcal{A}$ is the largest integer $n$ such that the family $\mathcal{A}$ contains $n+ 1$ sets with a non-empty intersection. If no such integer exists, the family has an infinite order.\\
A cover $\mathcal{B}$ is a refinement of the cover $\mathcal{A}$, if for every $B\in \mathcal{A}$ there exists an $A\in\mathcal{A}$ such that $B\subset A$. 
\end{Def}
 
\begin{Def}[\v{C}ech-Lebesgue dimension or covering dimension 1933]\label{leb-cech}
let $X$ be a normal space, the covering dimension of $X$, denoted by $dim(X)$, can be defined as follows:
\begin{enumerate}
\item $dim(X)\leq n$, with $n = - 1, 0, 1, 2, ... ,$ if every finite open cover of the space $X$ has a finite open refinement of an order $\leq n$.
\item $dim(X)=n$ if $dim(X)\leq n$ and the inequality $dim(X)\leq n - 1$ does not hold;
\item $dim(X) =\infty $ if the inequality $dim(X) \le n$  does not hold for each $n=-1,0,1,2...$.
\end{enumerate} 
\end{Def}
For the definition of Alexandroff's metric dimension, it is based on the same principle with the additional precision concerning the fixed diameter $\epsilon$ of the covers, for more details see \cite{eng}. These notions of dimension are all invariant by homemorphism. We must point out that, by the the \textit{coincidence theorem} based on the \textit{compactification theorem} we have $dim(X)=ind(X)=Ind(X)$ for every separable metric space $X$ (see \cite{eng}). We should also point out that in general, these dimensional notions do not respect the union property, i.e. we do not have $Ind(X\cup Y)=\max(Ind(X),Ind(Y))$. For example, we have $Ind(\mathbb{Q}\cap [0,1])=Ind((\mathbb{R}\backslash\mathbb{Q})\cap [0,1])=0$, while $Ind([0,1])=1$. This is also the case for $ind$ and $dim$. 

\subsection{Topological dimensions related to an axiomatization and homeomorphism: \small{Riesz(1880-1956), Fréchet(1878-1973) and Banach(1892-1945)}}\label{3.3}
As pointed out in \cite{aul}, Frigyes Riesz was one of the first mathematicians to suggest how to axiomatize the point set topology in his two articles of 1906 and 1908. He came to this in connection with his research on how to capture what the mathematical continuum is \cite{rod}. For example, he succeeded in correctly formulating the concept of connectedness in 1906, Nels Johann Lennes had also succeeded in obtaining the same formulation at the same time and independently of Riesz \cite{aul}. Today, Riesz is best known as one of the founders of functional analysis. His contributions are decisive, one can quote among others the representation theorem of Riesz, formulated simultaneously and independently by Fréchet in 1906 \cite{aul}. Or again, his theorem stating that any topological Hausdorff vector space is of finite dimension if and only if it is locally compact. This is less well known, but Riesz had started his research by being interested in the problem of dimension. We mentioned above that Riesz was critical of both the conception of space and the way of introducing the notion of dimension in Poincaré's 1903 article \cite{poincare1}. Through his 1905 paper \cite{rie}, Riesz was the first mathematician to propose an axiomatization of the notion of dimension. His objective was to base geometry on the concept of dimension, which he quickly abandoned in view of the difficulty of constructing an adequate theory of dimension \cite{john2}.

The short 1905 article by Riesz in question is split into two parts. In the first part, he will demonstrate that any discontinuous set in the plane is part of a continuous curve in the same plane and without multiple points. He precise that this remains true in any dimension. He intends by discontinuous set that any set having no connected subset, which means totally disconnected sets in the modern sense. He ends this first part by stating, without demonstration, that the consequence of his theorem is that all perfect discontinuous (totally disconnected) sets are homeomorphic. It is only in the second part that he will introduce his notion of dimension which we can formulate as follows: 
\begin{Def}[Riesz' number of dimensions 1905]\label{riesz}
Let $E$ be a set, it can be assigned a number of dimension $d(E)$ to $E$ if the function $d$ meets the following conditions:
\begin{enumerate}
\item $E\subset E' \Rightarrow d(E)\leq d(E')$; (Inclusion) 
\item We do not have simultaneously $d(E\cup E')>d(E)$ and $d(E\cup E')>d(E')$. Which is equivalent to $d(E\cup E')=\max(d(E),d(E'))$; (Union) 
\item If $d(E)=n$ and $d(E')=m$ then $d(E\times E')=n+m$; (Cartesian product)
\item Let $f$ be a homemorphism from $E$ into $f(E)=E'$, then $d(E)=d(E')$; (Invariance by homeomorphism) 
\item The dimension of the interval $[0,1]$ is one. (Concordance condition)  
\end{enumerate} 
\end{Def}
Riesz adds to this definition in the text the convention that a point is of dimension zero. Riesz was surely aware of the open question of the time posed by Dedekind concerning the non-existence of a homemorphism between $\mathbb{R}^n$ and $\mathbb{R}^{n+p}$ for $p\geq 1$. Thus, this axiomatization shows his strong conviction concerning the validity of the Dedekind hypothesis. It cannot be otherwise, because if a mathematician manages to show the contrary, it would mean that his definition is completely false and contradictory. Indeed, conditions 3. and 5. of his definition impose that $\forall n\in\mathbb{N}^{\ast}$, the dimension of $\mathbb{R}^n$ is equal to $n$. Now condition 4. imposes the equality of the dimensions in case of existence of a homemorphism between two sets. In other words, this is a real risk. This risk would perhaps be minimally controlled if Riesz was aware of the only result existing at the time and due to Jacob L$\ddot{u}$roth in 1878 \cite{lur}, stating that $\mathbb{R}$ and $\mathbb{R}^2$ are not homeomorphic. 

In the rest of his article, based on his axiomatization, he will show that every discontinuous (totally disconnected) set is necessarily of dimension zero. According to the conditions in his definition of dimension, the demonstration is very easy. Basically, the $n$ copies (Cartesian product) of any totally disconnected sets is a totally disconnected set and homemorphic to a strict subset of the interval $[0,1]$. Using the postulates 1., 3., 4. and 5. of his definition, we obtain the result. Note that this result is not true according to Fréchet's definition. We want to emphasize here that Riesz's result is essentially due to his condition 3. which turns out to be too rigid. In other words, in general, the dimension of a Cartesian product of sets is not equal to the sum of the dimensions of the sets in question. It is worth noting that there have been other attempts at axiomatizations, notably that of Menger in 1929 and Alexandroff in 1932 \cite{aul}, both retain the condition on invariance by homemorphism, and neither of them contain the condition on the Cartesian product. 

Let us now turn to dimension in the Fréchet sense. First of all, we can say that the two mathematicians Riesz and Fréchet have in common a real orientation towards the maximum of abstraction \cite{john2}. They have thus contributed to found the analysis and topology of the twentieth century in the framework of completely abstract spaces. Fréchet is known, for example, for having given the theorems on the real line a wide scope in abstract topological spaces. He succeeded in extracting the essence of these theorems to make a more general theory. Among other things, he introduced the notions of complete metric space, compactness and uniform convergence. In 1910, he published an article entitled \textit{Les dimensions d'un ensemble abstrait} exposing his conception of the notion of dimension. Like Riesz, he used the notion of homemorphism to formalize his notion of dimension. In his article, he does not quote Riesz and we did not find any reference indicating whether he had been aware of Riesz's work or not. As pointed out in \cite{arbo}, Fréchet returned to the question of dimension and the questions it generates until the end of his career, even if from 1928 he turned to probability and its applications in relation to the position he obtained at the Sorbone.

As the first two paragraphs of Fréchet's article indicate, he recognizes the importance of Cantor's theory of cardinals. Nevertheless, he thinks that the degree of abstraction is so high that important information is lost. For him, as we also pointed out at the end of the sub-section \ref{2.3.2}, comparing sets only via bijective correspondences is not enough, one must take into account the mutual relations between the elements contained in each sets. This is why he introduced a comparison between sets based on continuity.     

\begin{quoting}[font+=bf,begintext= ,endtext=]
\textit{L'introduction de la notion de puissance d'un ensemble a été d'une importance capitale pour la théorie des ensembles abstraits. Mais en fait les seuls types de puissances infinies qui interviennent dans les applications sont celle d'un ensemble dénombrable et celle du continu linéaire. Cela tient à ce que la définition de la puissance d'un ensemble comporte un si haut degré d'abstraction qu'elle ne fait intervenir en aucune façon les relations mutuelles des divers éléments de l'ensemble.\\
Il y avait donc lieu de chercher à établir une comparaison moins grossière des ensembles, une comparaison où l'on tienne compte de ces relations mutuelles sans préciser pour cela la nature des éléments afin de pouvoir
l'appliquer encore aux ensembles abstrait. C'est à quoi l'en arrive en tenant compte de la continuité, et en introduisant la notion plus précise du nombre de dimensions d'un ensemble. Je consacrerai ce mémoire à l'étude
des ensembles dont le nombre de dimensions est fini.}
\end{quoting}
Let us now present the concept of dimension according to Fréchet: 

\begin{Def}[Fréchet' type of dimension 1910]\label{frech}
Let $E$ and $E'$ be two sets, we denote by $dE$ and $dE'$ respectively the type of dimension of $E$ and $E'$: 
\begin{enumerate}
\item If $E$ is homemorphic to a part of $E'$ then $dE\leq dE'$;
\item If $dE\leq dE'$ and there is no homemorphism of $E'$ in $E$ or in any part of $E$ then $dE< dE'$;
\item If $E$ is homemorphic to a part of $E'$ and $E'$ is homemorphic to a part of $E$ then $dE = dE'$;
\item If there is no homeomorphism between the two sets, we say that their types of dimension are not comparable.
\end{enumerate} 
\end{Def}

Fréchet is recognized as the first mathematician, apart from Bolzano of course, to formalize a precise and achieved notion of dimension which is free from the notion of coordinate. Here is what Borel wrote about Fréchet's notion of dimension \cite{arbo}:

\begin{quoting}[font+=bf,begintext= ,endtext=]
\textit{C'est parce qu'il avait préalablement conçu la notion d'espace topologique abstrait qu'il a pu donner, pour la première fois, une définition du nombre de dimensions indépendante de la notion de coordonnées ou de paramère numérique.}
\end{quoting}

However, his notion of dimension is a totally comparative notion in the sense that it is not intended to calculate the dimension of a set but rather to compare the dimensions of sets. Moreover, as suggested by point 4. of his definition and exposed in his article \cite{fre} via examples, unlike Cantor's theory of cardinals, the order in the types of dimension is not total. Let us quickly analyze some of his results that we consider important for the following. Thanks to this definition, Fréchet will succeed in giving a partial answer to Dedekind's conjecture. Indeed, he will show that $d\mathbb{R}^n\leq d\mathbb{R}^{n+p}$ and by using the result of L$\ddot{u}$roth \cite{lur}, which he quotes in his article, he deduces that $d\mathbb{R}<d\mathbb{R}^{2}\leq d\mathbb{R}^{3}. ...\leq d\mathbb{R}^{n}...$, but he did not know the result of L$\ddot{u}$roth \cite{lur1} implying that $d\mathbb{R}^{2}<d\mathbb{R}^{3}$ as reported in \cite{john2}. In any case, this makes him distinguish only two possibilities: either $d\mathbb{R}^n$ have the same type of dimension from a certain rank $n\geq 2$, or we have $d\mathbb{R}<d\mathbb{R}^{2}< d\mathbb{R}^{3}...< d\mathbb{R}^{n}...$. Fréchet, therefore assumes the second case to be true, which is more reasonable.  

 We want to emphasize that Fréchet will bring two completely new, counter-intuitive, original and even unthinkable ideas to his 1910 article. This is the kind of idea that inspires and advances the mind. The first one being the possibility of existence of non-integer dimensions. We pointed out in subsection \ref{2.3.2} that Cantor had hypothesized the need for a notion of dimension outside the continuum framework. To our knowledge, he was the first and only one to go that far. We do not know if Fréchet was aware of Cantor's hope, but he will succeed in going much further than Cantor expected. Fréchet will succeed in explicitly constructing an infinite uncountable sequence of sets whose dimensional types are distinct and ordered in the interval $[0,1]$.

\begin{quoting}[font+=bf,begintext= ,endtext=]
\textit{Je vais maintenant montrer que l'on pent former explicitement une suite ordonndée de types de dimension inférieurs à $d\mathbb{R}_1$, tous distincts et formant un ensemble non dénombrable.}
\end{quoting}

He will thus not only take the notion of dimension out of the continuum, he will even take it out of the cardinality of the continuum in the sense that his notion of dimension is also applicable to countable sets. In fact, Fréchet's notion of dimension is based on the accumulation points of a set. So, to build his sets of different types of dimensions with values in $[0,1]$, he will start with a single accumulation point through the set $\{1,\frac{1}{2},\frac{1}{3},\frac{1}{4},...,\frac{1}{n},...\}\cup\{0\}$ and build sets with more and more accumulation points. With his result, the notion of dimension can be not only non-integer but even irrational. However, all totally disconnected compact metric sets which have no isolated points such as the Cantors, have the same type of dimension according to Fréchet's definition, because they are all homeomorphic to each other. Fréchet has clearly prepared the ground for Hausdorff and Bouligand at the conceptual level to introduce their notions of measured dimensions which can take non-integer values. Moreover, both authors quote Fréchet in their articles where they introduce their notions of dimension. This explains the arrows in the diagram (figure \ref{dia1}) going from Fréchet's cell to the measured dimensions. We must point out that there was a resistance of some mathematicians to the notions of dimensions that can take non-integer values like that of Fréchet, Hausdorff and Bouligand, preferring to them the notions of inductive and covering dimensions, as testified by the following passage in a letter that Fréchet sent to Kazimierz Kuratowski in 1933 \cite{arbo}:  
\begin{quoting}[font+=bf,begintext= ,endtext=]
\textit{Certains mathématiciens n'acceptent de considérer comme définition du nombre de dimensions ni la mienne ni celle de Hausdorff et Bouligand. C'est un point de vue que je ne partage pas, mais que je considère
comme soutenable, c'est un peu une affaire de tempérament.}
\end{quoting}

The second great and impressive idea in his 1910 article is the possibility of comparing infinite dimensional spaces with each other via their types of dimension. In other words, his notion of dimensions implies that there is a rigorous way to measure how infinite dimensional a set is. Through the following passage, he underlines the importance of this idea:      

\begin{quoting}[font+=bf,begintext= ,endtext=]
\textit{Nous ne nous sommes occupés à peu près exclusivement que des types de dimension finis. Mais c'est surtout l'étude des types infinis qui est intéressante au point de vue des applications. Elle est même indispensable si l'on veut établir un rapprochement utile entre le Calcul Fonctionnel et la Théorie des Fonctions d'une infinité de variables.}
\end{quoting}

Fréchet will demonstrate, for example in 1924 \cite{arbo}, that $d\mathcal{C}([0,1],\mathbb{R})=d\mathcal{C}([0,1]^n,\mathbb{R})$, where $\mathcal{C}([0,1],\mathbb{R})$ and $\mathcal{C}([0,1]^n,\mathbb{R})$ denote respectively the spaces of continuous functions from $[0,1]$ and $[0,1]^n$ with values in $\mathbb{R}$ endowed with the uniform convergence norm. There is a lot to say about Fréchet's work, we will do it in a future work. In any case, this idea of the possibility of quantifying the infinite dimension will be taken up by Urysohn, Menger, Banach and Kolomogorov later on. For example, Stefan Banach well familiar with Frechet's works on the notion of type of dimension, he then introduced the notion of linear dimension in 1932 in \cite{ban} as follows: 

\begin{Def}[Banach' linear dimension 1932]\label{bana}
Let $E$ and $E'$ be two topological vector spaces, we denote by $dim_lE$ and $dim_lE'$ respectively the linear dimension of $E$ and $E'$: 
\begin{enumerate}
\item If $E$ is isomorphic to a closed linear subspace of $E'$ then $dim_lE\leq dim_lE'$;
\item If $dim_lE\leq dim_lE'$ and there is no isomorphism from $E'$ to any closed linear subspace of $E$ then we posit $dim_lE< dim_lE'$;
\item If $E$ is isomorphic to a closed linear subspace of $E'$ and $E'$ is isomorphic to a closed linear subspace of $E$ then $dim_lE = dim_lE'$;
\item If there is no isomorphism between the two sets, we say that their linear dimensions are not comparable.
\end{enumerate} 
\end{Def}
Of course this definition of the linear dimension is only of interest in infinite dimension. This definition will be taken up by Kolmogorov in the Bourbaki seminar in 1958 \cite{K2}. This justifies the arrow going from the Banach cell to the Kolmogorov cell in the diagram (figure \ref{dia1}). Concerning Uryshon, he had underlined since the beginning of his work on the dimension in his 1925 dissertation \cite{ury1} the possibility of introducing transfinite dimensions because the dimension is an ordinal number. Before adding that this extension seemed to him devoid of interest, for the time, since there is a problem to include the Hilbert space in the classification:
\begin{quoting}[font+=bf,begintext= ,endtext=]
\textit{On peut introduire au lieu d'une dimension infinie des dimensions transfinies (pour tous les nombres de la deuxième classe, et même, peut-être au delà), car la dimension est (d'après sa définition) un nombre ordinal. Cette extension me semble d'ailleurs, au moins pour le moment, privée d'intérêt, d'autant plus qu'il resterait, à ce qu'il semble, même dans ce cas des ensembles ne rentrant pas dans la classification (les domaines de l'espace Hilbertien).}
\end{quoting}
This idea will be taken up later by Hurewicz and Wallman in \cite{hu1}.

\subsection{Measured dimensions related to measures: \small{Hausdorff(1868-1942), Bouligand(1889-1979) and Billingsley(1925-2011)}}\label{3.4}
We have already identified the link between the notion of measure and that of dimension going back to the Greeks in the sub-section \ref{2.1.1}. In his search for an adequate notion of dimension, as he had promised, Cantor tried in 1884 in \cite{cant5} to formalize a general notion of volume which he considered indispensable to obtain a notion of dimension independent of the notion of Cartesian coordinates. What he means by a general notion of volume is a notion which coincides with the ordinary notion of volume in the case of continuous sets, but which is applicable also in the case of non-continuous sets. This will give rise to the notion of Cantor-Minkowski measure. Concerning the link between dimension and measure, he points out for example that a unit square has a measure of zero as a three-dimensional geometric object and a measure equal to one as a two-dimensional geometric object \cite{cant5}: 

\begin{quoting}[font+=bf,begintext= ,endtext=]
\textit{Un carré p. e. dont le coté est égal à l'unité, a sa grandeur égale à zéro lorsqu'il est considéré comme partie eonstituante de l'espace à trois dimensions, mais il à la grandeur égale à 1, lorsqu'on le regarde comme partie d'un plan à deux dimensions. Cette notion générale de volume ou de grandeur m'est indispensable dans les recherches sur les dimensions des ensembles continus, que j'ai promises dans Acta mathematica...}
\end{quoting}

Contrary to the so-called topological dimensions, the introduction of measured dimensions was not motivated by the resolution of the Dedekind conjecture. Moreover, they are not invariant by homemorphism. Rather, it was due to the development of measure theory, which was to constitute a favorable ground for the formalization of notions of dimension on the basis of the dimension/measure link. Thus, in the line of Emile Borel and Constantin Carathéodory, Felix Hausdorff will introduce in 1918 in \cite{hau} his famous notion of measure, on which he will introduce the notion of dimension that bears his name. Of course, the work of Fréchet, which Hausdorff cites in his 1918 article, concerning non-integer dimensions had opened the way for him on the conceptual level. The main idea of measured dimensions is the fact that the volume of an object $X$ is proportional to its size to the power of its dimension $volume(X)\sim (size(X))^{dimension(X)}$. Let's introduce the well known notion of Hausdorff measure and dimension, we follow the presentation made in the reference book \cite{falco}:

\begin{Def}[Hausdorff measure and dimension 1918]\label{haus}
Let $(X,d)$ be a metric space, for $U\subset X$ we denote by $|U|=\sup\{d(x,y); x, y \in U\}$ the diameter of $U$. Let $F\subset X$ and $\{U_i\}$ a countable cover of $F$ with sets of diameter at most $\delta$, i.e. $F\subset \bigcup\limits_{i=0}^{\infty} U_i$ with $0\leq|U_i|\leq \delta$. For a non negative number $s$ we define   
\begin{equation}\label{meshaudelta}
H_{\delta }^{s}(F)=\inf \left\{\sum_{i=1}^{\infty }(|U_{i}|^{s}:\bigcup_{i=1}^{\infty }U_{i}\ is\ a\ \delta-cover\ of\ F \right\},    
\end{equation}
the $s$-dimensional Hausdorff outer measure is defined by
\begin{equation}\label{meshau}
H^s(F)=\lim\limits_{\delta\to 0}H_{\delta }^{s}(F).   
\end{equation}
The Hausdorff dimension of $F$ denoted $dim_H(F)$ is then defined as the critical value of $s$ such that:
\begin{equation}\label{dimhau}
dim_H(F)=\inf\{s\geq 0; H^s(F)=0\}=\sup\{s;H^s(F)=\infty\}.   
\end{equation}
\end{Def}
This definition follows Cantor remark with the additional sentence: the unit square has a measure of infinite as a one-dimensional geometric object. Thus, the Hausdorff dimension of a set is the critical dimension for which the Hausdorff measure loads the set in question. It is a metric notion of dimension completely indexed on the measure; it is precisely defined where it coincides with the Hausdorff measure. The ideal case being of course when $s=dim_H(F)$ then $0<H^s(F)<\infty$. Still, it is possible that we have $s=dim_H(F)$ and $H^s(F)$ equal to zero or infinite as for the Brownian motion, in which case we should consider the $\phi$-Hausdorff measure by replacing $|U_i|$ by $\phi(U_i)$ (with $\phi$ monotone increasing set function with $\phi(\emptyset)=0$) to find the adequate growth scale, or by using the Generalized Besicovitch dimension introduced by Larman in 1967 in \cite{lar}. The Hausdorff dimension does not only concern continuous sets, it also concerns sets of Lebesgue measure equal to zero. For example, the Hausdorff measure also loads Cantor sets. It does not however see countable sets. The Hausdorff dimension has the desirable properties of a dimension according to \cite{falco} i.e.:
\begin{itemize}
    \item Monotonicity. If $E \subset F$ then $dim_H(E)\leq dim_H(F)$.
    \item Stability. $dim_H(E \cup F) = \max(dim_H(E), dim_H(F))$.
    \item Countable stability. $dim_H(\bigcup\limits_{i=0}^{\infty}F_i)=\sup\limits_{0\leq i\leq \infty}dim_H(F_i)$. 
    \item Countable sets. $dim_H(F) = 0$ if $F$ is finite or countable.
     \item Open sets. If $F$ is an open subset of $\mathbb{R}^n$ then $dim_H(F) = n$.
    \item Smooth manifolds. $dim_H(F) = m$ if $F$ is a smooth $m$-dimensional manifold.
    \end{itemize}
These properties are, for the most part, inherited from the properties of the measure. These are the good properties wanted by mathematicians, but the two reference books \cite{tri} and \cite{falco} agree on the great disadvantage of the difficulty to compute or to estimate the measure and the Hausdorff dimension in practice. Claude Tricot goes even further in \cite{tri} to argue that this excellent theoretical tool will never be useful in practice: 
\begin{quoting}[font+=bf,begintext= ,endtext=]
\textit{C'est la raison pour laquelle nous ne parlons jamais de la dimension de Hausdorff, cet excellent outil de théorie de la mesure, dont nous croyons qu'il ne servira jamais à rien, tel qu'il est, pour l'étude des courbes provenant de la physique, de la biologie ou de l'ingénierie.}
\end{quoting}
 There is a link between the topological dimensions and the Hausdorff dimension dating from 1937 (see \cite{sz} and \cite{hu}). Let $X$ be a metric separable space, then the lower bound of the Hausdorff dimensions on all spaces $X'$ homemorphic to $X$ is a natural number and is equal to the topological dimension of $X$
 \begin{equation}\label{bornhau}
 \inf\limits_{X'\ homemorphic\ to\ X}(dim_H(X'))=dim(X).
     \end{equation}

We now turn to the notion of dimension introduced by Bouligand in 1928 \cite{bouli} without having been aware of the notion of dimension of Hausdorff \cite{john2}. He relies on the notion of Cantor-Minkowski's measure introduced in 1884 by Cantor \cite{cant5} and taken up by Minkowski in 1901 in \cite{min}. It is a notion of 'primitive' measure, as Tricot qualifies it in \cite{tri}, which starts by thickening the sets in order to facilitate the calculation of their measures as ordinary geometric objects, and then by performing a passage to the limit to obtain their measure. This thickening is called the \textit{Minkowski sausage}. For a detailed discussion of the subject, we refer the reader to the reference book \cite{tri}.       

\begin{Def}[Bouligand's dimensional number 1928]\label{boulig}
Let $X$ be a set assumed to be immersed in the Euclidean space $\mathbb{R}^n$. We denote by $B(x,\epsilon)$ the euclidean ball centred in $x$ with radii $\epsilon$. We note by $X(\epsilon)$ the set $X$ thickened or the Minkowski sausage of $X$, defined by:
\begin{equation}\label{saucisse}
X(\epsilon)=\bigcup\limits_{x\in X}B(x,\epsilon).    
\end{equation}
Then the dimensional number also called Minkowski-Bouligand dimension of $X$, denoted $\Delta(X)$ is defined by 
\begin{equation}\label{dimbouli}
\Delta (X)=\lim\limits_{\epsilon \to 0}\big(n-\frac{\ln (Vol_n(X(\epsilon)))}{\ln \epsilon}\big),    
\end{equation}
where $Vol_n$ denote the conventional Lebesgue $n$-dimensional measure.
\end{Def}
Of course, when the limit does not exist one can take the upper or lower limit. The Minkowski-Bouligand dimension has the same properties listed above for the Hausdorff dimension, except: the Countable stability and Countable sets. It is a notion of dimension like Fréchet's, which can assign a dimension to countable sets. A disadvantage of this notion of dimension according to \cite{falco} is the fact that $\Delta(\overline{X})=\Delta(X)$ which assigns the dimension one to $\mathbb{Q}\cap [0,1]$. However, the two references \cite{tri} and \cite{falco} underline the practical computational aspect of this notion, especially when its value coincides with the Hausdorff dimension. Tricot goes even further by affirming that this notion is so simple and natural that it must surely have been introduced rather in history. Moreover, he points out that later equivalent formulations come from different horizons \cite{tri}:         
\begin{quoting}[font+=bf,begintext= ,endtext=]
\textit{Sa définition paraît si simple et naturelle qu'on en retrouvera peut-être un jour des apparitions antérieures. En  tous cas, ses réincarnations subséquentes seront nombreuses, et proviendront de voies diverses. }
\end{quoting}
The H$\ddot{o}$lder exponent of a measure $\mu$ at a point $x$ is another quantity which intervenes naturally in the calculations of the Minkowski-Bouligand dimension in connection with $\frac{\ln (Vol_n(X(\epsilon)))}{\ln \epsilon}$. This type of quantity already appears in Bouligand's 1928 article \cite{bouli}. The first work on this quantity in connection with the Hausdorff dimension is due to Billingsley in 1960 in \cite{bil} and after in \cite{bil1}. The denomination H$\ddot{o}$lder's exponent appeared well after, maybe because of the link that this notion has with the H$\ddot{o}$lder's exponent of a function.  
\begin{Def}[Billingsley's  H$\ddot{o}$lder's exponent of a measure 1960]\label{bilexpo}
Let $\mu$ be a finite measure on $\mathbb{R}^n$ .i.e. $0 < \mu(\mathbb{R}^n) < \infty$. The H$\ddot{o}$lder's exponent of the measure $\mu$ at the point $x$ denoted by $dim_{\mu}(x)$ is defined as follows: 
\begin{equation}\label{expohold}
dim_{\mu}(x)=\lim\limits_{\epsilon \to 0}\frac{\ln (\mu(B(x,\epsilon)))}{\ln \epsilon},    
\end{equation}
where $B(x,\epsilon)$ denote the euclidean ball centred in $x$ with radii $\epsilon$.
\end{Def}
Of course, when the limit does not exist one can take the upper or lower limit.
\subsection{Measured dimensions related to compactness: \small{Pontrjagin(1908-1988)-Schnirelmann(1905-1938) and Kolmogorov(1903-1987)-Tikhomirov(1934)}}\label{pont}

The notions of dimensions that we will now present are both topological and measured in our view. Even if they are metric notions, we can consider them as topological because they use the notion of compactness, hence the arrows going from compactness to Box dimensions in the diagram (figure \ref{dia1}). Moreover their first introduction, as we will see, is related to the notions of topological dimensions. They are also measured, because there is a notion of tacit measure in their definition in connection with the Minkowski-Bouligand dimension. So let us start by presenting the notion called box dimension which is a notion equivalent to the notion of Minkowski-Bouligand dimension but which emerged from different considerations. The first introduction concerns the lower box dimension in a 1932 article by Lev Pontrjagin and Lev Genrikhovich Schnirelmann \cite{pontr}. In their article, there is no reference to either the Hausdorff dimension or the Minkowski-Bouligand dimension. However, they do cite the work on the dimension of Brouwer, Urysohn and Menger. The objective of their article was to construct a method based on "measurement standards" to compute the topological dimension in the framework of compact metric spaces. The dimension they wanted to obtain was rather the Alexandroff metric dimension, but all topological dimensions are equivalent in this setting.  

\begin{Def}[Pontrjagin and Schnirelmann's  metric order or lower Box dimension 1932]\label{pontbox}
Let $(X,d)$ be a compact metric set. Then, thanks to the Borel-Lebesgue theorem, for every $\varepsilon>0$ we can cover it by a finite number of balls of radius $\leq \epsilon$. Let $N_{\epsilon}(X)$ the minimal number of that balls. The lower Box dimension denoted by $\underline{dim}_B$ can be defined as follows:
\begin{equation}\label{lowbox}
\underline{dim}_B(X)=\liminf\limits_{\epsilon \to 0}\frac{\log(N_{\epsilon}(X))}{\log\frac{1}{\epsilon}}.
\end{equation}
\end{Def}
The authors call $N_{\epsilon}(X)$ the volume function of $X$, and note that if $X\subset \mathbb{R}^n$ of non-zero measure, this function is asymptotically equal to $\frac{c}{\epsilon^n}$. One can see this function as a measure of the degree of compactness of a set. In any case, the main objective of their article was to show that:  
\begin{equation}\label{bornbox}
 \inf\limits_{d\ metric\ on \ X}(\underline{dim}_B(X))=dim(X),
     \end{equation}
the lower bound of the lower Box dimensions for all metrics $d$ which does not alter the topological properties of the compact spaces $X$ is a natural number and is equal to the topological dimension of $X$. It turns out that this notion of dimension is equivalent to the lower Minkowski-Bouligand dimension (see \cite{tri}), even if it is introduced for completely different considerations, as we just saw.

Let us now turn to the notion of upper box dimension introduced by Andrei Kolmogorov and Vladimir Mikhailovich Tikhomirov in 1959 in \cite{KT}. These authors quote the article by Pontrjagin and Schnirelmann \cite{pontr}; nevertheless, the introduction of the upper box dimension is, as they explain, due to the notion of $\epsilon$-entropy inspired by Shannon's information theory introduced in 1948 \cite{Sh}. Indeed, Shannon had introduced the notion of entropy in order to quantify the quantity of information contained or delivered by a discrete information source. Consider a discrete random object $\xi=(X,\mathcal{F},P)$ (probability space) such that $X$ is formed by a finite number of elementary events $(x_i)_{i=1,2,..m}$, $\mathcal{F}$ a $\sigma$-algebra on $X$ and $P$ a probability on $X$ such that $P(x_i)=p_i$. The logarithm function in the formula is used to ensure the additivity of the information. More precisely, the information contained in two independent events must be equal to the sum of the information contained in each event. Then, according to Claude Shannon, the amount of information or entropy of this random object can be computed by the expression:
\begin{equation}\label{formuleshannon}
H(\xi)=\sum_{i=1}^{m}p_ilog_b(1/p_i)=-\mathbb{E}\big(log_b(P(X))\big).
\end{equation}
Here $\mathbb{E}$ represents the mathematical expectation and $\log_b(x)=\frac{\ln(x)}{\ln(b)}$ is the logarithm in base $b$. In general $b=2$ because we use bits which have two values $0$ or $1$. In the case where $X$ is of cardinal $N$ with equiprobability of realization of $x_i$, $P(x_i)=1/N$ we obtain
\begin{equation}\label{shannon}
H(\xi)=log_b(N).
\end{equation} 
It is obvious that in the case of a countable number of elementary events, the entropy can be infinite ($N\rightarrow\infty$ in (\ref{shannon}) for example). Nevertheless, for particular distributions whose mathematical expectation in the formula (\ref{formuleshannon}) is finite, we can obtain a finite entropy. Similarly, we have a natural equivalent to Shannon's formula in the continuous case, $H(\xi)=\int_X f(x)\log(f(x))dx$, called \textit{differential entropy}, here $f$ is the probability distribution on $X$. In any metric space, we do not necessarily have a probability distribution. This is why Kolmogorov \cite{K1} was interested exclusively in the uniform distribution. Thus, concerning a continuous source of information, Kolmogorov had proposed an alternative definition by introducing the notion of information with $\epsilon$ precision. More precisely, consider a random object $((X,d),\mathcal{F},P)$ where $X$ is both a probability space and a compact metric space. Since it is compact, we can cover it by a finite number $N(X,\epsilon)$ of balls of radius $\epsilon$. We suppose moreover that there is equiprobability to fall on one of these balls, this probability will be $\frac{1}{N(X,\epsilon)}$. Thus, by analogy with the formula (\ref{shannon}), we can say that the $\epsilon$-entropy of this object is defined by:
\begin{equation}\label{kolmo}
H_{\epsilon}(X)=\log_2(N(X,\epsilon)).
\end{equation}
They notice that in finite dimension $n$, $H_{\epsilon}(X)\approx n\log_2(\frac{1}{\epsilon})$. Based on this quantity, Kolmogorov and Thikomirov introduce in addition to the lower box dimension the upper box dimension. When the two coincide they call it the metric dimension \cite{KT}.  
\begin{Def}[Kolmogorov and Thikomirov's metric dimension or Box dimension 1959]\label{kolmobox}
Let $(X,d)$ be a compact metric set. Let $N_{\epsilon}(X)$ the minimal number of balls with radii $\leq \epsilon$ that cover $X$. The upper Box dimension denoted by $\overline{dim}_B$ can be defined as follows:
\begin{equation}\label{upperbox}
\overline{dim}_B(X)=\limsup\limits_{\epsilon \to 0}\frac{\log(N_{\epsilon}(X))}{\log\frac{1}{\epsilon}}.
\end{equation}
If $\underline{dim}_B(X)=\overline{dim}_B(X)$, one can define the box dimension by:
\begin{equation}\label{box}
dim_B(X)=\lim\limits_{\epsilon \to 0}\frac{\log(N_{\epsilon}(X))}{\log\frac{1}{\epsilon}}.
\end{equation}
\end{Def}
It is possible to construct sets for which $\underline{dim}_B(X)\neq \overline{dim}_B(X)$ (see \cite{tri}). We must emphasize that $\underline{dim}_B(X)$ a big disadvantage compared to $\overline{dim}_B(X)$ and $dim_B(X)$  because we do not have in general the equality $dim_B(X\cup Y)=\max(dim_B(X); dim_B(Y))$ for the lower box dimension. It is due to the well known fact: when we have two sequences $u_n$ and $v_n$ then in general the equality $\liminf\max\{u_n;v_n\}=\max\liminf\{u_n;v_n\}$ is not true. Again, we can see this notion of dimension as a kind of measure of the compactness of a set. More precisely, we can see it as follows: the higher the dimension, the more balls are needed to cover the set and the worse the compactness is until we reach the infinite dimension and there, the number of balls becomes infinite too and the compactness is lost. Nevertheless, since Kolmogorov was familiar with the idea of the notion of dimension in infinite dimension through his work concerning the linear dimension \cite{K2}, initiated by Banach in connection with the Frechet type of dimension, Kolmogorov and Thikomorov will estimate the $\epsilon$-entropy of some functional spaces of infinite dimension. The idea being that even if the set is of infinite dimension, it can be compact for a certain metric, and thus one can estimate the number of balls needed to cover  For example, we can obtain the growth $H_{\epsilon}(X)\approx (\log_2(\frac{1}{\epsilon}))^s$, for what they call $s$ the \textit{functional dimension}. This explains the arrow starting from the cell of linear Banach dimension and arriving at the cell of Kolmogorov and Thikomirov.              
 
To conclude this section, we refer the reader to the books \cite{tri} and \cite{falco} concerning the packing dimension. This notion of dimension, which combines the formalism of the Hausdorff dimension and the Box dimension, was introduced by Tricot in 1979 during his PhD thesis \cite{tri1}. We want also to mention the dimension of Patrice Assouad introduced in his doctoral thesis in 1977 \cite{assouad}, which is a notion of dimension related to the box dimension. Finally, we would like to mention a very interesting approach based on the link between the notion of dimension and the notion of capacity of sets in a Falconer article from 2021 \cite{falco1}. 
\vspace{-0.2cm}
\section{An introduction to the point-dimension theory}\label{pointdim}
In order to better understand the path of our reflection, let's start by quickly recounting the initial motivation that led us to do this long-term work. We had started to work on the notion of dimension through the notions of Kolmogorov's $\epsilon$-entropy in finite and infinite dimension to quantify the dimension of the attractors given by partial differential equations in \cite{GM} and \cite{maar0}. We had therefore started to think about a theory of dimension in a general framework going from finite to infinite dimension. In \cite{maar}, the first author introduced the idea of the extended fractal dimension, without giving a precise definition of it because  more hindsight on the issue was needed. As we wanted to use the notion of $\epsilon$-entropy, and this for several reasons that we will explain in the sequel, to estimate the dimension of infinite dimensional spaces and sets. It was necessary to localize in order to recover the compactness at least in a certain sense. Therefore, because it was clearly not possible to use a global notion of dimension, we quickly came to the conviction that a local notion was more appropriate to proceed. This localization mechanically brings up the need to deal also with the cases of spaces or sets which were not necessarily dimensionally homogeneous. Thus convinced of the necessity of localization, the natural question that arose for us was:  how far should we localize?

The greatest difficulty we had to face in carrying out this work was dealing  with the preconceived ideas. Indeed, we arrived, not without difficulty, at the conclusion that it was necessary to localize the notion of dimension at the point. This means that we have to attribute a non zero dimension to the point, but we have always learned and found adequate the fact that the point is of dimension zero! This conviction that it is inappropriate to attribute a dimension to the point, was shared by many mathematicians that we have met. Mathematically speaking, we succeeded rather quickly in formalizing a definition of the dimension of the point by localizing the box dimension. The most difficult step was to accept that it was necessary to proceed in this way. This leads inevitably to the question of what a point is mathematically. At the beginning we tried, as much as possible, to avoid philosophical questions like: does the point really exist? In the meantime, we had done a work \cite{maar1} combining philosophy and science through which we could see the potential richness of this way of proceeding. Then we came to the conclusion that we had to start from scratch; that is to say, to trace and analyze both the mathematical and philosophical history of the notion of dimension, because we realized that they are inseparable. 

We were pleasantly surprised that there was a lot of material available, as there has been a lot of debate about the concept of dimension. It has captured the attention of many mathematicians, philosophers and some historians/philosophers of mathematics. Even if we are well aware that the time when mathematicians wrote articles that were both mathematical and philosophical is over, we have decided to proceed in this way.
Before undertaking this work, we had arrived at our own conception, which has evolved, enriched and strengthened throughout the writing of this article. In the beginning of this work, we were already aware of some notions of localized dimensions such as the H$\ddot{o}$lder exponent and a passage in the book \cite{tri} concerning the localization of the Minkowski-Bouligand dimension. As we progressed in our research, we realized that the need to formalize a notion of local dimension or to localize a notion of dimension had arisen for some mathematicians long before us. Thus, the main objective of this section is to discuss the existing localizations, to ask why there is still a problem, conceptually speaking, of assigning a dimension to the point, and to explain the motivation that drives us to use the definition we introduce. In any case, it is curious that to extend the notion of dimension to the infinite dimension, we have arrived at the need to start by locating it at the point!

\subsection{Local dimensions and underlying issues}
\subsubsection{An overview of existing local dimensions}\label{4.1.1}
We will start by giving a brief overview of the different local notions of dimension that we have analyzed in order to better introduce our own. This explains the different arrows going from the cells of the local dimensions to the cell of our notion of dimension. We have already (see subsection \ref{2.3.1}) pointed out that when Bolzano wanted to formalize his notion of dimension, he had to arrive at the level of points. Of course he did not give a local scope to his definition but the spirit of locality was already there (see our modification of his definition \ref{mbdim}). Fréchet also in his 1910 article, through his comparative definition of dimensions, seems to realize that it is not technically appropriate to compare the dimensions of sets in blocks. That is why his definition concerns the homeomorphisms existing between the subsets of the considered sets. It is only in 1923 \cite{tie} that Heinrich Franz Friedrich Tietze introduces the idea of localizing the notion of Fréchet's type of dimension to points (as mentioned in \cite{arbo}). Subsequently, Fréchet introduced the notion of the local type of dimension in 1928 in \cite{fre2}. The definition states that:  
\begin{Def}[Fréchet's local type of dimension 1928]
Let $X$ and $Y$ two metric spaces, $x\in X$ and $Y\in Y$. We denote by $d_xX$ \textbf{the local type of dimension of the set $X$ at the point $x$}:
\begin{enumerate}
\item If for all neighbourhood $V_y$ of $y$ in $Y$, ther exists a neighbourhoud $U_x$ of $x$ in $X$ such that $U_x$ is homemorphic to a part of $V_y$ then $d_xX\leq d_yY$;
\item If we have $d_xX\leq d_yY$ but $d_yY\leq d_xX$ is not true then  $d_xX< d_yY$ ;
\item If we have simultaneously $d_xX\leq d_yY$ and $d_yY\leq d_xX$ then $d_xX=d_yY$;
\item If we have neither $d_xX\leq d_yY$ nor $d_yY\leq d_xX$ then we say that the local types of dimension at the points $x$ and $y$ are not comparable.
\item A set $X$ is said to be homogeneous if at all two points $x_1$ and $x_2$ of $X$, we have the equality $d_{x_1}X=d_{x_2}X$.
\end{enumerate} 
\end{Def}
We will notice that as soon as the notion of dimension is localized, the notion of dimensionally homogeneous set is automatically identified. This will be the case for all notions of local dimensions. While conceiving sets of different dimensions is not at all complicated, like the example of a disk with a segment that stands out from it that we gave in subsection \ref{2.3.1}. This aspect should normally be taken into consideration in any introduction of a notion of dimension. The second remark concerns the semantics used which will also be the same for all the other notions of dimension localized at the point : \textbf{the dimension of a set $X$ at point $x$}.

Before Fréchet introduced this notion of dimension, there were others. First, Brouwer in his 1913 article \cite{brou}, he accompanies his notion of Dimensionsgrad by its localization at the point. In general, we can say that all the localizations at the point of the notions of dimension follow the same natural principle. We refer, for example, to Dowker's article of 1955 \cite{dow} where he localizes the notions $ind$, $Ind$ and $dim$. More precisely, let $X$ be Hausdorff topological space and $Dim$ any notion of dimension capable of computing the dimension of open sets. Let $x\in X$ then the dimension of the set $X$ at the point $x$ denoted by $Dim_x(X)$ is the smallest of the dimensions of the neighbourhoods of $x$. We can then define a dimension of the whole set $X$ based on this localization by taking the largest value of $Dim_x(X)$ for $x\in X$, that Dowker denotes by $loc\ ind$, $loc\ Ind$ and $loc\ dim$. It is important to note that the dimensions thus obtained by localization of the global notions and then by reglobalization of these localizations are not in general equal to the initial global notions. In general, we only have the inequality $loc\ Dim(X)\leq Dim(X)$. For example, Dowker exhibits a set $X$ for which $loc\ dim(X)<dim(X)$. However, he easily and unconditionally shows that for any set $X$, $loc\ ind(X)=ind(X)$. This is due to the local origin, which we will explain below, of the notion $ind$. 

On a similar note, we can find the notion of dimension introduced by Larman in 1967 in \cite{lar}, based on the principle of Hausdorff dimension and neighborhoods formed by balls, or again the local Tricot dimension which can be defined using the notations of the Definition \ref{boulig} by: Let $X$ be a metric space then the local dimension of $X$ at the point $x\in X$ can be computed by the following formula    
\begin{equation}\label{bouligloc}
\Delta(X,x)=\lim\limits_{r \to 0}\Delta(X\cap B(x,r)).    
\end{equation}
This notion was introduced by Tricot in 1979 \cite{tri1} in order to search for points of maximal dimension. In \cite{tri}, Tricot points out that it is difficult to find a first reference of this notion because it appears most of the time in an implicit way: 
\begin{quoting}[font+=bf,begintext= ,endtext=]
\textit{Il est difficile de trouver une première référence claire sur la notion de dimension locale. Elle apparaît souvent de façon implicite.}
\end{quoting}
This passage from Tricot highlights another essential point; namely, that in general the use of notions related to the dimension at a point are primarily of technical use. In reality, we can distinguish two types of dimensions at the point. The notions of dimensions that are localized, such as those we have just mentioned, and those that originally have a local definition, such as those we will develop in the following. Before moving on to the latter, we would like to make the following remark: for us, the notions of dimension respecting the property $D(A\cup B)=\max(D(A);D(B))$ are more adequate to localize because they already contain in them a local character since the dimension of a set can be calculated through the dimensions of its subsets \textit{ad infinitum}. This is not the case for example for $Ind$ and $dim$. Concerning $ind$, there is a subtlety that we will point out in the following subsection.

Let us now turn to the notions of dimension that have been conceptualized and formulated initially starting from the dimension at the point. These are not localizations of global definitions of dimension at the point. We have already analyzed the concept of Bolzano, even if the spirit of its definition is local, he has given it an exclusively global scope. It is nevertheless easily transformable into a definition starting from the point dimension as we have done in Definition \ref{mbdim}. As already mentioned, Bolzano's goal was to be able to define correctly the different basic geometric objects like the line, the surface and the solid from their dimensions. It so happened that the question of finding definitions for these basic geometric objects came to the fore again in the early twenties of the last century. Indeed, mathematicians had already started to rid the notion of dimension from the notion of coordinates on the one hand, and formalized a new abstract mathematical discipline transcending the usual Euclidean space framework, called point-set topology, on the other hand. The fashionable question at the time was to formalize intrinsic topological definitions of a curve, a surface, a volume and more generally an $n$-dimensional geometric object in the framework of abstract topological spaces. The two young mathematicians who tackled this problem at the same time but independently were Urysohn and Menger.

The challenge is therefore to be able to identify the most general sets that we must still be able to call line, surface or volume, as Urysohn summarizes very well in his 1925 article \cite{ury1}: 
\begin{quoting}[font+=bf,begintext= ,endtext=]
\textit{Indiquer les ensembles les plus généraux qui méritent encore d'être appelés lignes, surfaces etc.}
\end{quoting}
In his magisterial article \cite{ury1} of 108 pages, Urysohn exposes both his philosophy by formalizing his way of proceeding and his mathematical results. Urysohn will then impose two conditions to treat the problem. The first one is to look for intrinsic definitions. The second one requires the use of local definitions as much as possible. In reality, whether for Urysohn or Menger, as for Bolzano before them, distinguishing the different geometric objects in an abstract framework can only be done through the notion of dimension. This notion is always, for them, badly defined as Urysohn underlines in this passage \cite{ury1}: 
\begin{quoting}[font+=bf,begintext= ,endtext=]
\textit{La notion (non définie jusqu'à présent) qu'on appelle dans le cas d'une multiplicité Jordanienne son nombre de dimensions, et dont la définition générale permettrait de discerner entre les lignes, les surfaces, les volumes etc... cette notion, on s'en rend compte aisément, une notion intégrale d'origine locale.}
\end{quoting}
In this passage Urysohn also specifies that, as long as he is concerned, the notion of dimension is an integral notion of local origin. At the bottom of page 36, he notes that he has always unconsciously and tacitly seen things in this way, acknowledging that it was Alexandroff who made him notice this aspect explicitly.

\begin{quoting}[font+=bf,begintext= ,endtext=]
\textit{C'est M. Paul Alexandroff qui attira mon attention sur le rôle que joue le principe des définitions locales dans mes recherches: inconsciemment je le respectais de tout temps. Mais je ne m'en apercevais pas.}
\end{quoting}
Although Urysohn and Menger formulated definitions of $ind$ localized at the point, in most of the reference books one will find the version of definition \ref{ury-men} presented above. This type of definition resembles Bolzano's definition, in the sense that the mind is local but the scope is global. So to speak, the dimension of the point is not clearly defined. Whereas in the original definition of Urysohn and Menger it is clearly the dimension of the point, they had succeeded, unlike Bolzano, in taking this conceptual step. The natural question that arises is: why is it more difficult to find the definition in its original localized version at the point than in its global version? Especially since, as we have just said, Urysohn insists that local definitions should be preferred. We will try to analyze this fact in the following subsection. We now present the version given in the book $\cite{hu1}$ respecting Urysohn's notations (simpler) and Menger's semantics (the most common): 
\begin{Def}[Original Urysohn-Menger dimension or small inductive dimension 1922-1923]\label{ury-men-loc}
let $X$ be a regular space, the small inductive dimension of $X$ at the point $x$, is denoted by $indim_x(X)$. Where $ind(X)$ denote the small inductive dimension of $X$:  
\begin{enumerate}
\item $ind(X) = -1 \iff X = \emptyset$;
\item $ind_x(X) \le  n\in\mathbb{N}$, if $x$ has arbitrarily small neighborhoods whose boundaries have dimension $\leq n-1$.\\ 
$ind(X) \le  n\in\mathbb{N}$ if $X$ has dimension $\leq n$ at each of its points. 
\item $ind_x(X)=n$ if it is true that $X$ has dimension $\leq n$ at $x$ and it is false that $X$ has dimension $\leq n-1$ at $x$.\\
$ind(X)=n$ if $ind(X)\leq n$ is true and $ind(X)\leq n-1$ is false. 

\item $ind(X) =\infty $ if $ind(X) \leq  n$ is false for each $n$. 
\end{enumerate} 
\end{Def} 

\subsubsection{Relativity and point-conception as a major difficulty of point localization}\label{4.1.2}
The notion of dimension $ind$ is a central notion in the dimension theory. It has partly succeeded in achieving the purpose for which it was created; namely, to distinguish different geometric objects through their dimension in a more abstract framework. For example, any curve within an ordinary Euclidean space contains a topological image of any one-dimensional subset contained in any separable metric space (\cite{men1}). It has also allowed a better understanding of key topological notions such as connectedness. Moreover, it has generated important questions that have allowed the development of topology in general. But the question that interests us in this subsection is of another order. Indeed, as we have noted above, not only did we take a long time to find a notion of dimension built on the dimension of the point, but we have found a serious reluctance concerning this approach. We were for example really surprised by the existence in the literature of definitions like the one above \ref{ury-men-loc}. Thus, there are two questions that arise: first, why is it that the global definition \ref{ury-men} the most common definition in the literature? Why this need for globalization, why not keep the original local version \ref{ury-men-loc}? Secondly, why does the notion of dimension of a point have a hard time to spread in the mathematical community?

For the first question, one could simply answer by saying that the two definitions \ref{ury-men} and \ref{ury-men-loc} are equivalent. Of course this is true for dimensionally homogeneous sets.  Nevertheless, one can oppose the fact that on the one hand, with definition \ref{ury-men}, the dimension at a point is not defined. On the other hand, it is indisputable that the spirit of the original definition of Urysohn and Menger is absent from the present version \ref{ury-men}. Thus, if one accepts that they are not exactly the same, the question becomes: is there a palpable mathematical interest in using \ref{ury-men} instead of \ref{ury-men-loc}? We think that it is rather the opposite since definition \ref{ury-men-loc} is more precise and complete than \ref{ury-men}. Indeed, if we can calculate the dimension of all the points of a set we have more information about it than when we try to calculate its dimension as a block. Thus, is it only a conceptual difficulty with respect to a certain \textit{a priori} judgement concerning the notion of dimension at a point? We think that this subject would deserve an in-depth epistemological analysis. For the scoop of our research, we will try to outline some elements that we think should be taken into consideration in this type of analysis. There is, in our opinion, the necessity of an adequate mathematical-philosophical conception of what a point is in general and of its dimension in particular. We do not pretend to bring this clarification here, but we want both to highlight the need for this debate and to bring a proposal for an alternative conception of mathematical space in the next section.

One of the most important problems to deal with when wishing to associate a dimension with a point is its relative aspect.  We are sure that Urysohn and Menger had to deal with this problem. We have already mentioned the difficulty of this relativity in subsection \ref{2.2.1}. More precisely, if a point belongs to a line $(\Delta)$, its dimension is one. But if this line $(\Delta$) is itself in a plane $(P)$, the points of this line are also points of the plane and their dimensions are then two relative to the plane $(P)$. There is a subtlety here that we have to be careful about.  Indeed, the dimension of ($\Delta$) taken as the maximum of the dimensions of the points relative to ($\Delta$) itself is one, but its dimension as the maximum of the dimensions of the points of ($\Delta)$ relative to $(P)$ is two. Another example showing the importance of precise dimensional relativity is the dimensional monotonicity relation. In fact, the following relation is always true $ind_{A\cup B}(A\cup B)=\max (ind_{A\cup B}(A),ind_{A\cup B}(B))$, this is due to the fact that we take the maximum of the dimensions of the points belonging to the set in question. The relation which is not always true is rather the following $ind_{A\cup B}(A\cup B)=\max (ind_{A}(A),ind_{B}(B))$. The last example concerns the inductive aspect of definition \ref{ury-men-loc}. More precisely, we start by computing the dimension of a point $x\in X$ by computing the dimension of the boundary of its neighborhood, whose dimension is computed through the dimension of its points relative to itself. In summary, our induction turns out to be one based on the relativity of the dimensions of the successive boundaries of the points relative to themselves. 

In the face of this difficulty of relativity, Urysohn and Menger employed two different semantics. Urysohn speaks of the dimension of a point $x$ relative to a set $C$ even if he uses the notation $dim_x(C)$ instead of $dim_C(x)$. For Menger, it is the dimension of the set $C$ at the point $x$. In short, points have variable dimensions: zero in the absolute or relative to themselves. Otherwise, it is the sets where they are located that give them their dimensions. It is inherited from the set where they are living. Urysohn probably had a particular conception of the notion of dimension in the back of his head (tacitly or explicitly). But he never could express it because he unfortunately died prematurely in 1925 at the age of 26, even before the publication of the article \cite{ury1}, that end with the sentence: to be continued. However, Menger explained how he concretely and intuitively conceived the notion of dimension on several times since his 1928 book \cite{men3}, then in \cite{men2} of 1943 and finally in \cite{men1} of 1954 through a simple example:
\begin{quoting}[font+=bf,begintext= ,endtext=]
\textit{To formulate the intuitive difference between lines, surfaces, and solids one can devise a simple experiment whose outcome depends upon the dimension of the object to which it is applied [5]. We cut out from the object a piece surrounding a given point. If the object is a solid we need a saw to accomplish this, and the cutting is along surfaces. If the object is a surface a pair of scissors suffices, and the cuts are along curves. If we deal with a curve we may use a pair of pliers and have to pinch the object in dispersed points. Finally, in a dispersed object no tool is required to perform our experiment, since nothing needs to be dissected. This characterization of dimension leads from $n$-dimensional to $(n -1)$-dimensional objects. It ends with dispersed sets, naturally called 0-dimensional, and, beyond these, with "nothing," in set theory called the "vacuous
set." It is, therefore, convenient to consider the latter as -1-dimensional.}
\end{quoting}
According to this passage, we can say that this is a practical or experimental conception of the notion of dimension. Menger clearly leaves the world of concepts to rely on the experience of the physical world to grasp the notion of dimension that he has introduced.
\subsection{Our conception of the mathematical space}
This section deserves a thorough analysis and discussion, which we will not do here for the sake of brevity. We were able to rely on several reading grids to better formulate our position. More precisely, we had to dive into the \textit{atomism} vs \textit{continuism} debate, into the \textit{rationalism} vs \textit{ empiricism} opposition, into the \textit{intuitionism} vs \textit{formalism} dispute, into the \textit{conventionalism} vs \textit{fictionalism} discussion and into \textit{Phenomenology} vs \textit{analytic philosophy}. We will try in the next subsection to be concise and succinct while hoping to have the opportunity in the future to come back with more detail analysis on some aspects.       

\subsubsection{Do the dimensions one and two not really exist? Dimensionad: The fundamental dimensional particle}\label{4.2.1}
As we have explained above, the notion of dimension transcends the only framework of the physical dimensions of space; we think, nevertheless, that this setting constitutes a good basis to better apprehend this notion since it is in this context that the latter manifests itself purely and simply. We must keep in mind, as Kant taught us, that even if reality exists, it remains inaccessible to us. We only have access to our experience of reality. In this sense we will never have access to the truth. Let us distinguish at this level three types of space: first, there is the \textit{physical space} which constitutes this unreachable objective physical reality. Secondly, there is the \textit{mathematical space} which is an attempt of subjective, formal and abstract mathematical representation of the physical space. This abstract character of this space which is cut off from our experience of reality, is indispensable if we wish to have an exact science not subjected to a continual revision, as Poincaré writes concerning geometry in \cite{poincare5}:
\begin{quoting}[font+=bf,begintext= ,endtext=]
\textit{Si la géométrie était une science expérimentale, elle ne serait pas une science exacte, elle serait soumise à une continuelle révision.}
\end{quoting}
In other words, mathematical space is not supposed to correspond to our sensible world. The third type is the \textit{empirical space} which is the subjective, tangible and observable space that we construct through our experience of physical space. The notion of dimension is what constitutes the common ground between these three types of space. A consequence of the subjectivity of the mathematical space is that the formalism that mathematicians posit for this space should be considered only as a convention. This is the current of thought called \textit{conventionalism}, an intermediate position between rationalism and empiricism introduced by Poincaré in \cite{poincare6}. Nevertheless, the conventions to be chosen concerning the structure of the mathematical space will have to be guided (not imposed) by experience i.e. by the empirical space. The choice made in this way will not be more true than another but only more convenient, as Poincaré explains very well in \cite{poincare4} about geometry:     
\begin{quoting}[font+=bf,begintext= ,endtext=]
\textit{Our choice is therefore not imposed by experience. It is simply guided by experience. But it remains free; we choose this geometry rather than that geometry, not because it is more true, but because it is the more convenient.}
\end{quoting}
By analyzing the spatial dimensions of the empirical space, we proceed first to reformulate the question of the title of this subsection. Is there any known real object in our physical world that is really (without approximation) of dimension one or two? At first glance, we might be tempted to say yes. We have many two-dimensional examples in front of us: a sheet of paper or a surface on a wall. But, on closer inspection, things are more complicated than they might seem. The sheet of paper cannot have the thickness of zero; otherwise, it would not exist. This could be explained by the fact that in a supposedly three-dimensional space the 3-dimensional measure of a 2-dimensional object is zero. Thus, by saying that the sheet of paper is two-dimensional, we make an approximation; we neglect its thickness. Concerning the surface on the wall, here the two-dimensional object is seen as the edge of a three-dimensional object. Is this enough to affirm the existence of a two-dimensional object? No. Because first, it does not exist independently of the three-dimensional object from which it is derived. Secondly, can we cut out this surface of the wall by taking a zero thickness? Of course not. 

At this point, only two possibilities remain: either there are real objects of dimensions one and two, but here they are inaccessible, unobservable and imperceptible to us, or there is no real object that has length but no width or thickness, or an object that has a surface without depth. We choose the second possibility which we think is more reasonable and conceptually richer. We are not the first to make the observation that the dimensions one and two do not really exist. This claim, has been made by many thinkers throughout history, as it is clearly expressed in the following passage from Aristotle \cite{gu}:
\begin{quoting}[font+=bf,begintext= ,endtext=]
\textit{There are some who, because the point is the limit and end of a line, the line of a surface and the surface of a solid, hold it to be inescapable that such natures exist.}
\end{quoting}
In other words, for Aristotle these entities exist in potency, but not in act. Another example of this type of reflection is given by the initiator of the philosophical conception of \textit{fictionalism}, the philosopher Hans Vaihinger in \cite{vai}:  
\begin{quoting}[font+=bf,begintext= ,endtext=]
\textit{The fundamental concepts of mathematics are space, or more precisely empty space, empty time, point, line, surface, or more precisely points without extension, lines without breadth, surfaces without depth, spaces without content. All these concepts are contradictory fictions, mathematics being based upon an entirely imaginary foundation...}
\end{quoting}
For us, this conception bears its foundation in the Nietzschian critique of logic. In our opinion, it would be a mistake to develop a kind of fanaticism towards concepts such as the point or the surface to the level of holding them as realities or even worse as truth. We are obliged to continuously question the concepts, to accept to change them or at least to test their modifications. The empirical space continually offers us new perspectives, and it is incumbent upon us to be broad-minded enough to seize these opportunities. Of course, we must keep in mind that empirical space is only the apparent part of the iceberg. It would be a mistake to believe that Logic alone would allow us to completely transcend our empirical space. Mathematical Logic, as it is conceived, is certainly not completely objective since it is subordinated to us. In other words, to paraphrase Nietzsche, logic is biological. This makes truth a subjective concept. It is a relationship to truth similar to that of the \textit{intuitionist} current. In this work, we have tried, to have a Husserlian \textit{phenomenological} approach of space. More precisely, we attempted to study the notion of dimension as it appears to our consciousness, in accordance with the famous phrase of Edmund Husserl:
\begin{quoting}[font+=bf,begintext= ,endtext=]
\textit{All consciousness is consciousness of something.}
\end{quoting}

The last example we will cite is that of Menger, who, in the following passage from his long article \cite{men1} of 84 pages, explains that the natural classification of objects can be done by their dimensions. For him, the various concepts of basic geometrical objects like solid, surface and curve correspond to a concrete differences in physical reality, before adding that rigorously speaking all physical objects are only solids:  
\begin{quoting}[font+=bf,begintext= ,endtext=]
\textit{Le premier principe de classification des objets géométriques est sans doute la considération de leur dimension. Dans l'espace euclidien ordinaire, nous distinguons des solides, des surfaces, des courbes et des objets sans cohésion. Ces concepts correspondent à des différences dans la réalité physique, quoique à parler rigoureusement, tout objet physique soit un solide.}
\end{quoting}
We have already pointed out at the end of the previous subsection that Menger was inspired by the empirical space to formulate his notion of dimension. In this passage, he indirectly admits that his conception of dimension does not account for the lived reality of the three-dimensionality of all physical objects. Thus, the question of this subsection becomes clearer. Indeed, if all real objects are three-dimensional, what do the dimensions two and one finally represent? 

\begin{figure}[h!]
    \centering
    \includegraphics[width=0.6 \textwidth]{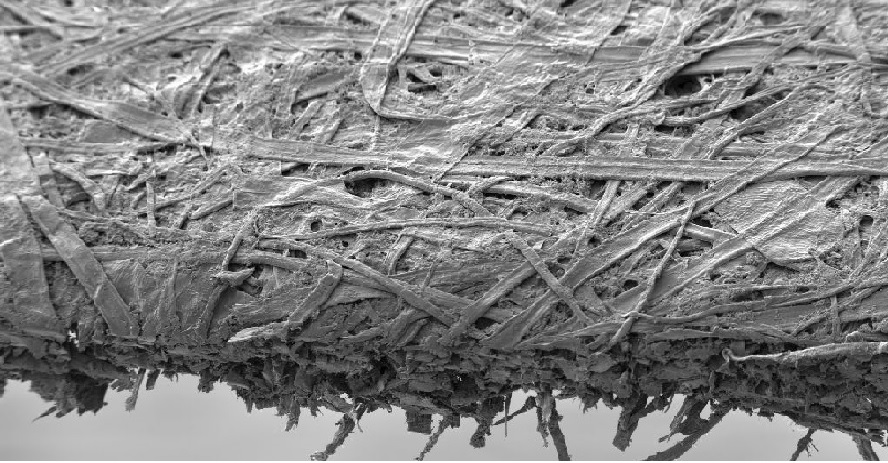}
    \caption{Scanning electron microscopy of a sheet of paper (from nanofab.uwo.ca)}
    \label{papier}
\end{figure}

At this level, we can say that in the empirical space, everything is three-dimensional. In our opinion, the empirical space in which we live suggests that the dimension two of the sheet of paper is simply the two-dimensional organization of the three-dimensional material that constitutes that sheet of paper. Specifically, the surface impression we get when looking at the sheet of paper depends strongly on the scale at which we look at. In other words, at the macroscopic scale, it is acceptable to consider the sheet of paper as a two-dimensional surface. But if we look at it through a scanning electron microscope (see figure \ref{papier}), the paper looks like a cluster of three-dimensional wood fibers, so it can be viewed as a three-dimensional object. Thus, to get around this difficulty, it is more convenient to see the dimension as a means of quantifying the organization of the material forming the object in question. In this sense, dimension is not presented as an intrinsic characteristic of the object, but rather as a characteristic of the relative organization or order of the three-dimensional matter forming the object in question. We must therefore pay attention to the relativity of dimension. In our example, the sheet of paper is of dimension three relative to the organization of the wood fibers. It is a three dimensional organization of three dimensional components. It can be seen of dimension two only if we consider its dimension relative to the organization of the small pieces of wood fiber clusters that have the same thickness of the paper sheet. At this point, the paper would be a two-dimensional organization of three-dimensional components.

Another example that corroborates this conception of dimension is \textit{graphene}, which is the thinnest two-dimensional physical object known. This material consists of a single layer of carbon atoms arranged as illustrated in figure (\ref{graph}) below. This makes it a perfect example of our view since it is a two-dimensional organization or order of three-dimensional atoms, if we assume that the atoms are three-dimensional. In fact, this example raises the question of the dimension of the atom. More generally, this approach poses the question of the dimension of the elementary components of the matter of the universe, should they exist. Indeed, by accepting the dimension as a quantification of the organization of the elements of the matter, one arrives inductively at the organization of the elementary constituents of the latter and their dimensions. In this sens, we suppose the existence of elementary constituents which are necessarily indivisible because otherwise they would be an organization of other constituents. According to this idea, it is these elementary components that give dimension to objects. In other words, we think that the dimension of our empirical and even physical space emerges from the dimension of elementary particles. The argument is the following: if these elementary particles exist, they are inevitably indivisible; their dimension does not emanate from the organization or order of their constituent because they do not have any. They are thus the only elements of this universe to which we should attribute an intrinsic dimension. At this stage, we can then formulate the following question: if the elementary constituents of matter exist, are they of dimension three?
\begin{figure}[h!]
    \centering
    \includegraphics[width=0.5 \textwidth]{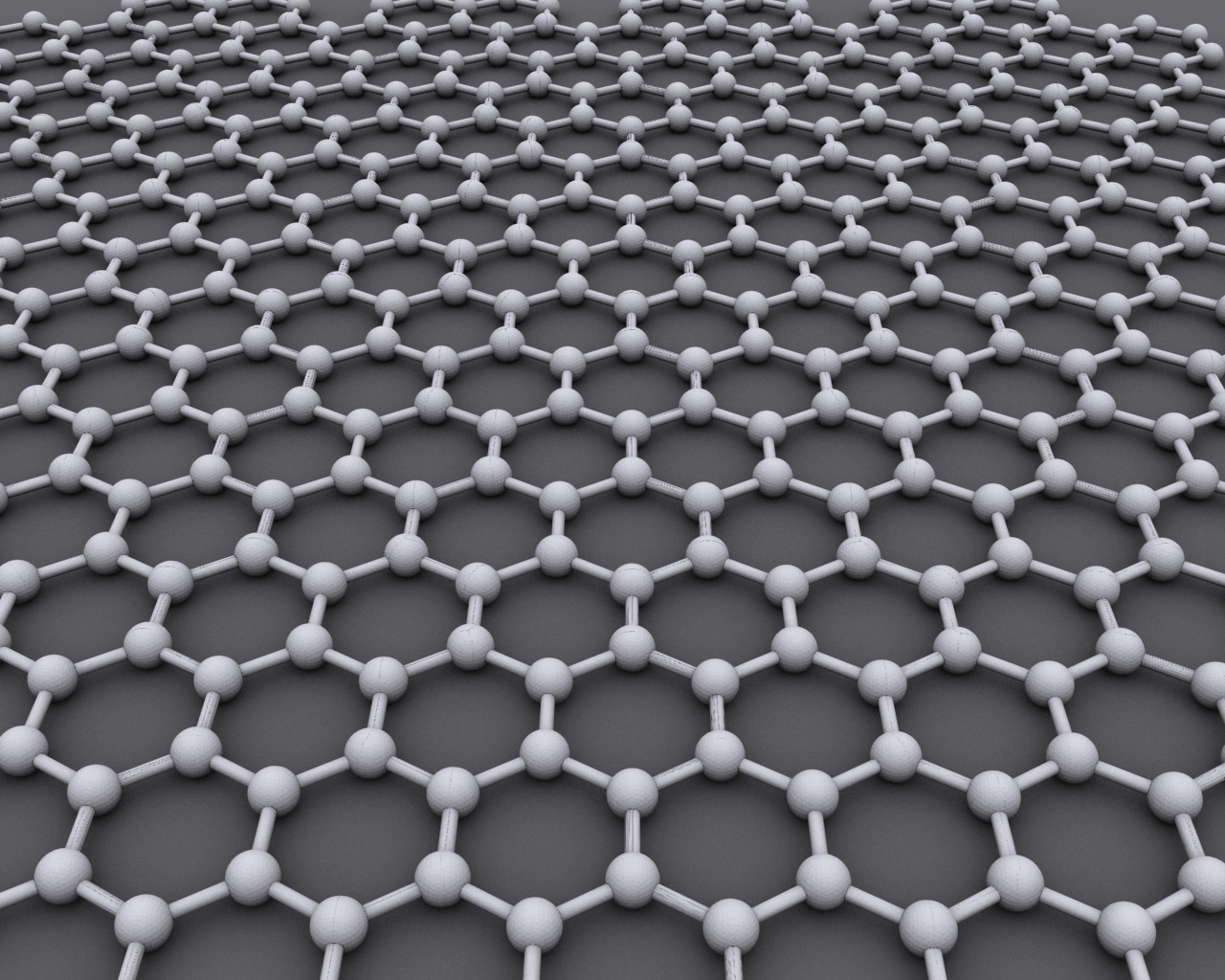}
    \caption{Graphene structure}
    \label{graph}
\end{figure}

In the history of atomism (i.e. the existence of an unbreakable element constituting matter), physics has taught us that limits are only temporary. Indeed, we have moved from the atom to the nucleus and electrons to the protons and neutrons to the quarks and bosons, which are supposed to be unbreakable for the moment. It is therefore very risky to make any advances on this ground. Nevertheless, even before the discovery of these elementary particles, Aristotle who was against the atomic conception, had clearly defended the idea that if the atom exists it should be three-dimensional \cite{ari1}.

\begin{quoting}[font+=bf,begintext= ,endtext=]
The starting point of all this is whether existing things are generated, alter and grow, and undergo the opposites of these, because the primary magnitudes are indivisible ones, or whether there is no indivisible magnitude: that makes a huge difference. And again, if these are magnitudes, are they bodies, as Democritus and Leucippus say, or planes, as in the Timaeus? Well, this very idea of resolving everything into planes is unreasonable, as I have said elsewhere. Hence it is more reasonable that there should be indivisible bodies. But these too produce much that is unreasonable.
\end{quoting}
This is reasonable since, \textit{a priori}, one cannot constitute a three dimensional object by two dimensional elements. This observation brings us to the last question of this subsection. If the elementary components of matter in the universe exist, they should not have a dimension less than three. Do they have a dimension greater than three? Indeed, our empirical space could very well be a three-dimensional organization of higher dimensional components. In other words, our physical space could be of a dimension greater than three and our empirical space would be a three-dimensional organization. This is plausible because in the same way that dimension one and two do not exist in our empirical space since they are only one or two dimensional organizations of three dimensional objects; our empirical space could also be a three dimensional organization of higher dimensional and even infinite dimensional components. 

We would like to conclude this subsection with three remarks:
\begin{itemize}
    \item The first one is the fact that this conception highlights a concrete and essential character of the inaccessible reality of our physical space, that is its dimensionality.
    \item The second one is the fact that this conception suggests that dimensionality springs from the elementary components of the universe. These elementary particles contain within them, in potency, the information of the real dimension of our universe. The manifestation of the different dimensions is nothing other than their organization in one way or another. Can we therefore succeed in detecting other spatial dimensions than the one, two and three dimensions? While we think so, we strongly believe that more work is needed before the conceptualization of experiments in this respect.
    \item The third one is the fact that the reasoning we have developed in this subsection suggests the hypothetical existence of an indivisible \textit{fundamental dimensional particle}, the \textit{Dimensionad}. Dimensionad is a name composed of the word dimension and the word \textit{monad} in reference to Leibniz's \textit{monadolgy}, one of the philosophies that inspired us the most in making this work. It is the particle which is at the origin of the dimension of space and objects, it has in it the potential to confer all the dimensions until the unknown dimension of physical space. Nevertheless, these dimensions only manifest themselves once they are related and organized with the other particles, as indicated the Leibnizian relationalism (see subsection \ref{2.2.2}) at the level of the dimension. If this particle exists, it must be a space-time particle since we can duplicate, by analogy, the reasoning that we had for the space dimensions to the time dimensions. In this way, it is this particle which confers the spatial and temporal dimensions to the space-time; without the dimensionad, space-time would not exist. Furthermore, if this fundamental dimensional particle exists, it should reasonably have an influence on all known elementary particles, and therefore it should be taken into account in the formulation of any physical theory. In particular, it would be necessary to understand the possible interactions between Dimensionad and the fundamental forces.
\end{itemize}
For some prospects in physics in connection with quantum mechanics, string theory, black hole and entropic gravity, we refer the reader to section \ref{persp}.
\subsubsection{Point-dimension theory}\label{4.2.2}
At this stage of the reflection, in order to  mathematize our view, we have several challenges to face. First, we want to maintain an abstract formalism of mathematical space that we want to augment and enrich with the conception presented in the previous subsection. Second, we have to reconcile, in some way, the atomism of empirical space with the continuism of mathematical space. In other words, we want to be guided by the conceptualization that we have managed to achieve in the discrete empirical space to choose the appropriate conventions for our mathematical space.

Unlike Bolzano, Urysohn and Menger, we will begin by abandoning to conceive the notion of dimension as that which allows us to fully characterize basic geometric objects such as line, surface and volume. We agree that it is important for any notion of dimension that these different objects are respectively of dimension one, two and three, but this is neither a necessary nor a sufficient condition. For example, if we consider the set $A=\mathbb{R}\times\mathbb{Q}$, we have $ind(A)=1$ whereas it is far from being a curve. Conversely, if we consider the Peano curve (or any space-filling curve) it is of dimension two even if it is constructed as a curve. In our opinion, this position is a valuable and unavoidable step in the search for an adequate notion of dimension, but the notion of dimension transcends this narrow framework. From our point of view, the first interest of a good notion of dimension is more generally to allow the formalization of adequate conception of the mathematical space. Indeed, since this notion constitutes the common point between any kind of mathematical space, it is consequently essential that this notion be taken into account in the conceptualization of these spaces.  

In the previous section concerned by physical and empirical space, we have succeeded in linking the notion of dimension to elementary particles, if they exist. In the mathematical space, we have to propose a conception where the dimension will be related to a mathematical entity which can be considered as the fundamental dimensional particle. Thus, in order to specify our conception of the mathematical space, we naturally arrive at the notion of the point. This notion, which seems simple at first sight, is much more complex than one might think, as Poincaré has expressed it on several occasions (see for example \cite{poincare1}):       
\begin{quoting}[font+=bf,begintext= ,endtext=]
Se contenter de cela, ce serait supposer que nous savons ce que c'est que l'ensemble des points de l'espace, ou même qu'un point de l'espace. Or cela n'est pas aussi simple qu'on pourrait le croire. Tout le monde croit savoir ce que c'est qu'un point, et c'est même parce que nous le savons trop bien que nous croyons n'avoir pas besoin de le définir.
\end{quoting}
More deeply, the debate on the elementary constituent of the continuum in particular and of mathematical spaces in general is far from being closed. More precisely, we think that the opposition between the infinitesimalist and pointillist conception is not completely outdated. It is true that the pointillist approach is particularly appreciated and approved through the formalism of the point-set theory, but it has not yet won the game. Even if Bolzano was able to make the ponctillist conception of the continuum accepted thanks to the revolutionary idea, for the time, of neighborhood (see subsection \ref{2.3}), this advance does not completely solve the problem of knowing how the points could form the continuum. In this sense, several philosophers and mathematicians opposed to the idea of a continuum formed by points have tried to conceive point-free formalism. For example, there is the continuum's conception of Charles Sanders Peirce \cite{peirce}, the \textit{point-free geometry} of Alfred North Whitehead \cite{white} and the \textit{pointless topology} initiated by Karl Menger in his book Dimensionstheorie \cite{men3}.

In this work, part of a long project, we want to start by exploring the possibilities offered by the pointillist vision, we hope to investigate the pointless approach in a future project. On the one hand, the point is considered to be without extension; in other words, the point is what has no part. On the other hand, contrary to the intuitionist conception \cite{sha}, we suppose that the point is prior to the continuum and more generally to set. In this sense, we have a relationalist approach inspired by Leibniz (see subsection \ref{2.2.2}). It is a position which is in line with the formalistic Bourbaki position of a set: a set is formed of elements sharing certain properties and having some relations between them, or with elements of other sets \cite{cartier}. In other words, the points pre-exist the set; there is no set or space without points, or more precisely without the relations between the points forming them. It is these relations between the points that create at the spaces and their dimensions at same time. 

Contrary to Menger and other authors (see subsection \ref{4.1.2}), we believe that the point dimension is not the dimension of a set $A$ at a point $x$ since we refuse to explain the prior by the posterior. It is not the set that gives the point its dimension, it is rather the opposite. More precisely, admitting that the point is prior to the set, it is the dimensions of the points that will fix the dimension of the set. Hence, for us, it is rather the dimension of a point $x$ relative to a set $A$. Accordingly, point-dimension theory is the building of the notion of dimension starting from the dimension of a point. In this construction, we have to accept that the point does not have a fixed dimension; it has a dimension that depends on the set that it forms. In other words, we cannot conceive of point-dimension theory without the relativity of the dimension of the point (see subsection \ref{4.1.2}).   

Our approach in this work could be seen, in some respect, as the pointillist analogue of Abraham Robinson's approach of introducing the nonstandard analysis. Specifically, what the infinitesimalist and pointillist conceptions have in common is that they are both based on two convenient fictions: the infinitesimal and the point. There are two significant differences between the two conceptions that we alluded to in subsection \ref{2.3.2}. The first difference is the fact that, unlike the infinitesimal, the point has a rigorous definition: that which has no extent, that which has no part. The vagueness of the definition of the infinitesimal $dx$ did nevertheless not prevent the development of the infinitesimal calculus. But it was the source of well-founded and virulent criticism of this conception, notably from George Berkeley. This resulted in the preference of mathematicians for the pointillist formalism with the use of the $(\epsilon,\delta)$-procedure through the notion of limit. Thus, the non-standard analysis introduced by Abraham Robinson's objective in the sixties \cite{robinson} was to introduce a rigorous definition of the infinitesimals. 

The second one is the fact that, contrary to the point, the infinitesimal has an imprint or a signature of the environment from which it comes; for example, curve, surface or volume. By contrast, the points are all similar; they do not carry in them any mark allowing to distinguish them one from the other (see section \ref{2.3}). In this sense, the point-dimension theory aspires to attribute dimensional identities to the points, which makes it possible to specify the dimension of the point in relation to the organization of which it is part. This identity attributed to the point brings a criterion to distinguish between points relatively to the sets to which they belong and to the relations which they form with the other points within the same set. 

More generally, we believe that the point should be augmented by adding attributes to it. In this way, we would like to, for the moment, to keep a pointillist conception of the sets, which means that we cannot augment the point in any manner. For example, we cannot attribute to the point an extent, a length, a surface or a volume; otherwise, the pointillism of the set will be lost. The notion of dimension makes it possible to realize this objective to enrich the point while preserving a pointillist approach, as long as one can have a conception which is compatible, pertinent and consistent, hence the introduction of the point-dimension theory. Another example of a notion that realizes this conception is the curvature which is a point-notion, but contrary to the notion of dimension it has a narrower scope of application. By augmenting the attributes and information of the points in the set, we mechanically augment the set itself. Indeed, in connection with the analysis performed in subsection \ref{2.3.2}, the points of the unit square $[0,1]\times [0,1]$ are not now just labels i.e. Cartesian coordinates $(x,y)\in [0,1]\times [0,1]$, although any point has both a label and a dimension, in this case two. In this sense, this dimensional attribute provides information about the mutual relations between the points, which is in line with Fréchet's objective (see the first quote in section \ref{3.3}).  

A case in point, by assigning a dimension to the points of the sets, we create a way to compare and classify them based on their points. More precisely, two sets will be said to be equivalent if we can show a one-to-one correspondence between the two, respecting moreover the dimensions of the paired points. This constitutes, in a certain way, the reverse path of Fréchet's approach to dimension (see subsection \ref{3.3}) since instead of comparing the dimensions of the sets through their topologies via homeomorphisms, we start by assigning dimensions to the points via the topology of their neighborhoods, which allows us to compare the topologies of the two sets by comparing them point by point. In summary, to compare the topologies of two sets we could rely on homeomorphism. The procedure described here gives another comparative way based on the enrichment of the points of a set by assigning dimensions to them. This shows that the dimension assigned to the point gives it an identity since we can compare two sets by comparing their points one by one. We are not the first to want to increase the point, but maybe not for the same considerations, we refer the reader to Pierre Cartier's article where he analyzes the evolution of the notions of spaces and symmetry from Alexandre Grothendieck to Alain Connes and Maxime Kontsevitch. 

To finish this subsection, we will clarify our conception of mathematical space. In connection with the previous subsection, the dimension of a set is a quantification, in a certain sense, of the organization of the points constituting it. The intrinsic dimension of the points constituting the space should be as large as possible in order to allow the existence of all dimensions. Thus, the mathematical space should be considered as a set of points potentially of infinite dimension. In other words, in some sense, we assume here that any set can be immersed in a space of infinite dimension; therefore, the point is initially of infinite dimension relatively to the whole space. The fact of considering the point relative to a given set imposes a constraint, brings information and thus reduces its dimension relative to the set in question. For example, this conception suggests that the points of a line are of infinite dimension, but the information on their organization through the Cartesian equation $y=ax+b$ will allow to assign them the dimension one relative to this line. However, a point relative to itself, which can be considered as an isolated point will of course have zero dimension. 

\begin{figure}[h!]
    \centering
    \includegraphics[width=0.45 \textwidth]{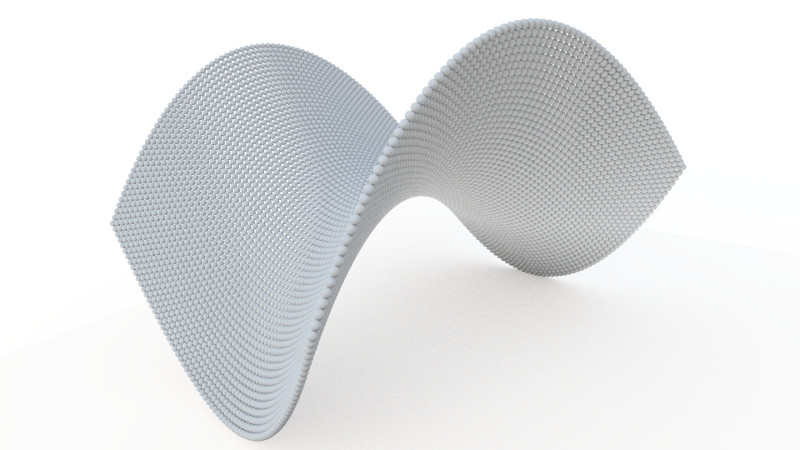}
    \includegraphics[width=0.45 \textwidth]{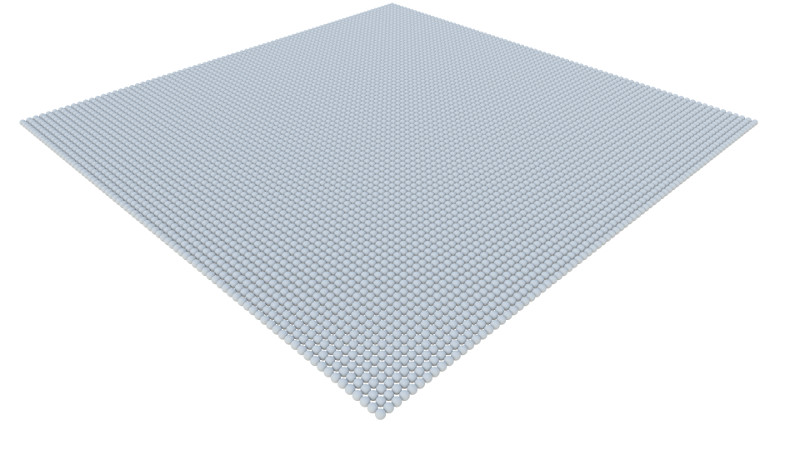}
    \caption{Dimensional representations of two surfaces formed by three-dimensional points}
    \label{anass}
\end{figure}

In other words, an object in this mathematical space is an organization of points of infinite dimension and it is this organization that will allow us to assign it a dimension. However, it is the link and the relations with the other points that will determine the dimension of the point. If a point is considered without relationship with other points, its dimension is zero. That is to say, an isolated point will have zero dimension. To illustrate this conception, the figure \ref{anass}, presents a dimensional representation of two surfaces formed by three-dimensional points. Indeed, the points composing these surfaces are supposed to be of dimension three, having the same dimension of the ambient mathematical space, but they form surfaces of dimension two. To go further into this conception, conceiving the dimension as an organization also allows to change the point by any other object. Indeed, what we quantify is the organization independently of the nature of the elements of this organization. A line formed by cars in the space of cars will be of dimension one.  

\subsection{On what notion of dimension should we base ourselves?}
\subsubsection{Topological or measured notion of dimension?}
Through all the analysis we have suggested so far, our leaning for a topological notion seems quite clear. We have underlined several times in subsection \ref{2.1} the persistent confusion between size and dimension, between null measure and zero dimension. More generally, any existing notion of dimension based on the notion of measure suffer from the fact that any negligible set, let alone non-measurable sets, is usually of zero-dimensional. For example, any countable set is systematically of zero dimension. In our opinion, an adequate notion of dimension should avoid the confusion size/dimension. Furthermore, this way of quantifying dimension would suggest that this notion only concerns sets whose cardinals are uncountable. However, we have learned (see subsection \ref{2.3.2}) through Cantor's investigation that for infinite sets the dimension has no connection with the cardinal. This is why it would be judicious to consider a notion of dimension, which also allows to attribute different dimensions to countable sets according to their organization. A notion of dimension as large as possible able to be extended to a wide class of sets. 

On the other hand, point-set topology seems to have a broader scope than measure theory; more particularly, most measures are built on the basis of a predefined topology on the set in question. Topology covers a very vast field in mathematics, it can be used in many different areas. We can find the employment of point-set topology almost everywhere in mathematics, even in branches of mathematics that are \textit{a priori} far from it. For example, one can find applications of topology in the purely algebraic branch of commutative algebra. Furthermore, point-set topology as a pointillist discipline seems to be more adequate for the introduction of a point-dimension theory.    

However, we will not use the classical topological dimensions presented in the two subsections \ref{3.1} and \ref{3.2}, for two reasons. First, these notions assign the zero dimension to a large class of totally disconnected sets, containing in particular the countable sets. This rigidity is due to the fact that these dimensions can only take integer values, because of the inductive nature of their definitions. Nevertheless, many of the authors who introduced these topological dimensions have constructed twisted sets, for which it would be more appropriate to assign non-integer dimensions. Bolzano's curve and Menger sponge are good example here. 

Secondly, we explained in section \ref{2.2} the brake that has persisted regarding the passage to higher dimensions. The acceptance was mainly due to the change of point of view of mathematicians and philosophers. From then on, geometry should be founded on logic and does not have to be based on sensible experience. In other words, higher dimensions are accepted as soon as geometry is dissociated from our three-dimensional empirical space. Since the beginning of the twentieth century, however, higher dimensions had to be accepted not only as an abstraction, but also as a fact. Physics has taught us that it is very likely that our space is higher than three dimensions. In the same sense, it would be a mistake to ignore non-integer dimensions. It is true that for the moment it is still difficult to have a precise idea of what these non-integer dimensions represent, but it may very well be that even the dimension of our physical space is non-integer. For that matter, there are already works on the possible \textit{fractality} of space-time initiated by Laurent Nottale \cite{not}. It is true that the assumptions of regularity on a uniform space-time allowing to write the equations and to formulate the physical theories remain unverified \textit{a priori}.  

Moreover, we think that a notion of non-integer dimension will be more adequate for a point-dimension theory building. Indeed, we will present in the following (see examples \ref{ex2} and \ref{ex21}), sets whose dimensions vary with the position, which would have been difficult to construct with notions of integer dimensions. Thus, our choice naturally fell on box dimension. We have succeeded in freeing this notion from the metric, in such a way that we can consider it fully as a topological dimension notion. As we pointed out in section \ref{3.4}, the notion of box dimension seems to be a notion with very appreciable properties except for one. Indeed, the box dimension of a dense subset is equal to the dimension of the set in question, in particular $dim_B(\mathbb{Q}\cap [0,1])=dim_B([0,1])$. This property, allowing to identify the dimension of a set and its dense subsets, seems to us acceptable and even adequate according to our conception of a topological dimension, which quantifies the organization of points. In particular, it seems relevant that the dimension is a kind of density measure of a set. We understand that one can find inappropriate that the dimension of $\mathbb{Q}\cap [0,1]$ be the same as the dimension of a curve, but we have already pointed out in subsection \ref{4.2.2} that we do not expect a notion of dimension that classifies the different basic geometric objects. Finally, once again, one should not see a link between dimension and cardinal. We can have sets with the same cardinal, but with different dimensions, just as we can have sets with different cardinalities but the same dimension.

Box dimension respects our wishes and even more. More precisely, in addition to being topological, it contains in it a link with the measure. Indeed, as an example in the original article of Pontrjagin and Schnirelmann \cite{pontr} they named the quantity $N_{\epsilon}(X)$ involved in the definition of the box dimension \ref{kolmobox}, \textit{volume function}. It is indeed understandable to consider this quantity as a kind of measure. It could be regarded as a metric measure with integer value with $\epsilon$ precision. It verifies for example the property: let $A$ and $B$ be two compact sets with $A\cap B=\emptyset$ then $N_{\epsilon}(A\cup B)=N_{\epsilon}(A)+N_{\epsilon}(B)$ for $\epsilon$ small enough. It could be seen as a sets compactness measure. The less compact the set $A$ is, or more precisely its closure $\bar{A}$, the faster $N_{\epsilon}(A)$ grows. In particular a closed interval will be "less compact" than a closed square. In this sense, $N_{\epsilon}(A)$ seems interesting to measure the compactness of sets in the infinite dimensional framework and thereby to deduce a quantity representing dimension even in this setting.        

We have already pointed out the link between the box dimension and the Minkowski-Bouligand dimension in section \ref{pont} which explains its connection with the measure. Nevertheless, we want to formally bring a precision about this relationship here. To do this, consider $(X,d)$ a compact set and $N(\epsilon)$ the minimal number of balls of radius $\epsilon$ needed to cover $X$. It is obvious that:

\begin{equation}\label{evident}
V(X)\approx N(\epsilon)V(\epsilon),
\end{equation}
 here $V(X)$ and $V(\epsilon)$ are respectively the volumes of $X$ and of a ball of radius $\epsilon$. We do not specify here the exact meaning of the volume $V$, it is a not specified measure.
We further assumed that 
\begin{equation}\label{evident1}
V(X)\propto (size(X))^{dimension(X)},
\end{equation}
 which constitutes the basis of the conception of the so-called measured dimensions (see subsection \ref{3.4}). Here, $\propto$ means proportional to. By assuming that $X$ is dimensionally homogeneous, that for enough small $\epsilon$ the dimension of $X$ is equal to the dimension of the balls $B(\epsilon)$ of radius $\epsilon$ covering $X$ and by combining relations (\ref{evident}) and (\ref{evident1}) for $B(\epsilon)$ we get 
\begin{equation}\label{evident2}
 V(\epsilon)\approx \frac{V(X)}{N(\epsilon)}\propto \epsilon^{dimension(X)}\propto (size(B(\epsilon)))^{dimension(X)} ,
\end{equation}
by computation, one can derive the formula of the box dimension. Thus, in summary, the derivation of the box dimension formula starts with measure-related considerations, which are completely discarded at the end. This makes it an interesting notion in the context of infinite dimension.  

\subsubsection{The difference between entropy and dimension}\label{4.3.2}
By analyzing the link between Cartesian coordinates and the notion of dimension in subsection \ref{2.2.1}, we have highlighted the connection between the latter and the quantity of information. Moreover, the fact that linking dimension to organization or order of matter presented in subsection \ref{4.2.1}, seems necessary to link dimension to the concept of entropy. In section \ref{pont}, we explained Kolmogorov's motivation to introduce the notion of $\epsilon$-entropy for continuous sets by analogy with Shannon's notion of entropy. Nevertheless, there are some subtleties related to our way of conceiving the notion of dimension that we wish to put forward in this subsection.    

Taking again the simple example of a set $X$ of cardinal $N$ with equiprobability of realization of $x_i$, $P (x_i) = 1/N$, then Shannon's entropy is the quantity $H_2(X)=\log_2(N)=\frac{\ln(|X|)}{\ln(2)}$ bits. The quantity $\ln(2)$ is related to bits $Y=\{0,1\}$. In this case, the entropy of the set $X$ is relative to the Base $2$, we can write $H(X)=\frac{\ln(|X|)}{\ln(|Y|)}$. In practice, if $X=\{a,b,c,d\}$ is an alphabet then $H_2(X)=2$ which means that we need exactly two bits to encode all the elements of the set $X$. For example, $a=00$, $b=01$, $c=10$ and $d=11$. In the case where the set $Y=\{1,2,3,4\}$, then $H_4(X)=1$ and we need only one symbol to encode each elements of $X$. More generally, we can consider any base $|Y|=b$, then $H_b(X)=\log_b(N)=\frac{\ln(|X|)}{\ln(|Y|)}$. In summary, the entropy defined in this way can be seen as the amount of information in the set $X$ divided by the amount of information in the channel $Y$. Entropy would thus be a comparison between two quantities of information.           

Let us now consider the famous Boltzmann's entropy which appeared in the context of statistical physics, defined by $S=k_B\ln{(\Omega)}$ (which reasonably does not depend on any radix), where $\Omega$ is the cardinal of all possible configurations of the constituent particles of matter and $k_B$ is the Boltzmann constant. It is obvious that the formulas of Boltzmann's entropy and Shannon's are very similar and even the same under certain conditions. Indeed, it is possible to take the Boltzmann constant equal to one ($k_B=1$) by normalizing, which gives a temperature in unit of energy (see \cite{boltz}). At this moment, the two formulations coincide if we use the natural unit of information \textit{nat}; namely, we take the logarithm in base $e$ in Shannon's formula.   

In fact, from our point of view, the notion of entropy should only concern formulas in base $e$ i.e. $H(X)=\ln(|X|)$. The quantity $H_b(X)=\frac{\ln(|X|)}{\ln(|Y|)}=\frac{\ln(N)}{\ln(b)}$ represents rather a dimension, which is the dimension of a finite set. In this case, contrary to sets of infinite cardinalities, the notion of dimension actually depends on the cardinal of the set considered. To better understand this way of seeing things, let $(X',d)$ be a metric space; if we go back to the definition of the box dimension: $dim_B(X')=\lim\limits_{\epsilon \to 0}\frac{\ln(N_{\epsilon}(X'))}{\ln\frac{1}{\epsilon}}$ we can rewrite this formula differently $dim_B(X')=\lim\limits_{\epsilon \to 0}\log_{\frac{1}{\epsilon}}(N_{\epsilon}(X'))$ where $\log_{\frac{1}{\epsilon}}$ denote the logarithm function of base $\frac{1}{\epsilon}$. At this point, it is either the box dimension is not a notion of dimension, but rather an asymptotic entropy in base $\frac{1}{\epsilon}$, or it is the Shannon entropy in base $b$ which is not an entropy but rather a dimension. We have chosen the second option for several reasons.

In summary, to be more precise, $H(X)=\ln(|X|)$ and $H_{\epsilon}(X')=\ln(N_{\epsilon}(X'))$ are entropies, while $H_b(X)=\log_b(|X|)$ and $dim_B(X')=\lim\limits_{\epsilon \to 0}\log_{\frac{1}{\epsilon}}(N_{\epsilon}(X'))$ are dimensions. The first is the comparison of the entropy of $X$ with the entropy of a set $Y$ with $|Y|=b$. The second is the comparison of the entropy of $X'$ with the $\epsilon$-entropy of an interval since $N_{\epsilon}(interval)\sim \frac{1}{\epsilon}$. In other words, if for the box dimension we had compared with the entropy of a surface ($\log_{\frac{1}{\epsilon^2}}(N_{\epsilon}(X'))$), the surfaces would then be of dimension one and the intervals of dimension $\frac{1}{2}$. This is another aspect of the relativity of dimension. Our analysis allows us to define the dimension in another way: the dimension of a set is a comparison (the ratio) between its entropy and the entropy of a reference set. This definition of dimension is in a certain sense similar to the concepts of dimension of Riesz and Fréchet exposed in the section \ref{3.3}. Indeed, the dimension appears from then on as a comparative notion. 

\begin{rem}
Rigorously speaking, we must underline the fact that since we consider the notion of dimension as a quantification of organization or order, we should rather have looked for the connection between dimension and the \textit{negentropy} or \textit{negative entropy}. This notion, highlighted among others by Erwin Schrödinger, quantifies order and organization, unlike entropy, which measures disorder and disorganization. Nevertheless, this does not change anything in our case since we have defined the dimension as the ratio between two entropies, which should be equivalent to the ratio between the negentropies. We prefer to use the notion of entropy because it is better known and much more used than the negentropy.  
\end{rem}

This approach also extends the notion of dimension to finite sets. In this finite frame, the dimension of a point is not useful, it will always be equal to zero. If we consider a countable set $X$ without accumulation point, its dimension will be infinite, $H_b(X)=+\infty$ for all $2 \leq b\in \mathbb{N}$. Furthermore, its point-box dimension will be zero because the quantity of information of $X$ is locally finite. If now $X$ is countable with at least one point $x$ of accumulation, its dimension $H_b(X)$ will always be infinite. Nevertheless, we can start talking about the dimension of the point $x$. Indeed, in the neighborhood of $x$ the quantity of information is locally infinite and therefore we can try to compare it to the quantity of information of a segment. In other words, the  point-dimension estimation of a set make sense if its entropy is locally infinite, at least in the neighborhood of a point. But even in this case, the entropy of the neighborhood of $x$ may be very small compared to the entropy of a segment. At this moment, it should be compared with entropies smaller than that of an interval (see example \ref{invisible}). In this case, the dimension will then be said to be invisible.

To finish this section, let us notice that this approach is compatible with the notion of vector space dimension. Let us start by the finite vector spaces case. Let $K$ be a finite field with $|K|=q$ a prime power and $E$ a finite $K$-vector space, then $|E| = |K|^n=q^n$. As explained above, since $|E|$ is finite then the dimension of $E$ must be based on the comparison of its cardinal and the cardinal of a reference set. Here, we can reasonably designate the field $K$ as the reference set. Then the dimension of $E$ can be computed by the formula:          
\begin{equation}
    H_q(E)=dim(E)=\frac{H(E)}{H(K)}=\frac{\ln(q^n)}{\ln(q)}=n.       
\end{equation}
The same goes for all linear subspaces $F$ of $E$, $H_q(F)=dim(F)=\frac{H(F)}{H(K)}=\frac{\ln(|F|)}{\ln(q)}$. In the case of an infinite $K$-vector space $E$ on an infinite field $K$, one can defined the dimension of a point. But since these spaces are dimensionally homogeneous, the dimension of $E$ can be identified by computing the dimension of the unit ball (see the next section). Then, we can posit that: 
\begin{equation}\label{entrodim}
    dim_B(E)=\lim\limits_{\epsilon \to 0}\frac{H_{\epsilon}(B_E(0,1))}{H_{\epsilon}(B_K(0,1))}=\lim\limits_{\epsilon \to 0}\frac{\ln(N_{\epsilon}(B_E(0,1)))}{\ln(N_{\epsilon}(B_K(0,1))},
\end{equation}
where $B_E(0,1)$ and $B_K(0,1)$ are respectively the unit balls of $E$ and $K$. In other word, we compare the $\epsilon$-entropy of $E$ with the $\epsilon$-entropy of the designated reference set $K$. This definition allows to recover the vector space dimensions. In particular, $dim_B(\mathbb{R}^n)=n$ and $dim_B(\mathbb{C}^n)=2n$ as $\mathbb{R}$-vector space and $dim_B(\mathbb{C}^n)=n$ as $\mathbb{C}$-vector space. This definition allows also to compute the dimensions of vector subspaces, if $F$ is a vector subspace then $dim_B(F)=\lim\limits_{\epsilon \to 0}\frac{H_{\epsilon}(B_F(0,1))}{H_{\epsilon}(B_K(0,1))}=\lim\limits_{\epsilon \to 0}\frac{\ln(N_{\epsilon}(B_F(0,1)))}{\ln(N_{\epsilon}(B_K(0,1))}$. 
\begin{rem}We want to emphasize the fact that the comparison between cardinals only works properly in the case of finite cardinals. Indeed, one could be tempted to extend the comparison of cardinals to the infinite case, for different order of infinities. But this approach is not adequate for us since one will always find infinity. Consider for example $\mathbb{R}$ as a $\mathbb{Q}$-vector space, one could formally write $dim_B(\mathbb{R})=\frac{H(\mathbb{R})}{H(\mathbb{Q})}=\frac{\ln(|\mathbb{R}|)}{\ln(|\mathbb{Q}|)}=+\infty$, which is the right dimension as a vector space. Algebraically speaking it is convenient since this case requires a base of infinite cardinal. But, it seems to us completely superficial to see $\mathbb{R}$ as a $\mathbb{Q}$-vector space. In this case of non complete field, the formula (\ref{entrodim}) does not coincide with the algebraic dimension because it gives dimension one to $\mathbb{R}$ as a $\mathbb{Q}$-vector space. It is a case where topology does not coincide with algebra since even Riesz's Lemma does not work in this case, the closed unit ball is compact while its algebraic dimension is infinite.  
\end{rem}

\section{Point-extended box dimension}\label{5}
In adequacy with our conception of the mathematical space as a topological organization of points, our aim to define the dimension arising from the point and the choice of the spirit of box dimension, we extend box dimension in many directions. First, to the general setting of topological spaces freeing it from the notion of metric. Second, localizing it to the point and finally capturing many dimension scales even infinite order. As observed in the introduction some minimal requirements are needed to balance geometry, topology and algebraic properties. Briefly, in order to formalize the notion that we are going to present in the following, we have been inspired by the box dimension, by Kolmogorov's contribution on the linear dimension \cite{K2} and on $\epsilon$-entropy \cite{KT}, and by the developments made by Gelfand \cite{gel} in connection with these notions.  

\subsection{Point-extended box dimension in the general setting of topological vector space}
In the absence of any  metric structure, we consider the largest possible framework of topological vector space to introduce this notion. On the one hand, more precisely, we usually compute the dimension of a set assumed be included in some kind of space. On the other hand, the ambient space should possess enough structure allowing to quantify the desired dimension. The focal point in our approach consists of, first of all, extending the notion of dimension to the point having memory from the ambient space. Hence, after careful consideration, we chose to use the \textit{topological vector space} as a framework avoiding any restriction to metric spaces.  The main advantage in  topological vector space is the fact that any neighborhood of zero provides a neighborhood at any point when translated.

We denote by $X$ a topological vector space and let $V$ be a neighbourhood of $0$. We say that $V$ is \textit{absorbing} if $\forall x\in X$, $\exists \lambda >0$ such that $cx\in V$ for all $c\leq \lambda$. Obviously, in a topological vector space since $ 0·x = 0$, so by continuity there exists  $\lambda > 0$ such that $cx \in V$ if $c \leq \lambda$ and hence every neighborhood of $0$ is absorbing. 

It is well known that every topological vector space $X$ has a basis ${\mathcal B}$ of absorbing closed neighbourhoods of zero. 

In the sequel, we denote 
 $\Phi =\{ \varphi =(\phi_1,\phi_2): \lim\limits_{x\to +\infty}\phi_i(x)=+\infty \mbox{ and } \lim\limits_{x\to +\infty}\frac{\phi_i(2x)}{\phi_i(x)}=1\}$  for the family of  function scales involved in our definitions.
 \begin{Def}[$\epsilon$-number]\label{salina}
 Let $E$ be a Hausdorff $K$ topological vector space and let $V$ be a subset of $E$. For any $U$ a neighbourhood of $0$ the $\epsilon$-number is given by  
$$ N(\epsilon,V,U)=\min\# \{v\in V : V\subset \bigcup\limits_{v\in V}v+\epsilon U \}$$
 \end{Def}
 We have the following immediate properties:
 \begin{rem}\label{neps} Under the notations above,
     \begin{itemize}
     \item For every $\epsilon>0$ and $r>0$, $N(\epsilon,V,rU)= N(r\epsilon,V,U)$.
     \item $\epsilon_1\leq \epsilon_2\Rightarrow N(\epsilon_1,V,U)\le N(\epsilon_2,V,U)$,
     \item $V_1 \subset V_2\Rightarrow N(\epsilon,V_1,U)\le N(\epsilon,V_2,U)$,
     \item $U_1\subset U_2\Rightarrow N(\epsilon,V,U_2)\le N(\epsilon,V,U_1)$.
 \end{itemize}
 \end{rem}
\begin{Def}[Point-extended box dimension]\label{originale}
Let $E$ be a Hausdorff $K$ topological vector space  with $K$ complete. For all $x\in E$, we denote by $\mathcal{U}(x)$ the neighbourhood basis of $x$. Let $V\in \mathcal{U}(x)$ and $U\in\mathcal{U}(0)$. The point-extended box dimension of $x \in E$, associated with $\varphi \in \Phi$, denoted below  $dim_{\varphi}\{x\}_E$, is defined by the following  
 $$dim_{\varphi}\{x\}:=dim_{\varphi}\{x\}_E=\sup\limits_{U\in\mathcal{U}(0)}\inf\limits_{V\in\mathcal{U}(x)}\overline{\lim\limits_{\epsilon \to 0}}\frac{\phi_1(N(\epsilon,V,U))}{\phi_2(\frac{1}{\epsilon})},$$
 Let $A\subset E$ be given.  The associated relative point-extended box dimension is given for $x\in E$ by  
 $$dim_{\varphi}\{x\}_A=\sup\limits_{U\in\mathcal{U}(0)}\inf\limits_{V\in\mathcal{U}(x)}\overline{\lim\limits_{\epsilon \to 0}}\frac{\phi_1(N(\epsilon,V\cap A,U))}{\phi_2(\frac{1}{\epsilon})}.$$
\end{Def}
 \begin{rem} We have the next general observations, 
     \begin{itemize} 
         \item For  $x\notin {\bar A}$, there exist a neighborhood $V$ of $x$ such that $V\cap A =\emptyset$. It is then clear from the definition that $dim_{\varphi}\{x\}_A=0$ in this case.
         \item The use of $\limsup$ instead of $\lim$ is due to the fact that the limit does exist not necessarily. Also, it is preferred to $\liminf$ to guaranty $dim_{\varphi}(A\cup B)= \max(dim_{\varphi}(A),dim_{\varphi}(B))$ as mentioned above.
         \item The supremum and the infinimum in our definition are motivated by its local spirit. Indeed, for every $x$ in a Hausdorff space $x=\bigcap\limits_{V\in \mathcal{U}(x)}V$ and $N(\epsilon,V,U)$ is increasing in $V$ and decreasing in $U$.  
     \end{itemize}
 \end{rem}

Using remark \ref{neps} we have the following:
\begin{rem}\begin{itemize}
    \item Since $N(\epsilon,V,U)$ is monotone in either $U$ and $V$, the above definition of dimension is independent of the choice of the basis of the neighbourhoods. Furthermore, the basis can consists in only open neighbourhoods or closed neighbourhoods. 
    \item If $x\in A$ is isolated then we have obviously $dim_{\varphi}\{x\}_A=0$. By convention we will also put $dim_{\varphi}\{x\}_{A}=dim_{\varphi}(x)_{A\cup \{x\}}=0$ if $x\notin \bar A$ .
    \item In the definition of $dim_{\varphi}\{x\}_A$, one can ask the question of the pertinence of the choice $x\in\bar{A}$. In fact, we have, with the notion we use, automatically $\forall x\in A$, $dim_{\varphi}\{x\}_A=dim_{\varphi}\{x\}_{\bar{A}}$, so we preferred to extend to $\bar{A}$ the definition of the dimension of $x$ relative to $A$ from the beginning. The consistency of our definition derives from the fact that $V \cap A \ne \emptyset$ for every $x\in \bar A$ and $V\in \mathcal{U}(x)$.   
     
    \end{itemize}
    \end{rem}
    We can define the point-extended box dimension of $A$ as follows:
\begin{Def}[Point-extended box dimension of sets]
Let $E$ be a topological vector space and $A\subset E$. The point-extended box dimension of $A$ is defined as  
$$dim_{\varphi}(A)=\sup_{x\in \bar A}dim_{\varphi }\{x\}_A=\sup_{x\in \bar A}\sup\limits_{U\in\mathcal{U}(0)}\inf\limits_{V\in\mathcal{U}(x)}\overline{\lim\limits_{\epsilon \to 0}}\frac{\phi_1(N(\epsilon,V\cap A,U))}{\phi_2(\frac{1}{\epsilon})}.$$
In particular, $$dim_{\varphi}(E)=\sup_{x\in E}dim_{\varphi}\{x\}=\sup_{x\in E}\sup\limits_{U\in\mathcal{U}(0)}\inf\limits_{V\in\mathcal{U}(x)}\overline{\lim\limits_{\epsilon \to 0}}\frac{\phi_1(N(\epsilon,V,U))}{\phi_2(\frac{1}{\epsilon})} .$$
\end{Def}
\begin{pro}Let $E$ be a topological vector space and $A\subset E$. Then for every $x\in E,$ we have $dim_{\varphi}\{x\}=dim_{\varphi}\{0\}$. It follows that, $$dim_{\varphi}(E)= dim_{\varphi}\{0\}.$$
\end{pro}
\begin{proof}
It suffices to see that given a neighborhood $V$ of zero, $x+V$ is a neighbourhoods of $x$ and that $N(\epsilon,x+V,U)=N(\epsilon,V,U)$, for every $\epsilon >0$.
\end{proof}
As a direct consequence
\begin{cor}Let $E$ be a topological vector space and $A\subset E$. If $A$ contains an open set, then $dim_\varphi(A)= dim_\varphi(E).$ 
\end{cor}
It suffices to see from the definition that for $x\in V\subset A,$ with $V$ open set, we have $dim_\varphi(\{x\}_A)= dim_\varphi(E).$ 


\begin{rem}\label{closed}
    \begin{enumerate}
    \item We have, for every vector topological space, $dim_{\varphi}\{x\}_E = 0$ in the case of the coarse topology and the discrete topology. In particular for every $A\subset E$, $dim_{\varphi}(A) =\sup\limits_{x\in\bar A}(dim_{\varphi}\{x\}_A) = 0$ and hence the localization is not useful in this case. However, for finite sets the discrete topology is the adequate one in connection with Shannon's entropy, see discussion in subsection \ref{4.3.2}.  
    \item In the previous definitions, the condition $K$ is complete allows to avoid some abnormal situations. For example, 
${\mathbb R}$ as a ${\mathbb Q}-$vector space has infinite algebraic dimension, while endowed with the usual Euclidian toology, $dim_{\ln}({\mathbb Q})=dim_{\ln }({\mathbb R})=1$.
Also, the algebraic dimension of ${\mathbb Q}$ as a discrete ${\mathbb Q}-$vector topological space is equal to $1$ while its point extended box dimension is $0$.
\item We point out at this level that the relativity of the expression, $dim_{\varphi}(A)=\sup_{x\in A}dim_{\varphi}\{x\}_A$, in term of $A$, do not affect the definition as a map on $\mathcal{P}(E)$. More precisely, if $A \subset F \subset E$, where $F$ is any topological subspace of $E$, then $dim_{\varphi}(A)$ is the same when $A$ is regarded as a subset in  $F$ or in $E$.
    \end{enumerate}
\end{rem} 
In the remainder of this article,  $K$ is a complete field. It is  either the real field, the complex field or the p-adic field. 

On  can extend the notion of box dimension to the topological vector space framework without going through the points in  the following way:
\begin{Def}[Extended box dimension]
Let $E$ be a topological vector space, $\varphi\in \Phi$ and $A\subset E$. The extended box dimension of $A$ is defined as  
$$Dim_{\varphi}(A):=\sup\limits_{U\in\mathcal{U}(0)}\overline{\lim\limits_{\epsilon \to 0}}\frac{\phi_1(N(\epsilon,A,U))}{\phi_2(\frac{1}{\epsilon})}.$$
\end{Def}

We have obviously $dim_{\varphi}(A)\leq Dim_{\varphi}(A)$, for every $A$. In the next example, we observe that the inequality may be strict in a general setting. 
\begin{ex}
Let $E=(\mathbb{R}^2,\mathcal{T}_d)$ where $\mathcal{T}_d$ is the discrete topology on $\mathbb{R}^2$ and an arbitrary $\varphi\in \Phi$. Clearly, $E$ is a topological vector space. For every infinite set  $A$, we have:
\begin{itemize}
    \item $dim_{\varphi}(A)=0$, since we use discrete topology.
    \item For the neighbourhood of zero $U=\{0\}$, we have $N(\epsilon,A,U)=+\infty, \forall \epsilon>0$, and then $$Dim_{\varphi}(A)=\sup\limits_{U\in \mathcal{U}(0)}\overline{\lim\limits_{\epsilon \to 0}}\frac{\phi_1(N(\epsilon, A,U))}{\phi_2(\frac{1}{\epsilon})}=+\infty.$$ 
\end{itemize}
\end{ex}
Following Dowker's approach \cite{dow} (see subsection \ref{4.1.1}), a localization of the extended box dimension is defined  as follows: $$Dim_{\varphi}\{x\}_{A}:=\inf\limits_{V\in \mathcal{U}(x)} Dim_\varphi(V\cap A) \: \mbox{ for } \; x\in \bar A.$$ 
The latter formula induces, according to Dowker's notation, the $locDim_\varphi$, by writing $$locDim_\varphi(A)= \sup\limits_{x\in \bar A}Dim_{\varphi}\{x\}_{A}.$$
From the previous example, $Dim_{\varphi}\{x\}=dim_{\varphi}\{x\}=0$ for every $x\in E=(\mathbb{R}^2,\mathcal{T}_d)$ and hence we have $loc Dim_{\varphi}(A)=0<Dim_{\varphi}(A)=+\infty$ for every infinite set $A$.  

It is also clear that in general $dim_{\varphi}\{x\}\leq Dim_{\varphi}\{x\}$ for every $x$. It is however natural to ask when the point-extended box dimension $dim_{\varphi}(A)$ of a set $A$ coincides with the localized extended box dimension $loc Dim_{\varphi}(A)$.  

We start with the next properties.
\begin{pro}\label{increasing}Under the previous notations, let $A\subset A', B\subset E$ and $x\in \overline{A\cup B}$. Then
\begin{enumerate}
\item  $dim_{\varphi}\{x\}_{A}\le dim_{\varphi}\{x\}_{A'}$.
    \item $dim_{\varphi}\{x\}_{A\cup B}= max(dim_{\varphi}\{x\}_{A},dim_{\varphi}\{x\}_{B})$, 
    \item $dim_{\varphi}\{x\}_{A}=dim_{\varphi}\{x\}_{\bar A}$,
    \item  If moreover $\phi_1(x.y)\leq \phi_1(x)+ \phi_1(y)$, for $A_1\subset E_1$ and $A_2\subset E_2$ and  $(x_1,x_2)\in \overline{A_1\times A_2}$ we have $$\max(dim_{\varphi}\{a_1\}_{A_1},dim_{\varphi}\{a_2\}_{A_2})\leq dim_{\varphi}\{(a_1,a_2)\}_{A_1\times A_2}\leq dim_{\varphi}\{a_1\}_{A_1}+dim_{\varphi}\{a_2\}_{A_2}.$$
\end{enumerate}
In particular, 
\begin{enumerate}
\item[1'] $dim_{\varphi}(A)\le dim_{\varphi}(A')$,
    \item[2'] $dim_{\varphi}(A\cup B)= max(dim_{\varphi}(A),dim_{\varphi}(B))$,
    \item[3'] $dim_{\varphi}(A)=dim_{\varphi}(\bar A)$.
    \item[4'] If $\phi_1(x.y)\leq \phi_1(x)+ \phi_1(y)$, we have $$\max(dim_{\varphi}(A_1),dim_{\varphi}(A_2))\leq dim_{\varphi}(A_1\times A_2)\leq dim_{\varphi}(A_1)+dim_{\varphi}(A_2).$$
\end{enumerate}

\end{pro}
\begin{proof} 
\begin{enumerate}
  \item Derives directly from the monotonicity of $N(\epsilon, V, U)$.
 \item  The inequality $dim_{\varphi}\{x\}_{A\cup B}\ge  max(dim_{\varphi}\{x\}_{A},dim_{\varphi}\{x\}_{B})$, derives from $(1)$, since $A\subset A\cup B$ and $B\subset A\cup B$. For the reverse inequality, let $x\in \overline{A\cup B}$, and $V$ be a neighbourhood of $x$. The union of an open covering of $V\cap A$ and an open covering of $V\cap B$ provides an open covering of $V\cap(A\cup B)$. It follows that
$$ N(\epsilon,V\cap(A\cup B),U)\le N(\epsilon,V\cap A,U)+N(\epsilon,V\cap B,U)\leq 2max(N(\epsilon,V\cap A,U),N(\epsilon,V\cap B,U)).$$
Since  the set of $\epsilon>0$ such that the maximum is attained  is infinite for either $A$ or $B$, we assume that it this the case in $A$, for example. We will deduce that 
$N(\epsilon,V\cap(A\cup B),U) \le 2N(\epsilon,V\cap A,U)$.  Which will give $dim_{\varphi}\{x\}_{A\cup B}\leq  max(dim_{\varphi}\{x\}_{A},dim_{\varphi}\{x\}_{B})$, by using the assumption $\lim\limits_{x\to +\infty}\frac{\phi(2x)}{\phi(x)}=1$.
\item  Derives in the same way as for the classical box dimension. We use the definition with closed neighborhood in such way that any finite covering of $A$ is also a finite covering of ${\bar A }$.
\item We endow $E_1\times E_2$ by the product topology of $E_1$ and $E_2$. We thus choose $U_1\times U_2$, where $U_1$ is an open set in $E_1$ and  $U_2$ is an open set in $E_2$,
as the basis of our  topology. For $(x_1,x_2)\in \overline{A_1\times A_2}$,  $V_1\times V_2$ a neighborhood of $(a_1,a_2)$ and $U_1\times U_2$ a neighborhood of $(0,0)$, we have 
$$ N(\epsilon,(V_1\times V_2)
\cap(A_1\times A_2),U_1\times U_2) = N(\epsilon,V_1\cap A_1,U_1)+N(\epsilon,V_2\cap A_2,U_2).$$
We get

$$\begin{array}{lll}
   \limsup\limits_{\epsilon \to 0}\frac{\phi_1(N(\epsilon,(V_1\times V_2)\cap(A_1\times A_2),U_1\times U_2)) }{\phi_2(\frac{1}{\epsilon})}  &=& \limsup\limits_{\epsilon \to 0}\frac{\phi_1(N(\epsilon,V_1\cap A_1,U_1)\times N(\epsilon,V_2\cap A_2,U_2))}{\phi_2(\frac{1}{\epsilon})} \\
     &\leq & \limsup\limits_{\epsilon \to 0}(\frac{\phi_1(N(\epsilon,V_1\cap A_1,U_1)}{\phi_2(\frac{1}{\epsilon})}+\frac{\phi_1(N(\epsilon,V_2\cap A_2,U_2)}{\phi_2(\frac{1}{\epsilon})}),\\ 
      &\leq & \limsup\limits_{\epsilon \to 0}\frac{\phi_1(N(\epsilon,V_1\cap A_1,U_1)}{\phi_2(\frac{1}{\epsilon})}+\limsup\limits_{\epsilon \to 0}\frac{\phi_1(N(\epsilon,V_2\cap A_2,U_2)}{\phi_2(\frac{1}{\epsilon})}. 
\end{array} $$

Finally,  $$\max(dim_{\varphi}\{a_1\}_{A_1},dim_{\varphi}\{a_2\}_{A_2})\leq dim_{\varphi}\{(a_1,a_2)\}_{A_1\times A_2}\leq dim_{\varphi}\{a_1\}_{A_1}\times dim_{\varphi}\{a_2\}_{A_2}.$$
 \end{enumerate}
The remaining parts of the proof follow from the expression $dim_{\varphi}(A)=\sup_{x\in \bar A}dim_{\varphi }\{x\}_A$.
\end{proof}

The next example shows that countable additivity is not satisfied in general.
\begin{ex} Let $A=\{\frac{1}{n+1}, n\ge 0\}$ be the harmonic sequence set and  $A_k=\{\frac{1}{n+1}, n\le  k\}$. For  $\varphi=(ln,ln)$, $dim_{\varphi}(A)$ is the classical box dimension and it is known that $dim_{\varphi}(A)=\frac{1}{2}$. On the other hand,   we have $dim_\varphi(A_k)=0$ as a finite set. It follows that   $$\frac{1}{2}=dim_{\varphi}(A)=dim_{\varphi}(\cup_kA_k)\ne max_kdim(A_k)=0.$$
\end{ex}
A countable addivitivity formula is given by
\begin{pro} Let $(A_k)_{k\ge 0}$ be a family of sets and denote $B_n=\bigcup\limits_{k\ge n+1}A_k$ . We have $$dim_\varphi(\bigcup\limits_{k\ge 0}A_k) =max(\max\limits_{k \ge 0}dim_\varphi(A_k),\lim\limits_{n\to \infty} dim_\varphi(B_n)). $$
\end{pro}
We prove this result by writing $\bigcup\limits_{k\ge 0}A_k=\bigcup\limits_{k\le n}A_k\cup B_n$ and by using finite additivity property together with $B_n$ is decreasing and then  $\lim\limits_{n\to \infty} dim_\varphi(B_n)$ exists.

Additional structural  condition on  topology allows to remove the supremum in the definition of the point-extended box dimension. To this aim, we introduce the next well known class of topological vector spaces.
\begin{Def} A topological vector space $E$ is locally bounded if it possess a bounded neighborhood of $0$. More precisely, if there exists $V\in \mathcal{U}(0)$ such that for every $ U\in \mathcal{U}(0)$ there exists $\alpha\in K$ satisfying $V\subset \alpha U$.  
\end{Def}
Let $B$ be a bounded neighbourhood of zero and $B_r=rB$ for $r>0$. A basis of neighbourhoods of zero is given by $(B_r)_{r>0}$. It follows that
\begin{pro}\label{boundedpro} Let $X$ be a locally bounded topological vector space, $x\in X$ and $\varphi \in \Phi$. Then 
\begin{equation}\label{bounded}
dim_\varphi\{x\}=dim_\varphi\{0\}=\sup\limits_{r\ge 0}\inf\limits_{s\geq 0}\overline{\lim\limits_{\epsilon \to 0}}\frac{\phi_1(N(\epsilon,B_s,B_r))}{\phi_2(\frac{1}{\epsilon})}=\inf\limits_{s\geq 0}\overline{\lim\limits_{\epsilon \to 0}}\frac{\phi_1(N(\epsilon,B_s,B))}{\phi_2(\frac{1}{\epsilon})}= \overline{\lim\limits_{\epsilon \to 0}}\frac{\phi_1(N(\epsilon,B,B))}{\phi_2(\frac{1}{\epsilon})}
\end{equation}
If  $x\in A\subset E$, the relative point-extended box dimension is given by
\begin{equation}\label{boundedx}
dim_{\varphi}\{x\}_A=\inf\limits_{s\ge 0}\overline{\lim\limits_{\epsilon \to 0}}\frac{\phi_1(N(\epsilon,(x+B_s)\cap A,B))}{\phi_2(\frac{1}{\epsilon})}.
\end{equation}
\end{pro}
\begin{proof} The equality $dim_\varphi\{x\}=dim_\varphi\{0\}$ is valid in the general setting of vector topological spaces.
For a fixed $s$, let $r\le 1$ be given,  then it is obvious that $N(\epsilon,B_s,B_r) = N(r\epsilon,B_s,B)$.
On the other hand, by induction,  $\lim\limits_{x \to +\infty}\frac{\phi_2(2^kx)}{\phi_2(x)}=1$ for every $k\in {\mathbb Z}$ and then 
$\lim\limits_{x \to +\infty}\frac{\phi_2(\alpha x)}{\phi_2(x)}=1$ for every $\alpha >0$. This yields 
$$ \inf\limits_{s\geq 0}\overline{\lim\limits_{\epsilon \to 0}}\frac{\phi_1(N(\epsilon,B_s,B_r))}{\phi_2(\frac{1}{\epsilon})}=\inf\limits_{s\geq 0}\overline{\lim\limits_{\epsilon \to 0}}\frac{\phi_1(N(r\epsilon,B_s,B))}{\phi_2(\frac{1}{\epsilon})} = \inf\limits_{s\geq 0}\overline{\lim\limits_{\epsilon \to 0}}\frac{\phi_1(N(\epsilon,B_s,B))}{\phi_2(\frac{1}{\epsilon})},$$
for every $r>0$.\\
For the last equality, it suffices to show that $N(\epsilon,B_s,B) = N(\frac\epsilon s,B,B)$. Indeed suppose  $B_s\subset\bigcup\limits_{i=1}^{N}x_i+\epsilon B$, with $x_i\in B_s$. We write $x_i=sy_i$, with $y_i\in B$ to get 
$B_s\subset\bigcup\limits_{i=1}^{N}sy_i+\epsilon B= s\big(\bigcup\limits_{i=1}^{N}y_i+\frac\epsilon sB\big)$. We deduce from the last expression that  $B\subset\bigcup\limits_{i=1}^{N}y_i+\frac\epsilon sB$.
The required result is then obtained.\\
The remaining part run similarly.
\end{proof}

 We also have. Under the notations above, 
\begin{pro}Let $E$ be a totally bounded vector space space and $A\subset E$. Under the notations above, the mapping  $ x \rightarrow dim_{\varphi}(\{x\}_A)$
is upper semi continuous.
\end{pro}

\begin{proof}
For $x\in \bar{A}$, using the definition of the infinimum and limsup, we derive that  for every $\eta>0$, there is  $ s_0>0$ and  $\epsilon_0 >0$ such that  for all $s<s_0$  and $\epsilon<\epsilon_0,$ we have    
$$dim_{\varphi}\{x\}_A+\eta>\frac{\phi_1(N(\epsilon,(x+B_s)\cap A,B))}{\phi_2(\frac{1}{\epsilon})}.$$
Now for every  $ y\in (x+Bs)\cap \bar A$, there is $ \delta >0$ such that  $y+B_t\subset x+B_s$ for every $t\le \delta$, and hence
$$dim_{\varphi}\{x\}_A+\eta>\frac{\phi_1(N(\epsilon,(y+B_t)\cap A,B))}{\phi_2(\frac{1}{\epsilon})},$$
for every $t\le \delta$ and $\epsilon<\epsilon_0.$ Thus
$$dim_{\varphi}\{x\}_A+\eta>\inf\limits_{t\geq 0}\overline{\lim\limits_{\epsilon \to 0}}\frac{\phi_1(N(\epsilon,(y+B_t)\cap A,B))}{\phi_2(\frac{1}{\epsilon})}=dim_{\varphi}\{y\}_A, $$
that is $x\to dim_{\varphi}(\{x\})_A$
is upper semi continuous.
\end{proof}
\begin{cor}\label{max} Let $E$ be a totally bounded vector space space and $C\subset E$ be  compact. There exists $a\in C$ such that   $$dim_{\varphi}(\{a\}_C)=dim_{\varphi}(C).$$
\end{cor}
\begin{proof}
Since  $dim_{\varphi}(C)=\sup_{x\in C}dim_{\varphi}\{x\}_C$, there exists $(x_n)_{n\in\mathbb{N}}$ such that $\lim\limits_{n\rightarrow \infty}dim_{\varphi}\{x_n\}_C=dim_{\varphi}(C)$. From $C$ is compact, we may suppose that  $(x_n)_{n\in\mathbb{N}}$  converges to some $a\in C$. Now using  $x\to dim_{\varphi}\{x\}_C$ is upper semicontinuous, we derive that $$dim_{\varphi}(C)= \lim\limits_{n\rightarrow \infty}dim_{\varphi}\{x_{n}\}_C \leq dim_{\varphi}\{a\}_C\le dim_{\varphi}(C).$$   
\end{proof}
We also have
\begin{pro}\label{kk} Let $E$ be a totally bounded vector space space and $C\subset E$ be compact. Then
$$dim_{\varphi}(C)= Dim_{\varphi}(C).$$ 
\end{pro}
\begin{proof}
We only have to show that $dim_{\varphi}(C)\ge Dim_{\varphi}(C).$ To this aim, we use the finite adiditivity property. 
 $$Dim_\varphi(\cup_{i=1}^{n}E_i)=\max_{i=1...n}\Big(Dim_\varphi(E_i)\Big).$$ 
Consider a finite open cover of
$C$ by  $x_{1,i}+B_{\frac{1}{2}}$, and denote $a_1=x_{1,i_1}$ such that 
$Dim_\varphi(C)= Dim_\varphi(C\cap(a_1+B_{\frac{1}{2}}))$.
Again, since $C_1=C\cap(a_1+B_{\frac{1}{2}})$ is  compact, Consider a finite open cover of
$C$ by $x_{2,i}+B_{\frac{1}{2^2}}$ and $a_2=x_{2,i_2}$ such that 
$Dim_\varphi(C)= Dim_\varphi(C\cap(a_2+B_{\frac{1}{2^2}}))$. We obtain inductively a decreasing sequence of compact sets $C_n=C\cap(a_n+B_{\frac{1}{2^n}})$ such that $Dim_\varphi(C)= Dim_\varphi(C_n)$. Now, writing,  $a=\lim a_n$, we claim that $dim_\varphi\{a\}_C=Dim_\varphi(C).$ Indeed, as before for every $\eta>0$, there is  $n_0>0$ and  $\epsilon_0 >0$ such that  for all $n\ge n_0$  and $\epsilon<\epsilon_0,$ we have   
$$dim_{\varphi}\{a\}+\eta>\frac{\phi_1(N(\epsilon,(a+B_{\frac{1}{2^n}})\cap C,B)}{\phi_2(\frac{1}{\epsilon})}.$$
Also, for $m$ large enough, we have   $C_m\subset (a+B_{\frac{1}{2^n}})\cap C$, and then
$$dim_{\varphi}\{a\}+\eta>\frac{\phi_1(N(\epsilon,C_m,B)}{\phi_2(\frac{1}{\epsilon})}.$$
From which we deduce that $$dim_{\varphi}\{a\}+\eta>Dim_\varphi(C_n)= Dim_\varphi(C).$$
\end{proof}
Since,  $dim_{\varphi}(A)\le loc Dim_{\varphi}(A)\le  Dim_{\varphi}(A),$ for arbitrary $A$, we get
\begin{cor}Let $E$ be a totally bounded vector space space and $C\subset E$ be compact. Then
$$dim_{\varphi}(C)=loc Dim_{\varphi}(C)= Dim_{\varphi}(C).$$ 

\end{cor}

\subsection{Point-extended box dimension in metric spaces}
We devote this section to the class of metric spaces. We abandon the structure of topological space in favor of the notion of distance. This will give us a locally more intuitive topology thanks to the discs around the points. This will allow us to give simpler definitions and offer more readable proofs. Kuratowski theorem states that any metric space is isometrically embedded in an infinite dimensional Banach space (that is in particular a topological space). However, no information can be obtained directly, since the problem of metric spaces that  can be embedded in a finite dimensional space remains open.

Let $(X, d) $ be a  metric space and denote   $B_r(a)=\{x\in E/ d(a,x)\leq r \}$ for $a\in X$ and $r>0$. Then $\mathcal{B}(a)=(B_{\frac{1}{n}}(a))_{n\in \mathbb{N}^*}$ is a basis of neighbourhood of $a$. As usual given a subset $A$, the diameter is given by $\delta(A) = sup\{d(x,y), x,y \in A\}$  and an $\epsilon$-covering of $A$  is a family of open sets $(U_i)_i $ in $(E, d) $ such that $\delta(U_i) \le \epsilon$ for every $i$ and $A\subset \cup_iU_i$. Finally, by the $\epsilon$-number, we denote
$$\begin{array}{llll} N(\epsilon, A)&=& \min\{ n  : \mbox{ such that there exists an }  \epsilon-\mbox{covering of } A  \mbox{  with } n  \mbox{ elements}\}\\
 &=& \min\# \{v\in A : A\subset \bigcup\limits_{v\in A}B_{\epsilon\over 2}(v) \}.\end{array}$$ 
By convention the minimum equals $+\infty$ if no $\epsilon$-finite cover exists.\\
In contrast with the topological vector space, coverings are not issued from neighborhoods of zero and the neighborhood base is fixed as a function of $x$. This explains the notation $N(\epsilon,V)$ instead of $N(\epsilon,V, U)$.
\begin{Def}[Point-extended box dimension for a singleton]
Let $(X, d) $ be a  metric space and $\varphi \in \Phi$.
 The point-extended box dimension of $x \in X$, denoted below  $dim_{\varphi}\{x\}_X$, is defined by the following  
 $$dim_{\varphi}\{x\}_X=\inf\limits_{V\in\mathcal{U}(x)}\overline{\lim\limits_{\epsilon \to 0}}\frac{\phi_1(N(\epsilon,V))}{\phi_2(\frac{1}{\epsilon})}.$$
 If  $x\in \bar A\subset X$, the relative extended box dimension is given by  
 $$dim_{\varphi}\{x\}_A= \inf\limits_{V\in\mathcal{U}(x)}\overline{\lim\limits_{\epsilon \to 0}}\frac{\phi_1(N(\epsilon,V\cap A))}{\phi_2(\frac{1}{\epsilon})}.$$
\end{Def}
The dimension of a set is defined in the same spirit as in vector topological spaces
\begin{Def}[Point-extended box dimension for a set]
Let $(X, d) $ be a  metric space and $\varphi \in \Phi$.
 The  extended box dimension of $A\subset X$, denoted below  $dim_{\varphi}(A)$, is defined by $$dim_{\varphi}(A)=\sup\limits_{x\in \bar A}dim_{\varphi}\{x\}_A.$$
 
 \end{Def}
 
 Since the family $\{B_r(x), r>0\} $ of is a basis of   neighborhoods of $x$, the following proposition is immediate.
 \begin{pro}
Let $(X, d) $ be a  metric space, $x \in X$ and $\varphi \in \Phi$. We have
 $$dim_{\varphi}\{x\}_X=\inf\limits_{r>0}\limsup\limits_{\epsilon \to 0}\frac{\phi_1(N(\epsilon,B_r(x)))}{\phi_2(\frac{1}{\epsilon})}=\lim\limits_{r\to 0}\limsup\limits_{\epsilon \to 0}\frac{\phi_1(N(\epsilon,B_r(x)))}{\phi_2(\frac{1}{\epsilon})}.$$
 If  $x\in \bar A\subset X$, we have
$$dim_{\varphi}\{x\}_A=\inf\limits_{r>0}\limsup\limits_{\epsilon \to 0}\frac{\phi_1(N(\epsilon,B_r(x)\cap  A))}{\phi_2(\frac{1}{\epsilon})}=\lim\limits_{r\to 0}\limsup\limits_{\epsilon \to 0}\frac{\phi_1(N(\epsilon,B_r(x)\cap A))}{\phi_2(\frac{1}{\epsilon})}.$$
\end{pro}
Proposition \ref{increasing} remains valid in the case of metric spaces using the same arguments. More precisely
\begin{pro}\label{increasingm}Let $(X,d)$ be a Hausdorff space. Under the previous notations, let $A\subset A', B\subset X$ and $x\in \overline{A\cup B}$. Then
\begin{enumerate}
\item  $dim_{\varphi}\{x\}_{A}\le dim_{\varphi}\{x\}_{A'}$.
    \item $dim_{\varphi}\{x\}_{A\cup B}= max(dim_{\varphi}\{x\}_{A},dim_{\varphi}\{x\}_{B})$, 
    \item $dim_{\varphi}\{x\}_{A}=dim_{\varphi}\{x\}_{\bar A}$,
    \item  If moreover $\phi_1(x.y)\leq \phi_1(x)+ \phi_1(y)$, for $A_1\subset X_1$ and $A_2\subset X_2$ and  $(x_1,x_2)\in \overline{A_1\times A_2}$ we have $$\max(dim_{\varphi}\{a_1\}_{A_1},dim_{\varphi}\{a_2\}_{A_2})\leq dim_{\varphi}\{(a_1,a_2)\}_{A_1\times A_2}\leq dim_{\varphi}\{a_1\}_{A_1}+dim_{\varphi}\{a_2\}_{A_2}.$$
\end{enumerate}
In particular, 
\begin{enumerate}
\item[1'] $dim_{\varphi}(A)\le dim_{\varphi}(A')$,
    \item[2'] $dim_{\varphi}(A\cup B)= max(dim_{\varphi}(A),dim_{\varphi}(B))$,
    \item[3'] $dim_{\varphi}(A)=dim_{\varphi}(\bar A)$,
    \item[4'] If $\phi_1(x.y)\leq \phi_1(x)+ \phi_1(y)$, we have $$\max(dim_{\varphi}(A_1),dim_{\varphi}(A_2))\leq dim_{\varphi}(A_1\times A_2)\leq dim_{\varphi}(A_1)+dim_{\varphi}(A_2).$$
\end{enumerate}

\end{pro}
 We give next examples highlighting the specificity of the case of metric spaces.
 
 \begin{ex}\label{invisible}
 We start with an example providing an infinite family of \textit{"invisible"} dimensions. Let $\alpha > 1$ and denote $A_\alpha=\{\exp(-n^{\alpha}), n\ge 0\}$. Since $0$ is the only accumulation point of $A$, we have $dim\varphi(A_\alpha)= dim_\varphi(\{0\})$. Let $\epsilon>$ and $N(\epsilon,B(0))$ be as above and consider $n_0$, such that $\epsilon\sim \exp(-n_0^{\alpha})$ or equivalently $n_0\sim (\ln\frac{1}{\epsilon})^{\frac{1}{\alpha}}$. We will have $\exp(-n^{\alpha})\in \epsilon B(0)$ for every $n\ge n_0.$ Moreover 
 $$\begin{array}{lll}
    \exp(-n^{\alpha})-\exp(-(n+1)^{\alpha})  & =& \exp(-n^{\alpha})(1-\exp(n^{\alpha}-(n+1)^{\alpha})\\
      & =&\exp(-n^{\alpha})(1-\exp(n^{\alpha})(1-(1+(\frac{1}{n+1})^{\alpha})\\
      &\sim &\exp(-n^{\alpha})(1-\exp(-n^{\alpha} \frac{\alpha}{n+1}))\\
       &\sim &\exp(-n^{\alpha})(1-\exp(-\alpha n^{\alpha-1}))>\frac{1}{2}\exp(-n^{\alpha}), 
 \end{array} $$
 for $n $ large enough. It follows that $N(\epsilon, B_0)\sim (\ln\frac{1}{\epsilon})^{\frac{1}{\alpha}}$ and then 
the box dimension $dim_{\ln}(A_\alpha)= \limsup \frac{\ln((\ln\frac{1}{\epsilon})^{\frac{1}{\alpha}})}{\ln\frac{1}{\epsilon}} = 0$ for every $\alpha>0$. On the other hands, putting $\varphi=(\ln,\ln\ln)$, we get
 $dim_\varphi(A_\alpha)= \limsup \frac{\ln((\ln\frac{1}{\epsilon})^{\frac{1}{\alpha}})}{\ln\ln\frac{1}{\epsilon}} = \frac{1}{\alpha}.$ Here, we have compared the entropy of the set $A$ with entropies of all finite box dimensional sets. This computation underlines that the set $A$ possess a kind of dimension less than all finite dimensional sets. In other words, finite box dimensional sets are infinite dimensional sets, in some sens, relatively to the sets $(A)_{\alpha\in\mathbb{R}^+}$. Conversely, these sets have invisible dimensions relative to the finite box dimension. 
     \end{ex}

 \begin{ex}\label{ex2} In this example, we exhibit a case where $dim_{\varphi}\{x\}$ vary with $x$ in $[0,1]$.
Let   $\varphi=(ln, ln)$ and $\beta =(\beta_n)_n$ be a sequence of non-negative numbers. We consider 
$$ A_{ \beta}=\{\frac{1}{ln(n)}+\frac{1}{m^{\beta_n}}, n,m\geq 1\}.$$
We have\\ $$dim_{A_{ \beta}}\{\frac{1}{ln(n)}\}= \frac{1}{\beta_n+1}.$$ Indeed, let $n\ge1$ be given.  First, notice that, there exists $  \delta(n) >0$ such that  $dist(\{\frac{1}{ln(n)}\}, \{\frac{1}{ln(p)}+\frac{1}{m^{\beta_p}, }, p\ne n\}) \ge \delta(n)$. This fact follows from,  $\lim\limits_{m\to \infty}|\frac{1}{ln(p)}+\frac{1}{m^{\beta_p}} -\frac{1}{ln(n)}| = |\frac{1}{ln(p)}-\frac{1}{ln(n)}|\ge \frac{1}{ln(n)}-\frac{1}{ln(n+1)} \ne 0$ and hence $ min|\frac{1}{ln(p)}+\frac{1}{m^{\beta_p}} -\frac{1}{ln(n)}|=\delta(n)>0$. 

Now for $\epsilon < \delta(n)$, we get 
$$A_\beta\bigcap ]\frac{1}{ln(n)}-\epsilon, \frac{1}{ln(n)}+\epsilon[ = (\{\frac{1}{ln(n)}\}\bigcup\{\frac{1}{ln(n)}+\frac{1}{m^{\beta_n}}, m\geq 1\})\bigcap ]\frac{1}{ln(n)}-\epsilon, \frac{1}{ln(n)}+\epsilon[.$$
It follows then, $dim_{A_{ \beta}}\{\frac{1}{ln(n)}\}= \frac{1}{\beta_n+1}.$  
\end{ex}
\begin{ex}\label{sin}  Let $A=\{(x,sin\frac{1}{x}): x>0\}$. Then
$$dim_{\ln}\{a\}_A=\left\{\begin{array}{lll}
   1  & if & a =(x,sin\frac{1}{x}) \mbox{with } x\ne 0 \\
    {3\over 2} & if & a=(0,t)\in {\bar A}, 0 \le t\le 1.
\end{array}\right.$$
To see this, for $a =(x,sin\frac{1}{x}) \mbox{with } x\ne 0 $, choose $\epsilon>0$ such that $B(a,\epsilon) \cap A$ is a simple curve in $A$. Which gives $dim_{\ln}\{a\}_A=1$. 

Let now $ a=(0,t)\in {\bar A}\setminus A$, with $t\ne 0$ and $0<r<1$. We have  $B(a,r)\cap A=\cup_n\Gamma_n$, where $\Gamma_n=\{(x,sin\frac{1}{x}) :  \frac{1}{(2n+1)\pi}\le x \le \frac{1}{(2n-1)\pi}$ and the Hausdorff distance $dist(B(a,r)\cap  \Gamma_n; B(a,r)\cap  \Gamma_{n+1})\propto \frac{1}{n^2}$ for  $r$  small enough. For $\epsilon>0$
 we denote  $B^i_n(\epsilon)$ an $\epsilon-$cover of $\Gamma_n$ so that for $\epsilon<\frac{r}{n^2}$. We clearly have 
$B^i_n(\epsilon)\cap   B^j_n(\epsilon))=\emptyset$.

We will get $$ N(\epsilon, B(a,r)\cap A)\sim \sum\limits_{n\le \frac{1}{\sqrt{\epsilon}}}N(\epsilon, B(a,r)\cap \Gamma_n)\sim \frac{1}{\sqrt{\epsilon}}\frac{r}{\epsilon}\sim \frac{r}{\epsilon\sqrt{\epsilon}}$$
from which we get $dim_{\ln}\{a\}_A={3\over 2}$.
\begin{rem}\label{out}
 Outlining the previous proof, we can obtain easily that for  $C=\{(x,xsin\frac{1}{x}): x\ne 0\}$, we have $dim_{\ln}\{a\}_C=1$ for every $a\ne 0$ and $ dim_{\ln}\{0\}_C={3\over 2}$. This is not in the spirit of Uryshon approach "a line is what has dimension one at all its points". Indeed, $C$ is considered as a curve but according to Uryshon conception it is not the case. It is a perfect example highlighting our motivation in freeing the classification of basic geometrical object from the notion of dimension detailed in subsection \ref{4.2.2}.     
 \end{rem}
\end{ex}
 \begin{ex}\label{ex21}   
In the next example $dim_{\varphi}\{x\}$ vary in $\mathbb{R}$ with $x$. For $\alpha >0$, denote $A_{ \alpha,k}= \{(\frac{1}{n^{\alpha}},x_2,\cdots,x_{k+1})_{x_2,\cdots,x_{k+1} \in {\mathbb R}} ,  n \geq 1\}$ and $B_k= \{(e^{-n},x_2,\cdots,x_{k+1})_{x_2,\cdots,x_{k+1} \in {\mathbb R}},   n \geq 1\}.$ We have First,  take  $\varphi=(ln, ln)$. The $dim_\varphi$ function can fill all values with simple sets. \\
 $$dim_\varphi(A_{\alpha,k})=k+\frac{1}{1+\alpha} \in {\mathbb R}\setminus{\mathbb N}, \: dim_\varphi(B_k) =k.$$
 Let  now   $\varphi(x)=(|ln|ln(x)|, |ln|ln(x)|)$.   Then 
 $$ dim_\varphi(A_{\alpha,k})= dim_\varphi(B_k)= 1.$$
 \end{ex}
\begin{rem}
Let $X$ be a metric space, in the case of $\varphi=(ln,ln)$, we recover the metric order discussed in \ref{pont}. In particular the infinimum of $dim_{\ln}(X)$ over all equivalent topologies on $X$ is an integer and coincides with the topological dimension of $X$.   
\end{rem}
The previous examples confirm the fact that the dimension is a point-notion since it can vary with the point. Therefore, we should specify what dimensional homogeneity can be. We follow Urysohn approach detailed in \cite{ury1}.  
\begin{Def} A set $X$ is said to be dim-homogeneous if for every $x$ in a dense subset of $A$, we have $dim_{\varphi}\{x\}_A=dim_{\varphi}(A)$. 
\end{Def}
\begin{rem}
The notion of dimensionally homogeneous set is a local notion although its definition has a global aspect. Indeed, the curve $C$ of the remark \ref{out} is not dimensionally homogeneous; nevrethless, removing only one point the set $C^{\ast}=C\setminus \{0\}$ becomes dimensionally homogeneous. It is possible to provide more twisted examples.
\end{rem}
\subsection{Two classical cases}
\subsubsection{The finite dimensional ${\mathbb R}-$vector spaces cases}
In the case of finite dimensional spaces on the real (also complex) field, it is clear that the topology is endowed with a norm and since  all norms are equivalent, we  assume that $E$ is Euclidean. Moreover, The contribution of the measure in the calculation of the dimension challenges us on its reality. Indeed, if we consider that dimension is a purely topological notion, it should not depend on the measure on the space. This pushes us to look for an adequate measure which does not conflict with our approach. We will take the Lebesgue measure  from the distance on Borelean $\sigma$-algebra. To recover the algebraic dimension as an extended box dimension $dim_{\phi}(\{x\})$, we first establish the following:

As for all topological spaces, we have     $dim_{\phi}(\{x\})_E= dim_{\varphi}(B)=dim_{\varphi}(E)$, for every ${\mathbb R}-$vector space $E$. Thus in order to show our results, we will often, without mentioning it, calculate $dim_{\varphi}(B)$.
\begin{pro}\label{fini}  Let  $B$ be the unit ball in $\mathbb{R}^{n}$ and $N(\varepsilon)=\# N(\epsilon, B)$ be as above. Then  
$$ \Big(\frac{1}{\varepsilon}\Big)^{n}\leq N(\varepsilon)\leq \Big(1+\frac{2}{\varepsilon}\Big)^{n}.$$
\end{pro}
\begin{proof}
For  $r=N(\epsilon)$ let  $x_1,x_2...x_r$  be such that  $B_{\mathbb{R}^{n}}(0,1) \subset \cup_{i=1}^rB_{\mathbb{R}^{n}}(x_i,\epsilon)= \cup_{i=1}^r (x_i+\epsilon B_{\mathbb{R}^{n}}(0,1))$. Using Lebesgue measure, it comes that  
$$
\lambda(B_{\mathbb{R}^{n}}(0,1))\leq \sum_{i=1}^r \lambda(\epsilon B_{\mathbb{R}^{n}}(0,1))=\epsilon^nN(\epsilon)\lambda(B_{\mathbb{R}^{n}}(0,1))$$ The first inequality is obtained. For the remaining inequality, we need an  auxiliary  lemma. For a given set $A$ in a metric space, $y_1,y_2,....y_s\in X$ are said to be $\epsilon$-distinguables if for every  $d(y_i,y_j)\ge \epsilon$ for every $i\ne j$. We have the following
  
\begin{lem}
Let  $(X,d)$ be a compact metric space and $s=N_1(\varepsilon)$ the maximal number of $\epsilon$-distinguables elements $y_1,y_2,....y_s\in B$. Then 
\begin{equation}\label{kef}
N_1(\varepsilon)\leq N(\frac{\varepsilon}{2})\leq N_1(\frac{\varepsilon}{2}).
\end{equation}
\end{lem}
Now, if  $s=N_1(\varepsilon)$, we obtain  $B(y_j,\frac{\varepsilon}{2})=y_j+\frac{\varepsilon}{2}B_{\mathbb{R}^{n}}(0,1)$  are disjoint sets in  $(1+\frac{\varepsilon}{2})B_{\mathbb{R}^{n}}(0,1)$ because $y_j\in B_{\mathbb{R}^{n}}(0,1)$. Hence 
$$\bigcup_{j=1}^s\Big(y_j+\frac{\varepsilon}{2}B_{\mathbb{R}^{n}}(0,1)\Big)\subset\Big(1+\frac{\varepsilon}{2}\Big)B_{\mathbb{R}^{n}}(0,1).$$ 
and then $$s\Big(\frac{\varepsilon}{2}\Big)^{n}\lambda(\Big(B_{\mathbb{R}^{n}}(0,1)\Big))\leq \Big(1+\frac{\varepsilon}{2}\Big)^{n}\lambda(\Big(B_{\mathbb{R}^{n}}(0,1)\Big)),$$
It follows that 
 $$s\leq \Big[\frac{2}{\varepsilon}\Big(1+\frac{\varepsilon}{2}\Big)\Big]^{n}=\Big(1+\frac{2}{\varepsilon}\Big)^{n}.$$ 
and finally 
$$ N(\varepsilon)\leq N_1(\varepsilon)=s\leq \Big(1+\frac{2}{\varepsilon}\Big)^{n}.$$  
\end{proof}
\begin{rem} \begin{itemize}
    \item Proposition \ref{fini} recovers the usual dimension of $\mathbb{R}^n$. Similar reasoning will assign dimension $2n$ to $\mathbb{C}^n$ as an $\mathbb{R}$-vector space. 
    \item Since dimension is independent from the choice of the basis of the neighbourhoods, the point-extended box dimension is norm independent in the finite dimensional setting, because all norms are equivalent. In particular the use of the supermmun norm will drastically simplify the proof since, using closed $\|.\|_\infty-$ balls, we get $N(\varepsilon)-(\frac{1}{\varepsilon})^n=o((\frac{1}{\epsilon})^{n+1})$.
   \item Consider $\mathbb{C}^n$ as an  $\mathbb{R}$ vector space. It is isometrically isomorphic to  $\mathbb{R}^2n$ and hence its point-extended box dimension is $2n$ and coincides with its algebraic dimension. This no more true when $\mathbb{C}^n$ is regarded as a $\mathbb{C}$ vector space. Indeed, in this case we have to compare the $\epsilon$-entropy of $\mathbb{C}^n$ with the $\epsilon$-entropy of the field $\mathbb{C}$ (see subsetcion \ref{4.3.2}).
   
\end{itemize}
\end{rem}
We have the following result
\begin{pro}\label{infinit}Let $E$ be a $n$-dimensional ${\mathbb R}-$vector space, then
$$dim_{\varphi}(E)=dim_{\varphi}\{0\}_E=\lim\limits_{\epsilon \to 0}\frac{\phi_1(\frac{1}{\epsilon^n})}{\phi_2(\frac{1}{\epsilon})}.$$
\end{pro}
\begin{proof} From the inequality $ \Big(\frac{1}{\varepsilon}\Big)^{n}\leq N(\varepsilon)\leq \Big(1+\frac{2}{\varepsilon}\Big)^{n}\le \Big(\frac{3}{\varepsilon}\Big)^{n}$ for $\epsilon$ small enough,  we get
$$\frac{\phi_1( (\frac{1}{\varepsilon})^{n})}{\phi_2(\frac{1}{\varepsilon})}\leq \frac{\phi_1( N(\varepsilon))}{\phi_2(\frac{1}{\varepsilon})} \le \frac{\phi_1((\frac{3}{\varepsilon})^{n})}{\phi_2(\frac{1}{\varepsilon})}= \frac{\phi_1((\frac{3}{\varepsilon})^{n})}{\phi_1((\frac{1}{\varepsilon})^n)}\frac{\phi_1((\frac{1}{\varepsilon})^{n})}{\phi_2(\frac{1}{\varepsilon})}.$$ Now we conclude since $\lim\limits_{\epsilon \to 0}\frac{\phi_1((\frac{3}{\varepsilon})^{n})}{\phi_1((\frac{1}{\varepsilon})^n)}=1$.
\end{proof}
As shown in the next example, the conclusion of Proposition \ref{infinit} is not valid in general if we omit, $F$ is a vector subspace.
\begin{ex} Let $A=\{\frac{n}{n+1}, n\ge 0\}$ and $\phi_1=\phi_2 =ln$. 
$dim_{\varphi}(A)=\frac{1}{2}$ but $dim_{\varphi}\{0\}_A=0.$ 
\end{ex}
We derive the next corollary
\begin{cor}Let  $E$ be a finite dimensional  ${\mathbb R}-$vector space, then
\begin{itemize}
    \item If $\varphi=(ln,ln)$, then $dim_\varphi(E)$ coincide with the algebraic dimension.
    \item If $\phi=(ln|ln|, ln|ln|)$, then  $dim_\varphi(E)=1$.
\end{itemize}

\end{cor}
\subsubsection{ The p-adic field case}

We are concerned in this part with the $p-$adic field $\mathbb{Q}_p$, where $p$ is a fixed prime number. The $p-$adic numbers, come from an alternate way of defining the distance between two rational numbers. In fact, any metric completion of the rational Field is either the real field $\mathbb{R}$ or the $p-$adic field $\mathbb{Q}_p$ for some prime number $p$. The standard distance function, the Euclidean absolute value, gives rise to the real numbers. While the real numbers are more natural to most of us, the  $p-$adic numbers is a competly counter intuitive organisation of real numbers. 

The $p-$adic valuation allows defining the $p-$adic r field $\mathbb{Q}_p$ is defined for  $ n\in \mathbb{Z} $  by 
$v_p(n)=max\{k \in \mathbb{N} \mbox{ such that } \frac{n}{p^k} \in \mathbb{Z}  \}$. For $r=\frac{a}{b}\in\mathbb{Q}$, the valuation is defined as $v(r)=v(a)-v(b)$.
\begin{Def}[$p-$adic]
For a rational number $r=\frac{a}{b}\in\mathbb{Q}$, the $p-$adic value is defined as $$\vert r\vert_p=p^{ -v_p(r)}.$$
The $p-$adic metric is then given by 
$d_p(r,s) =\vert r-s\vert_p$. We denote 
$Q_p$ and  $\mathbb{Z}_p $ for the closure of $\mathbb{Q} $ and $\mathbb{Z} $ under the  $\vert.\vert_p$ $p-$adic metric respectively.     Clearly, we have $\mathbb{Z}_p=\{r\in \mathbb{Q} \mbox{ such that } v(r)\ge 0\}.$
\end{Def}
 Notice   that $d_p$ is even an ultrametric, since $d_p(x,y) =\vert x-y\vert_p\leq \max(\vert x\vert_p,\vert y\vert_p)$.
Also,  as the metric is defined from a discrete valuation, every  ball is a clopen set. 
More precisely, for every $r>0$;  there is  a unique integer $v$ such that  $\displaystyle p^{-v}\le  r <\displaystyle p^{-v+1}$, and then $$  \{y\mid d_{p}(x,y)<r\}=\{y\mid d_{p}(x,y)\leq p^{-v}\} \subset \{y\mid f_{p}(x,y)\le r\}=  \{y\mid d_{p}(x,y)< p^{-v+1}\}.$$
This shows that every point  $a\in \mathbb{Q}_p$  admits a basis of neighborhood consisting of clopen balls. In particular  $\mathbb{Q}_p$ is totally disconnected which also says  that its small and large inductive dimensions are zero. For this reason, we will focus in this subsection on point-extended box dimensions of  the $p$-adic field $\mathbb{Q}_p$.

Is is also immediate to see  that the unit ball $ \bar B_{1}(0) =\mathbb{Z}_p$ is a compact set.

 We start by recalling the main classical properties of the ultra-metric space $(\mathbb{Q}_p, d_p). $ 
 \begin{pro}\label{pprop} Let $x\in \mathbb{Q}_p$ and $r\in\mathbb{Q}$. We have the following
 \begin{enumerate}
 \item $d(rx,ry)=|r|_pd(x,y)$.  
 \item $B_1(0)=\bigcup\limits_{k=0}^{p-1}(k+B_{\frac{1}{p}}(0))=\bigcup\limits_{Q\in ({\mathbb Z}/p{\mathbb Z})_{n-1}[X]}(Q(p)+B_{p^{-n}}(0))$. Here $({\mathbb Z}/p{\mathbb Z})_{n-1}[X] $ is the ring of polynomial with coefficient in ${\mathbb Z}/p{\mathbb Z}$ and of degree less or equal to $n-1.$
     \item $B_{p^{-n}}(0) = p^{-n}B_1(0)$.
     \item $B_r(x)=B_r(y)$ for every $y\in B_r(x)$. 
     \item  If $r\le r'$, then $B_r(x)\cap B_{r'}(y)\ne \emptyset \Rightarrow B_r(x)\subset B_{r'}(y)$.
     \item $\sum\limits_{n\ge 0} a_n$ converges if and only if
     $\lim\limits_{n\to \infty}a_n=0$.
 \end{enumerate}
 
 \end{pro}
 Every $p-$adic number $x$ has a $p-$adic expansion  $x=\sum\limits_{v_p(x)\le n} a_n(x)p^n$, where $0\le a_n(x)\le p-1$ are given integers. It follows that $d_p(x,y)=p^{-min\{n \mbox{ such that } a_{n}(x)\ne a_{n}(y)\}}$ and in particular $d_p(x,y)=p^{-min\{v_p(x),v_p(y)\}}$  if $v_p(x)\ne v_p(y)$ or $v_p(x)= v_p(y)$ and $a_{v_p(x)}(x)\ne a_{v_p(y)}(y)$.\\
 The next corollary is a  direct consequence of Proposition \ref{pprop}
 \begin{cor} Let $m>n>0$ be integers,  $r\in [p^{-n},p^{-n+1}[$ and $\epsilon \in [p^{-m},p^{-m+1}[$. Then
 $$ N(\epsilon,B_r) =p^{m-1-n}$$
 
 \end{cor}
 Indeed, 
 $$ N(\epsilon,B_r) =  N(p^{-m+1},B_{p^{-n}})=  N(p^{-m+1},p^{-n}B_{0})=N(p^{n-m+1}, B_{0})=p^{m-1-n}$$
 Since balls around $x\in \mathbb{Q}_p $ are translation of balls around zero, we are able to show the next theorem.
\begin{thm}     For every $x\in \mathbb{Q}_p$, we have $$dim_{\varphi}(\{x\}_{\mathbb{Q}_p})= \limsup\limits_{ m \to \infty}\frac{\phi_1(p^{m})}{\phi_2(p^{m})}.
$$
\end{thm}
\begin{proof} We first remark that $dim_{\varphi}(\{x\}_{\mathbb{Q}_p})=dim_{\varphi}(\{0\}_{\mathbb{Q}_p})$ and  using the previous corollary, we get
$$dim_{\varphi}(\{x\}_{\mathbb{Q}_p})= \inf\limits_{r> 0}{\limsup\limits_{\epsilon \to 0}}\frac{\phi_1(N(\epsilon,B_r))}{\phi_2(\frac{1}{\epsilon})}= \lim\limits_{n\to \infty}{\limsup\limits_{ m \to \infty}}\frac{\phi_1(p^{m-1-n})}{\phi_2(p^{m})}={\limsup\limits_{ m \to \infty}}\frac{\phi_1(p^{m-1-n})}{\phi_1(p^{m})}\frac{\phi_1(p^{m})}{\phi_2(p^{m})}
$$
We conclude because of  ${\lim\limits_{ m \to \infty}}\frac{\phi_1(p^{m-1-n})}{\phi_1(p^{m})} = 1$.

\end{proof}
As a corolary, we obtain
\begin{cor}  If $\phi_1=\phi_2$; then for every $x\in \mathbb{Q}_p$, we have $dim_{\varphi}(\{x\}_{\mathbb{Q}_p})=1$. In particular the box dimension ($\varphi=(ln,ln)$ ) and   the functional dimension ($\varphi=(ln|ln|,ln|ln|)$ ) of the $p-$adic field are equal to one. 

\end{cor}
Recall that two norms $\vert .\vert_1$ and  $\vert .\vert_2$ on a field $K$ are said to be equivalent if and only if  $ \exists \delta>0$ such that $\vert x\vert_1=\vert x\vert_2^{\delta}$  for every $x\in K$. Clearly equivalent norms generate the same family of neighborhood and hence provide the same topology. It follows that, for every $\alpha >0$,  the topology of ${\mathbb Q}_p$ is generated  by  $|.|^\alpha_p$. Using the same arguments as before, we obtain
\begin{pro} Let $\alpha >0$ and  $({\mathbb Q}_p, |.|^\alpha_p)$ be the metric space of $p$-adic numbers, then for $\varphi \in \Phi,$ we have $$dim_{\varphi}(\{x\}_{{\mathbb Q}_p})=\limsup\limits_{t\to \infty}\frac{\phi_1(p^{\alpha m})}{\phi_2(p^m)}.$$
\end{pro} 
\begin{rem} For the Box dimension, we obtain  $dim_{\varphi}(\{x\}_{{\mathbb Q}_p})=dim_{\varphi}({{\mathbb Q}_p})= \alpha$. We have $\inf\limits_{\alpha >0}dim_{\varphi}({{\mathbb Q}_p})=0$ which coincides with its topological dimension. This is in line with the result of Pontrjagin-Schnirelmann \cite{pontr} detailed in subsection \ref{pont}.
\end{rem}

\begin{rem}  Let $E=({\mathbb Q}_p)^n$ be an $n$ algebraic dimensional ${\mathbb Q}_p$-vector space equipped with the supermum norm $\|(r_1,\cdots,r_n)\|=\max\limits_{i=1}^n|r_i|_p^\alpha$, for some $\alpha>0.$ We have  $B_r(r_1,\cdots,r_n)=\prod\limits_{i=1}^nB_r(r_i)$, where $B_r(a)$   as usual is  the ball centered at $a$ with radius $r$. It follows that 
 $N(\epsilon, B_r(0))=(N(\epsilon, B_r(0))^n $ and then 
$$dim_{\varphi}(\{x\}_{{(\mathbb Q}_p)^n})=\limsup\limits_{t\to \infty}\frac{\phi_1(p^{\alpha nm})}{\phi_2(p^{m})}.$$
In the case of  the usual box dimension, we retreive the expression  $dim_{\varphi}(\{x\}_{({\mathbb Q}_p)^n})=ndim_{\varphi}({{\mathbb Q}_p})=n\alpha. $
\end{rem}
 
\subsection{Point-extended box dimension in infinite dimensional spaces}
\subsubsection{General framework}
In this section we focus on the case of infinite dimensional Banach vector spaces. In this case, the unit ball $B$ is not compact due to the Riesz theorem. Thus, the number $N(\epsilon,B,B)$ is always infinite in the formula $dim_\varphi\{0\}=\overline{\lim\limits_{\epsilon \to 0}}\frac{\phi_1(N(\epsilon,B,B))}{\phi_2(\frac{1}{\epsilon})}$ of proposition \ref{boundedpro} for every $\varphi=(\phi_1,\phi_2)$. Hence, going back to the original expression $dim_{\varphi}\{0\}_E=\sup\limits_{U\in\mathcal{U}(0)}\inf\limits_{V\in\mathcal{U}(x)}\overline{\lim\limits_{\epsilon \to 0}}\frac{\phi_1(N(\epsilon,V,U))}{\phi_2(\frac{1}{\epsilon})} $ in Definition \ref{originale},  we modify the family of covering sets $U$ of the unit ball $B$ in order to recover compactness. To this aim, we choose $\mathcal{U}(x)$ as the basis of neighborhoods in the weak topology. More precisely, we estimate the quantity $dim_\varphi\{0\}=\overline{\lim\limits_{\epsilon \to 0}}\frac{\phi_1(N(\epsilon,B,B^w))}{\phi_2(\frac{1}{\epsilon})}$, where $B^w$ denotes the weak unit ball. 

Even by using this setting of strong/weak topology, in the case of an infinite dimensional Banach space $X$, we can not expect that the box dimension of $X$ to be finite. In other words, even in this framework $\overline{\lim\limits_{\epsilon \to 0}}\frac{\ln(N(\epsilon,B,B^w))}{\ln(\frac{1}{\epsilon})}=+\infty$ (see Proposition \ref{boxinfinie}). Since for any finite dimensional subspace, the weak topology  and  the strong topology coincide, the same occurs for the corresponding dimensions. We derive that the box dimension of any infinite dimensional space is infinite.  Hence, we cannot expect any reduction to finite dimension, which seems to be adequate. This implies that we have to choose a $\varphi$ different from $(\ln,\ln)$ in the definition. More precisely, in this section we develop about $\varphi=(\ln\ln,\ln\ln)$, which represent the functional box dimension. In this setting, all finite dimensional sets have the same functional box dimension equal to $1$. They are all put in the same class. This class contains also some infinite dimensional objects as highlighted below. The spirit of the functional box dimension is to compare in one shot the entropy of a set to all entropies of sets of all finite dimensions, rather than just comparing with the entropy of sets of dimension 1, as is the case for the box dimension.

We devote this section to the functional dimension of the Banach space, $l^p= \{ {\bf x}= (x(k))_{k\ge 1}\in {\mathbb C}^\infty : \|{\bf x}\|=: (\sum_{k\ge 1}|x(k)|^p)^{\frac{1}{p}}<\infty \}$. We endow $l^p$ in the sequel with  the canonical basis in $l^p$ given by $({\bf e}_n)_{n\geq 1}$, where 
${\bf e}_n = (\delta_{k,n})_{k\ge 1}$ and $\delta_{k,n} $ stands for the classical Kronecker symbol.

As usual 
  $B({\bf x},r)=\{{\bf y} \in l^p : \|{\bf y}-{\bf x}\|_p\le r\}$ is the ball centered at ${\bf x}$ with radius $r$  and  $B=B(0,1)$ denotes the unit ball. Denote also $l^p_n= span\{{\bf e}_1,\cdots,{\bf e}_n\}$ for the $n$ dimensional space generated by ${\bf e}_1,\cdots ,{\bf e}_n$ endowed with the usual inherited norm, $P_n:l^p\to l^p_n$ the canonical  projection on $l^p_n$ and $B_n({\bf 0},1)$ the unit   ball of $l^p_n$. We clearly have $P_n(B({\bf x},r))=B_n(P_n({\bf x}),r)$ and if moreover $ {\bf x}\in l^p_n, $  we get $ B_n({\bf x},r) =B({\bf x},r)\cap l^p_n$.
  
   Recall that, on a Banach space $X$,  the weak topology on $X$  is given by:  $x_n$ converges weakly to $x$ ( generally denoted,  $x_n \rightharpoondown  x$) if and only if for every $u$ in the dual space $  X^*$, we have $<x_n,u>\to <x,u>$.  Here $<,>$ is the pairing product between  $X$ and  $X^*$. Since $l^p$ is a reflexive space, it follows from Kakutani theorem that the unit ball is weakly compact. On the other hand, $l^p$ is separable and hence the unit ball endowed with the weak topology is metrizable. For every $1 \le p<\infty$, the next Lemma exhibit a metric defining the weak topology on unit ball of $l^p$.

\begin{lem} Let $p>1$, ${\bf x}\in l^p $ and ${\bf x}_n$ be a bounded sequence in $l^p$. The following are equivalent
\begin{enumerate}
    \item ${\bf x}_n \rightharpoondown {\bf x}$;
    \item $ d_p({\bf x}_n,{\bf x}) =: (\sum\limits_{k\geq 1} \frac{|{\bf x}_n(k)-{\bf x}(k)|^p}{4^k})^\frac1p\to 0.$
\end{enumerate}
In particular,  $d_p$ defined on $l^p$, endows the weak topology on bounded sets. 
\end{lem}
Proof.  $1) \Rightarrow 2)$ Assume  ${\bf x}_n\in l^p$ is weakly converging to  $ {\bf x}$ and write $\|{\bf x}_n\|\le M$. For every $k\ge 1$ we have  ${\bf e}_k\in (l^p)^*$, and then
$${\bf x}_n(k)= <{\bf x}_n,{\bf e}_k> \to <{\bf x},{{\bf e}_k}>={\bf x}(k).$$ To show that $d_p({\bf x}_n,{\bf x})\to 0$, consider $\epsilon >0$  and  $N$ large enough to get  
$ {4^{1-N}(2M)^p\over 3}\le \frac{\epsilon}{2}$. We will have, 
 $$
 d^p_p({\bf x}_n,{\bf x})=\sum_{k\ge 1} \frac{|x_n(k)-x(k)|^p}{4^n} = \sum^{N-1}_{k= 1} \frac{|x_n(k)-x(k)|^p}{4^k}+\sum_{k\ge N} \frac{|x_n(k)-x(k)|^p}{4^k}.
 $$
 on  one  hand,
 $$ \sum_{k\ge N} \frac{|x_n(k)-x(k)|^p}{4^k}\le   {4^{1-N}\over 3}{(2M)^p} \le \frac{\epsilon}{2}.$$
On the other hand,  since $x_n(k)\to x(k)$ for every k, we can choose $n$ big enough such that  $\sum^{N-1}_{k= 1} \frac{|x_n(k)-x(k)|^p}{4^k} \le \frac{\epsilon}{2}$, which concludes the proof of the first part.\\ $2) \Rightarrow 1)$. It is clear that under $2)$, we have $x_n(k)\to x(k)$ for every $k\ge 1$ and since ${\bf x}_n$ is bounded, it is weakly converging to  $ {\bf x}$. Indeed, for $u=(u(k))_{k\geq 1}\in (l^p)^{*}=l^q$ where $q$ is the conjugate number of $p$, we have 
$$\begin{array}{llll}
  <\bf{x}_n-x,u>&=&\sum\limits_{k\geq 1}(\bf{x}_n(k)-\bf{x}(k))u(k)\\
  &=&\sum\limits^{N}_{k= 1}(\bf{x}_n(k)-\bf{x}(k))u(k)+\sum\limits_{k\geq N+1}(\bf{x}_n(k)-\bf{x}(k))u(k)\\ &\leq & \sum\limits^{N}_{k= 1}(\bf{x}_n(k)-\bf{x}(k))u(k)+(\sum\limits_{k\geq N+1}|(\bf{x}_n(k)-\bf{x}(k))|^p)^{\frac{1}{p}}.(\sum\limits_{k\geq N+1}|{u}(k)|^q )^{\frac{1}{q}}\\
  &\leq & \sum\limits^{N}_{k= 1}(\bf{x}_n(k)-\bf{x}(k))u(k)+2M.(\sum\limits_{k\geq N+1}|{u}(k)|^q )^{\frac{1}{q}}\to 0.
\end{array}$$
To avoid any confusion, for  ${\bf x}\in l^p $ and ${ P_n({\bf x})}={\bf x'}\in l_n^p $ ,  let us denote  $B^w({\bf x},r)=\{{\bf y}\in l^p \mbox{ such that } d_p({\bf y},{\bf x})\leq r\}$ and 
$ B^w_n({\bf x'},r)=\{{\bf y}\in l^p_n \mbox{ such that } d_p({\bf y},{\bf x'})\leq r\}$ for  the "metric" balls in $l^p $ and in $  l^p_n$ respectively. 
As for norm balls, we clearly have $P_n(B^w({\bf x},r))= B^w_n(P_n({\bf x}),r)$ and if moreover $ {\bf x}\in l^p_n, $  we get $ B^w_n({\bf x'},r) =B^w({\bf x},r)\cap l^p_n$.
\subsubsection{Functional box dimension of $l^p$ spaces for $1< p <+\infty$}

\begin{pro}\label{boxinfinie}
$(B,d_p)$  has infinite box dimension.
\end{pro}
\begin{proof}
For every $n$ denote $\|{\bf x}\|'=(\sum^n_{1} \frac{|x(k)|^p}{4^k})^{\frac{1}{p}}$ defined on $l^p_n$. We clearly have $\|.\| \equiv \|.\|'$, from which it follows that $dim_{\ln}(B_n) =n$. Now using $dim_{\ln} (B_n)\leq dim_{\ln} (B)$ for every $n$, we deduce the claim.
\end{proof}
Hence, even using weak balls, the box dimension of $B$ remains infinite. There is a need of another $\Phi$, the adequate one is $\Phi= (\ln\ln,\ln\ln)$, corresponding to the notion of functional dimension.
 
\begin{Def}
Let $E=l^p$ and $N_{\epsilon}$  be  the minimum number of metric balls $B^w({\bf x},\varepsilon)$
recovering the unit ball $B=B(0,1)$. The functional box dimension of $E$ is defined as
$$dim_f(E):= dim_f(B)= \limsup\limits_{\epsilon \to 0} \frac{\ln\ln N_{\epsilon}}{\ln \ln\frac{1}{\epsilon}}.$$
\end{Def}
Let us state the main theorme of this subsection.
\begin{thm} For every $p\in [1,+\infty[$, the  limit $\lim\limits_{\epsilon \to 0} \frac{\ln \ln N_{\epsilon}}{\ln\ln \frac{1}{\epsilon}}$ exists and we have
$$ dim_f(B)=\lim\limits_{\epsilon \to 0} \frac{\ln \ln N_{\epsilon}}{\ln \ln \frac{1}{\epsilon}} =2.$$
\end{thm}
\begin{proof}
To compute the functional dimension of $B$, notice first  that for every $N$, we have 

\begin{equation}\label{1}
d_p(P_N{\bf x}, P_N{\bf y})\leq d_p({\bf x},{\bf y}).
\end{equation}
Which gives that for every cover $B \subset \cup_{j=1}^{N_{\epsilon}} B^w(y_j, \varepsilon), $ we have \begin{equation}\label{2} P_N(B)= B_N \subset \cup_{j=1}^{N_{\epsilon}} B^w_N(P_N{\bf y}_j, \varepsilon).
\end{equation}
Using Lebesgue measure $\lambda$, it comes 
\begin{equation}\label{3}
\lambda_N(B_N)\leq  \sum_{j=1}^{N_{\epsilon})} \lambda_N(B^w_N(P_N{\bf y}_j, \varepsilon)).
\end{equation}

Since moreover, by translation invariance, and 
by the change of variable   $y_j=\frac{x_j}{\epsilon4^j}$, we get   $$\lambda_N(B^w_N(P_N{\bf y}, \varepsilon))=\varepsilon^N \lambda_N(B^w_N)=\varepsilon^N4^{\frac{N(N+1)}{2}}\lambda_N(B_N).$$

Hence 

\begin{equation}\label{40}
1\leq   N_{\epsilon} \varepsilon^N 4^{\frac{N(N+1)}{2}}.
\end{equation}
We minimise $\varepsilon^N 4^{\frac{N(N+1)}{2}}$ the right hand  in (\ref{40}) by differentiating to obtain
$    2N+1 = \frac{-\ln \varepsilon}{\ln 2}$, and then $N= -\frac{\ln \varepsilon+ \ln 2}{2\ln 2}$. It follows that 
\begin{equation}\label{41}
\begin{array}{lll}
\varepsilon^N 4^{\frac{N(N+1)}{2}} &=& \exp(N\ln \varepsilon +(N^2+N)\ln 2)\\
&= &\exp (-\frac{(\ln \varepsilon+ \ln 2)}{2\ln 2}\ln \varepsilon +   (-\frac{\ln \varepsilon+ \ln 2}{2\ln 2})^2+ -\frac{\ln \varepsilon + \ln 2}{2\ln 2})\ln 2)\\&=&\exp [ \frac{-1}{4\ln 2}(\ln \varepsilon )^2+(\ln 2-1 )\ln \varepsilon -  \frac{\ln 2}{4}]

\end{array}
\end{equation}
Replacing in  (\ref{40}); it follows that

\begin{equation}\label{7}
 1\le N_{\epsilon}\exp[ \frac{-1}{4\ln 2}(\ln \varepsilon)^2+(\ln 2-1 )\ln \varepsilon -  \frac{\ln 2}{4}]
\end{equation}

Applying  the logarithm function twice, we derive that: 

\begin{equation}\label{8}
2 \leq \liminf \frac{\ln \ln N_{\epsilon}}{\ln (-\ln \varepsilon)}.
\end{equation}

To establish the reverse inequality in (\ref{8}), let let us denote  $\nu_N( \varepsilon)$ for the minimum number of metric balls in $l^p_N$ of radius $\epsilon$,   recovering the unit  ball in $l^p_N$. Notice first that, 

\begin{equation}\label{9}
d({\bf x},{\bf y})^p=d(P_N{\bf x},P_N{\bf y})^p+d({\bf x}-P_N{\bf x},{\bf y}-P_N{\bf y})^p.
\end{equation}

\noindent  For every  ${\bf x}$ in  $B$, we have 

\begin{equation}\label{10}
d({\bf x}-P_N{\bf x},0)^p\leq \sum_{n=N+1}^{\infty} 4^{-n}= 4^{-N-1}\frac 4 3=\frac{1}{3.4^N}.
\end{equation}

Now choosing  $N$ such that  $\frac{1}{3.4^N}=(\frac{\varepsilon}{2})^p$, we get $d(Q_N{\bf x},0)\le \frac{\varepsilon}{2}$.
It follows that any $\epsilon-$covering of $B$ derives from an $\frac{\varepsilon}{2}-$covering of  $B_N$. Then

\begin{equation}\label{11}
N_{\epsilon}\leq \nu_N(\frac \varepsilon 2).
\end{equation}
Where $\nu_N(\epsilon)$ denotes the minimum number of $\epsilon$-balls needed to cover $B_N$.

Moreover, from $d_p({\bf x},{\bf y})\leq ||{\bf x}-{\bf y}||_p$  we derive that $B({\bf x},r)\subset B^w({\bf x},r)$ for every ${\bf x}$ and hence, if $  \{B({\bf x}_i,\epsilon), i=1,2,\cdots, k\} $ covers 
$B_N$ so do $  \{B^w({\bf x}_i,\epsilon), i=1,2,\cdots, k\} $.
In particular  $\nu_N(\frac \varepsilon 2)\leq (\frac \varepsilon 2)^{-N}$
and hence 
\begin{equation}\label{12}
\ln N_{\epsilon}\leq  -N \ln \varepsilon+ N \ln 2.
\end{equation}
From $\frac{1}{3.4^N}=(\frac{\varepsilon}{2})^p$, we derive $N=-({p\over \ln 4}(\ln \varepsilon-\ln 2) +\frac{\ln 3}{\ln 4})$ and replacing in (\ref{12}), we obtain
\begin{equation}\label{13}
\ln N_\epsilon\leq  \frac{p(\ln \varepsilon)^2}{\ln 4}-\frac{2p\ln 2-\ln 3}{\ln 4}\ln \varepsilon -\frac{\ln 3-p\ln 2}{\ln 4}.
\end{equation}
Finally, 
\begin{equation}\label{4}
\limsup \frac{\ln \ln N_\epsilon}{\ln (-\ln \varepsilon)}\leq 2.
\end{equation}
Finally, the limit exists and
\begin{equation}
\limsup \frac{\ln \ln N_\epsilon}{\ln (-\ln \varepsilon)}= 2.
\end{equation}

\end{proof}
\begin{rem}\begin{enumerate}
    \item It is clear in the previous computation, the metric  $  d_p({\bf x}_n,{\bf x}) = (\sum\limits_{k\geq 1} \frac{|{\bf x}_n(k)-{\bf x}(k)|^p}{4^k})^\frac1p\to 0$ that we used to describe weak topology of $l^p$ can be replaced by 
$ d_p({\bf x}_n,{\bf x}) = (\sum\limits_{k\geq 1} \frac{|{\bf x}_n(k)-{\bf x}(k)|^p}{\lambda^k})^\frac1p\to 0$ with any $\lambda>1$. It will be interesting to find the minimal conditions on a sequence $(\lambda_k)_{k\ge 1}$ so that $ d({\bf x}_n,{\bf x}) =: (\sum\limits_{k\geq 1} \frac{|{\bf x}_n(k)-{\bf x}(k)|^p}{\lambda_k})^\frac1p\to 0.$ For example 
$ \sum\limits_{k\geq 1} \frac{1}{\lambda^q}$ converges, where $q$ is the conjugate number of $p$.
\item Since every separable Hilbert space $\mathcal{H}$ is isometrically isomorphic to $l^2$ it follows that $dim_f(\mathcal{H})=2$.
\end{enumerate}

\end{rem}

\subsubsection{Functional box dimension of  Hilbert cubes.} 
In the previous section, we have shown that the functional box dimension of all $l^p$ with $1<p<\infty$ is $2$. The question whether if this result remains true for $p=1$ or $p=+\infty$ has not been tackled. Through this subsection we investigate the $l^{\infty}$ case.

Let $l^\infty = l^\infty(\mathbb{R})$ be the Banach space of bounded sequences endowed with the supermum norm and $I^\infty =\prod\limits_1^\infty[0, 1]$, its unit ball. Since $l^\infty$ has infinite dimension, its unit ball is not compact, and hence its point-extended box dimension is infinite. 

In the framework of $l^p$ we made the choice of weakening the topology of covering in order to unsure compactness of the unit ball in the strong topology. In the current setting of $l^{\infty}$, since the weak topology is not metrisable, for instance, we prefer to proceed in the reverse way. Hence, we keep the strong topology as a covering topology and we change the unit ball by a family of compact sets. Now, for ${\bf x}\in l^{\infty}$ we have ${\bf x}=\lim\limits_{n \to +\infty} P_n({\bf x})$ if and only if $\lim\limits_{k \to +\infty} x_k=0$. Using this observation we consider an adequate compact modification of the unit ball $I^\infty$. In this sense, let $\lambda=(\lambda_n)_n\in l^\infty$ be a decreasing sequence and denote $I_\lambda=\prod\limits_1^\infty[0, \lambda_n]$. We have the following Lemma.

\begin{lem} $I_\lambda$ is compact if and only if $\lim\limits_{n\to \infty}\lambda_n=0$.
\end{lem}
\begin{proof}
By using Bolzano-Weierstrass theorem, it suffices to extract a converging sequence from any sequence in $I_{\lambda}$. By a diagonal argument, let $(x_n(k))\in I_\lambda$, $ x_{\xi_1(n)}(1)$ a converging sub sequence of $x_n(1)$, $ x_{\xi_2(n)}(2)$ converging sub sequence of $ x_{\xi_1(n)}(2)$ and by induction
$ x_{\xi_k(n)}(k)$ converging sub sequence of $ x_{\xi_{k-1}(n)}(k)$. Finally the sequence $ x_{\xi_{n}(n)}(n)$ is a converging sub sequence of $x_n(k)$. To end the proof, It suffices to see that for $\epsilon$ and $n_0$  such that   $\lambda_{n_0}\le \epsilon$, we have 
$\sup\limits_{n\ge n_0}|x_{\xi_{n}(n)}-x_{\xi_{m}(m)}|\le 2\epsilon$ and  $x_{\xi_k(n)}(k)$  converges for every $k\leq n_0$. For the reverse implication since $\lambda_n$ is decreasing we argue by contradiction assuming that $\lim\limits_{n \to +\infty}\lambda_n=\eta>0$. It follows that $\eta I^{\infty}\subset I_{\lambda}$ and hence $I_{\lambda}$ because its interior is nonempty.     
\end{proof}

To compute $N(\epsilon,  I_\lambda, I^\infty)$ from Definition \ref{salina}, let $n_\epsilon$ be the smallest number such that   $\lambda_{n_\epsilon}\le \epsilon$. Clearly, every   $\epsilon-$cover of $\prod\limits_1^{n_\epsilon}[0, 1]$ will provide an $\epsilon-$cover of  $I_\lambda$. Thus,

\begin{pro}\label{hilbertcubepro}
\begin{equation}\label{hilbertcube}
N(\epsilon, I_\lambda, I^\infty)=\prod\limits_1^{n_\epsilon}\frac{\lambda_k}{\epsilon}=\frac{1}{\epsilon^{n_\epsilon}}\prod\limits_1^{n_\epsilon}\lambda_k.
\end{equation}
\end{pro}
Proposition \ref{hilbertcubepro} allows to provide a wide class of compact Hilbert cubes with different functional dimensions. 
Indeed it follows from the previous proposition that
$$\ln(N(\epsilon,  I_\lambda, I^\infty))= \ln(\frac{1}{\epsilon^{n_\epsilon}}\prod\limits_1^{n_\epsilon}\lambda_k)= n_\epsilon\ln(\frac{1}{\epsilon})+\ln(\prod\limits_1^{n_\epsilon}\lambda_k).$$
Now for different choices of $\lambda$, we will obtain different associated dimensions. We discuss next several cases of Hilbert cubes.
We only consider the box dimension,  $\varphi=(\ln,\ln)$, 
the functional dimension, $\varphi=(\ln\ln,\ln\ln)$  and the mixed one, $\varphi=(\ln\ln,\ln)$.
\begin{enumerate}
    \item {\bf The polynomial case $\lambda_n=\frac{1}{n^p}$ for some $p>0$.}  We will get $n_\epsilon \sim (\frac{1}{\epsilon})^{\frac{1}{p}}$ which gives by replacing  in (\ref{hilbertcube}),  $N(\epsilon, I_\lambda, I^\infty)=\prod\limits_1^{n_\epsilon}\frac{\lambda_k}{\epsilon}=\frac{1}{\epsilon^{n_\epsilon}n!^p} $ and then 
$$ \ln(N(\epsilon,  I_\lambda, I^\infty))= (\frac{1}{\epsilon})^{\frac{1}{p}}\ln(\frac{1}{\epsilon})-p\ln(n_\epsilon!)$$
By using Stirling formula, ($n!\sim \sqrt{2\pi n}e^{-n}n^n$), we get $\ln(n_\epsilon!)= \frac{1}{2}(\ln 2\pi + p\ln(\frac{1}{\epsilon}))- (\frac{1}{\epsilon})^{\frac{1}{p}}+(\frac{1}{\epsilon})^{\frac{1}{p}}\ln(\frac{1}{\epsilon}).$ Thus
$$ \ln(N(\epsilon,  I_\lambda, I^\infty))= -\frac{1}{2}(\ln 2\pi + p\ln(\frac{1}{\epsilon}))+ (\frac{1}{\epsilon})^{\frac{1}{p}}
\sim (\frac{1}{\epsilon})^{\frac{1}{p}}.$$
We deduce that 
\begin{itemize}
    \item $\limsup\limits_{\epsilon\to 0}\frac{\ln(N(\epsilon,  I_\lambda, I^\infty))}{\ln(\frac{1}{\epsilon})}=+\infty$, infinite box dimension,
    \item $\limsup\limits_{\epsilon\to 0}\frac{\ln\ln(N(\epsilon,  I_\lambda, I^\infty))}{\ln\ln(\frac{1}{\epsilon})}=+\infty$, infinite functional box dimension,
    \item $\limsup\limits_{\epsilon\to 0}\frac{\ln\ln(N(\epsilon,  I_\lambda, I^\infty))}{\ln(\frac{1}{\epsilon})}=p$.
\end{itemize}

\item {\bf The geometric  case $\lambda_n=a^n$ for some $0<a<1$}. We will get $n_\epsilon \sim \frac{\ln\epsilon}{\ln a}$ and then $$\ln(\prod\limits_1^{n_\epsilon}\lambda_k)= \ln(\prod\limits_1^{n_\epsilon}a^n)\sim\frac{\frac{\ln\epsilon}{\ln a}(\frac{\ln\epsilon}{\ln a} +1)}{2}\ln a= \frac{(\ln \epsilon)^2 +\ln a\ln \epsilon}{2\ln a}.$$ We derive that
$$ \ln(N(\epsilon,  I_\lambda, I^\infty))\sim \frac{(\ln \epsilon)^2}{2\ln a}.$$
We deduce as before for the geometric case  
\begin{itemize}
    \item $\limsup\limits_{\epsilon\to 0}\frac{\ln(N(\epsilon,  I_\lambda, I^\infty))}{\ln(\frac{1}{\epsilon})}=+\infty$, infinite box dimension,
    \item $\limsup\limits_{\epsilon\to 0}\frac{\ln\ln(N(\epsilon,  I_\lambda, I^\infty))}{\ln\ln(\frac{1}{\epsilon})}=2$, finite functional box dimension,
    \item $\limsup\limits_{\epsilon\to 0}\frac{\ln\ln(N(\epsilon,  I_\lambda, I^\infty))}{\ln(\frac{1}{\epsilon})}=0$.
\end{itemize}
\item {\bf The exponential  case $\lambda_n=e^{-e^n}$.} We will get $n_\epsilon \sim \ln\ln\frac{1}{ \epsilon}$ and then $$\ln(\prod\limits_1^{n_\epsilon}\lambda_k)= \ln(\prod\limits_1^{n_\epsilon}e^{-e^k})= -{\sum\limits_{k=1}^ne^k} = -\frac{e^n-1}{e-1}e\sim -\frac{e}{e-1}e^n\sim -\frac{e}{e-1}\ln\frac{1}{\epsilon}.$$ We derive that
$$ \ln(N(\epsilon,  I_\lambda, I^\infty))\sim \frac{e}{e-1}\ln\frac{1}{\epsilon}.$$
We obtain in a similar manner as in the geometric case  
\begin{itemize}
    \item $\limsup\limits_{\epsilon\to 0}\frac{\ln(N(\epsilon,  I_\lambda, I^\infty))}{\ln(\frac{1}{\epsilon})}=+\infty$,
    \item $\limsup\limits_{\epsilon\to 0}\frac{\ln\ln(N(\epsilon,  I_\lambda, I^\infty))}{\ln\ln(\frac{1}{\epsilon})}=1$, 
    \item $\limsup\limits_{\epsilon\to 0}\frac{\ln\ln(N(\epsilon,  I_\lambda, I^\infty))}{\ln(\frac{1}{\epsilon})}=0$.
\end{itemize}
\end{enumerate}
\begin{rem} \begin{enumerate}
    \item We notice that any compact Hilbert cube $I_\lambda$ has empty interior in $l^\infty$, and $P_n(I_\lambda)$ has non empty interior in $l_n^\infty$. This says in particular that $dim_{\ln }(I_\lambda)=\infty$ and $dim_{\ln \ln}(I_\lambda)\ge 1$, for every $\lambda$. The exponential case in the previous examples are optimal in this sense.
    \item In this enquiry, we did not calculate the dimension of $l^{\infty}$, we only considered some compact objects closer and closer to the unit ball of $l^{\infty}$. After this investigation, it seems in this case that the first strategy developed in the previous section is more adequate. Thus, it would make sense to cover the unit ball of $l^{\infty}$ with the weak topology even in the absence of the metric.
\end{enumerate}

\end{rem}

\section{Conclusions and research prospects}\label{persp}
In \cite{maar1} we have introduced the concept of \textit{Experimental Philosophy of Science} in order to reach new methods concerning the \textit{Scientific Prediction}. We have tried to position the philosophy of science as a stakeholder in the progress of science itself, and even argued in favor of its ability to contribute to new discoveries. This approach proved to be enriching for the philosophy of science as well, since it allowed us to introduce the view of \textit{Extensive Structural Realism} \cite{maar1}. In the same spirit, we meant to place this work in a broader framework that aims to make the history and philosophy of mathematics an integrated and driving part of the study of science. In other words, we would like to make of the epistemology of mathematics an active actor in the development of mathematics, and not to confine to a posterior analysis. 

Currently, our research program is guided by the notion of dimension. We think that it is a fundamental and central notion for mathematics and philosophy, in particular, and for science, in general. Through this first article, we have tried to better understand this notion by starting to trace its history at the philosophical as well as the mathematical level. It turns out that this intuitive notion, as we have seen, is not only indispensable, but also unavoidable for mathematics; though we concluded, it proved difficult to formalize. This is why, this notion has remained for a long time subjugated to the development of other mathematical theories. 

It is well known that the question of space and time is central for most philosophers. Now, it happens that the notion of dimension is intimately linked to this question. Descartes with his coordinates gave to the notion of dimension a new breath by allowing it formally to take on values higher than the three dimensions of our sensible space. For mathematicians, taking this step was very difficult since it took more than two centuries. More particularly, it was necessary to accept the gap that could exist between reality \textit{per se} and our experience of reality as Kant pointed out. It was also necessary to accept that mathematics could be a field of pure abstraction governed by pure logic, as Grassmann persisted.  

Bolzano was an \textit{avant-gardist} who came to investigate the question of dimension in connection with his obsession to define concepts correctly. He was able to instruct a modern definition, which will be independently rediscovered a century later. In his quest, Bolzano laid the cornerstone of point-set topology. It is only after the problem raised by Cantor that mathematicians have considered the question of dimension in the twentieth century. We can thus reasonably argue that the birth and the development of point-set topology are a direct consequence of the investigation of several mathematicians into the notion of dimension.  

Analyzing the history of the different mathematical definitions of the concept of dimension introduced at the beginning of the twentieth century, it turns out that this question was addressed by the most imminent mathematicians. This historical review allowed us to have the necessary hindsight to both better avoid some persistent confusions (as the confusion between dimension/measure) and to be able to extend the notion of dimension properly to other frameworks (as invisible and infinite dimension cases). Thus, the relationship that we have identified between the notion of dimension and the quantity of information has led us to link the notions of dimension and entropy. 

The need for a point-conception of the dimension notion presented itself to us long before the writing of this article; however, the need to create the general, adequate and elaborate framework of the point-dimension theory only became evident during the writing. Indeed, we feel strongly that the lack of an appropriate conceptualization of a mathematical space designed as a result of this vision has surely been a hindrance to the development of point-dimension notions. That is why we deemed it necessary to detach ourselves, as much as possible, from the idealization of concepts such as the point, in order to carry out this undertaking.

More precisely, introspectively, in this article we have essentially proceeded according to three grids of reflection:

\begin{enumerate}
    \item The first grid consists of questioning basic concepts such as the point, the line or the surface. It is, for the moment, a careful and measured attempt to deconstruct these concepts in order to achieve a better reconstruction.
    \item The second one is a reconstruction based on the analysis of these concepts as they present themselves to our consciousness. It is a matter of starting by combining intuition with lived experience, leaving abstraction aside. In this sense, we have employed a \textit{phenomenological} approach.  More concretely, it is indisputable that the point, the curve or the surface are indispensable mathematical objects, at least for geometry; even though they are too idealized. This idealization has been underlined in many ways by different thinkers through the ages. For example, in Aristotle, the point, the curve or the surface are concepts that exist in potency and not in act. In the same way, we started from the fact that there are only bodies, consenting that any other types are only idealizations. One can reasonably argue that this is how the different real objects present themselves to our collective consciousness since this perception is a way of seeing that is intuitively accessible to any ordinary person.
    \item The third grid relies on the analysis of the language for essentially two reasons. On the one hand, a particular attention ought to be paid to language accuracy to avoid misleading shortcuts leading to confusing meaning. On the other hand, we need to constantly on the lookout for inherent ideas that could be generated by the language use itself. In fact, there are certain ideas that appeared to us essentially while writing this article through certain connections born within language formulations. Thus, this way of proceeding could be called the \textit{phenomenological approach to language}.
\end{enumerate}

In this regard, this work presents interesting prospects at the philosophical level, which we hope to develop more deeply in the future. Mathematically speaking, by succeeding in constructing the notion of dimension starting from the point, we have augmented the points and consequently the sets themselves. This enhancement, through the attribution of dimensions to the points, revealed a new and different way of apprehending the sets; that is what we have named the \textit{point-dimension theory}. This enrichment has helped open up new prospects, such as for the criteria for classifying sets, since one is able to compare two sets point by point. Moreover, this notion permits not only to finely analyze sets, but also to create more sophisticated sets where the dimension may depend on the position, the direction or even the time.  

The notion of dimension, as we have tried to define in this article, underlines the relative character of this notion in several respects. Indeed, in addition to the relativity due to its point-conception, there is also the fact that our definition, as we have explained, is based on the comparison between the entropies. This relativity gives rise to many possibilities and ways of calculating the dimension of a set. In finite dimension, this problem arises less since we have the possibility to gauge in reference to classical sets, while in infinite dimension it is more complicated. In this article, we have presented one way of quantifying the dimension in infinite dimension. We remain conscious that there are other ways that we will expose in future works.      

Furthermore, the point-dimension theory that we have developed in this article allowed us to extend the notion of dimension to several frameworks. Our premise is that we already perceive the need of an axiomatization in order to convey certain ideas that are otherwise difficult to reach. For the moment, we try as much as possible to avoid this axiomatization by using the framework of point-set topology. Hence, for the scope of our research, the most important planned continuation of this work at the mathematical level is rather to focus on the possible connections between this theory and other theories. While continuing to investigate other existing ways of perceiving the notion of dimension, we have a clear idea of the different mathematical developments that we will undertake in future projects.         

Our findings suggest that placing the notion of dimension in a fundamental particle (the Dimensionad) and establishing the link between the dimension and the entropy will reasonably have profound consequences in physics. Indeed, since the notion of dimension is a notion that lies at the crossroads of physical theories, it is therefore natural that any research concerning the notion of dimension should automatically have an impact on physical science. Concretely, all existing physical formalisms contain a parameter associated with the notion of dimension; this parameter is now quantified through a particle. In this way, we share a list of possible connections that we hope to investigate in depth in the future:

\begin{enumerate}
    \item The Hilbert space framework seems to have imposed itself as an exclusive mathematical space for quantum mechanics. In fact, even Paul Dirac, the initiator of the \textit{bra} and \textit{ket} notation, finds the Hilbertian framework restrictive (see \cite{dir}, 1930) :  
 \begin{quoting}[font+=bf,begintext= ,endtext=]
\textit{The bra and ket vectors that we now use form a more general space than a Hilbert space.}
\end{quoting}   
    Here, we invite the reader to consult \cite{dela} and \cite{bow} for a discussion concerning this topic; more specifically the use of a broader setting; namely, the \textit{rigged Hilbert space} (also named Gelfand triplet). By assigning a dimension to the points in the general framework of topological vector spaces, we have succeeded in freeing the notion of dimension from the notion of vector basis or Hilbert basis. Thus, the superposition property of quantum states could be expressed, in a more general setting than Hilbert spaces, directly located at the level of the infinite dimensional points of the quantum states' space.

    \item This hypothetical fundamental particle Diemensionad could be an alternative path to string theory in order to unify quantum mechanics and Einstein's general relativity theory. On the one hand, the Dimensionad should be a very small particle because according to our conception it should be indivisible, which makes it a quantum particle. On the other hand, the Dimensionad is the particle that is supposed to generate space-time and could even be its elementary constituent. It is therefore reasonable to think that if we manage to better understand this particle, we could obtain a good basis to unify these two physical fields.  
    
    \item Since the Dimensionad is supposed to confer dimension to space-time; reasonably, this particle should be involved at the beginning of the Big Bang. In this sense, a better understanding of this particle could eventually contribute to the formalization of the Grand Unified Theory stating that at high energy the three-gauge interactions of the standard model (electromagnetic, weak and strong forces) merge into one force. More precisely, we believe that the Dimensionad could trace the link between fermions and bosons, which could constitute an alternative path to supersymmetry and thus to superstring theory.
    
    \item We can also establish a connection that seems natural with black holes, in particular, concerning the calculation of their dimensions. Indeed, for one, the black hole is supposed to be a point; in this work we have established a framework for the calculation of the dimension of the points. In this respect, the works of Stephen Hawking and Jacob Bekenstein allow us to attribute an entropy to some black holes by analogy with thermodynamics. More precisely, they succeeded in calculating the quantity of information contained in the black hole with the convention that an elementary particle contains one bit of information. However, in this work, we have highlighted the link between the entropy and the dimension. We remain confident that with more time, we will be able to calculate the dimension of some black holes, which we conjecture to be larger than the four dimensions of space-time. 
    
    \item More generally, our conception of dimension could serve in the theory of entropic gravity linking gravity with thermodynamics by describing gravity as an entropic force. We think in particular of the work of Erik Verlinde (see \cite{verl}) combining the thermodynamic approach and the holographic principle to describe gravity.  
    
    \item To bring this article to a close, we have reached a more twisted connection conclusion, adding one more turn of the screw into the research in connection with dimension theory: knowing that the entropy of the universe is increasing and linking the entropy to the dimension, the real question then becomes: did the universe start with a zero dimension and is its dimension in perpetual increase?
\end{enumerate}

\section*{Acknowledgment}  
Many thanks go to Professor Olivier Goubet, from the university of Lille, for the demonstration of the estimation of the functional dimension in the case of Hilbert space. We would like to extend deepest gratitude to Professor Najib Mokhtari, from UIR Sc Po school, for his annotated reading of the article and for his textual edits. We also, would like to share our special appreciation to Dr. Mehdi Najib, from UIR ESIN, for his precious help.

\end{document}